\documentclass[alpha-refs]{wiley-article}

\usepackage{siunitx}
\usepackage{mathtools}
\usepackage{graphicx}
\usepackage{natbib}
\usepackage{xcolor}
\usepackage{multirow}
\usepackage{longtable}

\newcommand{\bigcell}[2]{\begin{tabular}{@{}#1@{}}#2\end{tabular}}

\papertype{Advanced Review}

\title{Text-based Question Answering from Information Retrieval and Deep Neural Network Perspectives: A Survey}

\abbrevs{QA, question answering; KB, knowledge base; CNN, Convolutional Neural Network;  RNN, Recurrent Neural Network; BERT, Bidirectional Encoder Representations from Transformers.}

\author{Zahra Abbasiantaeb, Saeedeh Momtazi}

\affil{Computer Engineering Department, Amirkabir University of Technology (Tehran Polytechnic), Tehran, 1591634311, Iran}

\corraddress{Saeedeh Momtazi, Computer Engineering Department, Amirkabir University of Technology (Tehran Polytechnic), Tehran, 1591634311, Iran}
\corremail{momtazi@aut.ac.ir}

\fundinginfo{}

\runningauthor{Abbasiantaeb et al.}

\begin{document}

\maketitle

\begin{abstract}
Text-based Question Answering (QA) is a challenging task which aims at finding short concrete answers for users' questions. This line of research has been widely studied with information retrieval techniques and has received increasing attention in recent years by considering deep neural network approaches. Deep learning approaches, which are the main focus of this paper, provide a powerful technique to learn multiple layers of representations and interaction between questions and texts. 
In this paper, we provide a comprehensive overview of different models proposed for the QA task, including both traditional information retrieval perspective, and more recent deep neural network perspective. We also introduce well-known datasets for the task and present available results from the literature to have a comparison between different techniques.  

\end{abstract}


\section{Introduction}
\label{intro}
Question Answering (QA) is a fast-growing research problem in computer science that aims to find short concrete answers. There are two major approaches for QA systems: text-based QA, and knowledge-based QA. 
Knowledge-based QAs rely on knowledge bases (KB) for finding the answer to the user's question. Freebase is one of the most popular KBs \citep{Bollacker:2008} which has been widely used as a benchmark in many recent works on knowledge-based QA. KBs include entities, relations, and facts. Facts in the knowledge base are stored in (subject, predicate, object) format where the subject and the object are entities and the predicate is a relation, indicating the relation between the object and the subject. For example, the answer to the question `In which city was Albert Einstein born?' could be stored in a fact like '(Albert Einstein, place-of-birth, Ulm)'. In this task, there are two types of questions: single-relation and multi-relation questions. A simple question is answered by one fact in KB, while the answer of a multi-relation question is found by reasoning over more than one fact in the KB \citep{Yu:2017}. SimpleQuestions and WebQSP are the major datasets of the single-relation and multi-relation questions, respectively.

In text-based QA, the answer to a candidate question is obtained by finding the most similar answer text between candidate answer texts. Consider the question {Q} and set of answers $\{A_{1}, A_{2}, ..., A_{n}\}$, the goal of this system is finding the best answer among these answers. Recent works have proposed different deep neural models in text-based QA which compares two segments of texts and produces a similarity score. In this paper, we focus on this type of QA and review the available methods on text-based QA. 

\begin{figure}[bt]
\centering
  \includegraphics[width=0.95\textwidth]{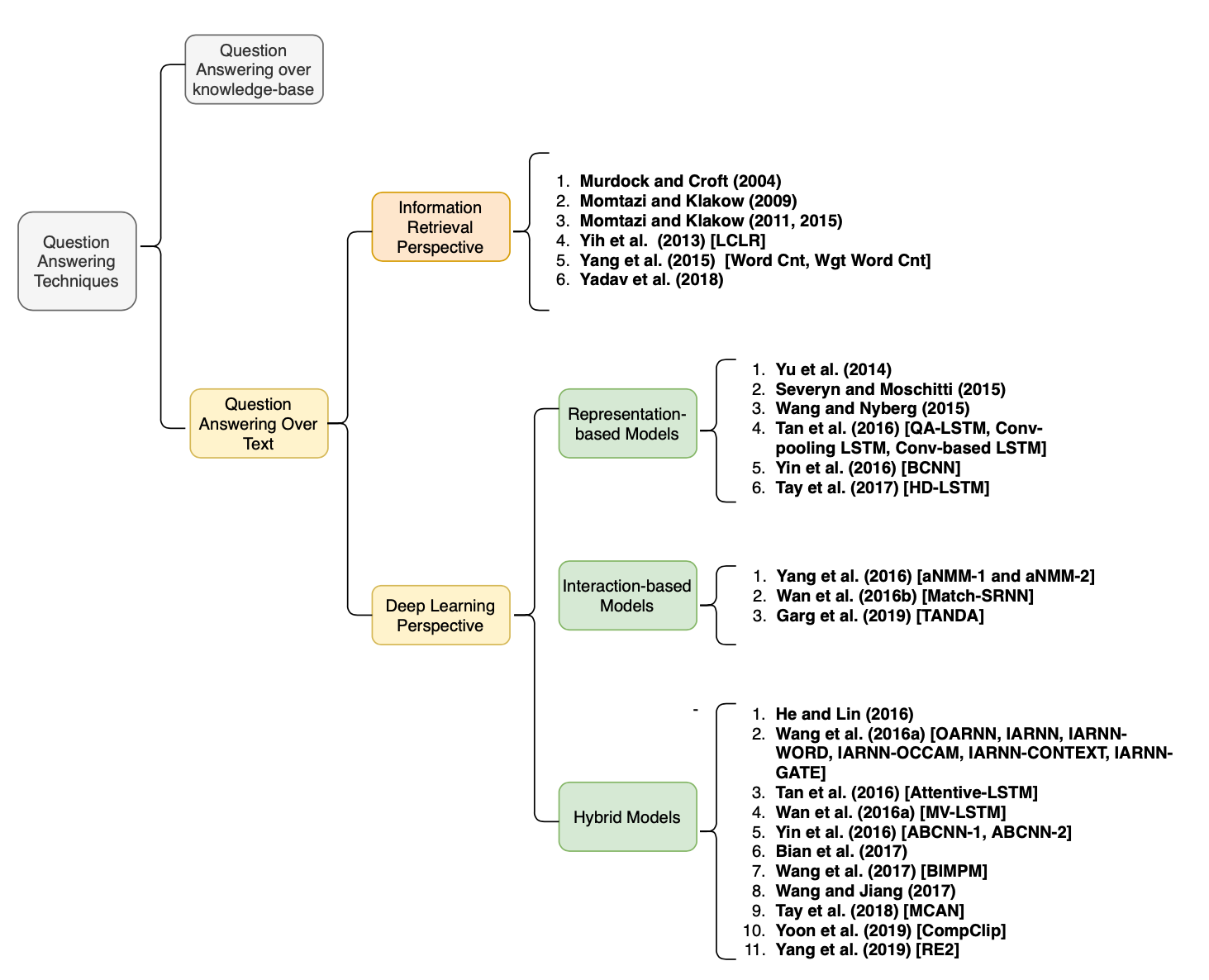}
  \caption{Taxonomy of QA techniques provided in this paper}
  \label{fig:Taxonomy}
\end{figure}

Figure \ref{fig:Taxonomy} presents an overall taxonomy of QA systems including the main representative models of each category.
 \citet{Diefenbach:2018} provided a survey that only covers QA over knowledge-base and contains the proposed models until 2017.  \citet{Soares:2018} is another survey article that does not discuss the proposed models and it also includes the works before 2018. \citet{Kodra:2017} named a wide variety of QA systems and only discusses a small number of proposed models in each type of QA while focusing on neural models. \citep{Wu:2019} is another recent survey which only covers the QA over knowledge-base. \citet{Dimitrakis:2019} provided an overview of different components of a QA system, but it does not discuss the architecture of the proposed models. \citep{Lai:2018} published a survey that provides a different perspective for classifying deep learning methods and does not cover the architecture of proposed models.
 
 Our paper provides a coherent and complete overview of the architecture of the representative models in QA over text from different perspectives, including the state-of-the-art models in this area. Also, noted features of proposed models are discussed in this paper. Moreover, in recent years, pre-trained contextualized language models, such as ELMO, BERT, RoBERTa, and ALBERT, has demonstrated great advances in Natural Language Processing (NLP) downstream tasks including QA, but none of the above-mentioned articles discussed these models.

This paper organizes as follows: In section 2, we present the architecture of QA systems. We discuss information retrieval-based models used for question answer similarity in section 3 and deep learning models in section 4, respectively. In sections 5 and 6, we introduce the most popular QA datasets and evaluation metrics used in QA, respectively. We report and compare the results of reviewed models in section 7 and discuss the paper in section 8. The paper is concluded in section 9.

\section{Architecture of Text-based Question Answering}

\begin{figure}[bt]
\centering
  \includegraphics[width=0.95\textwidth]{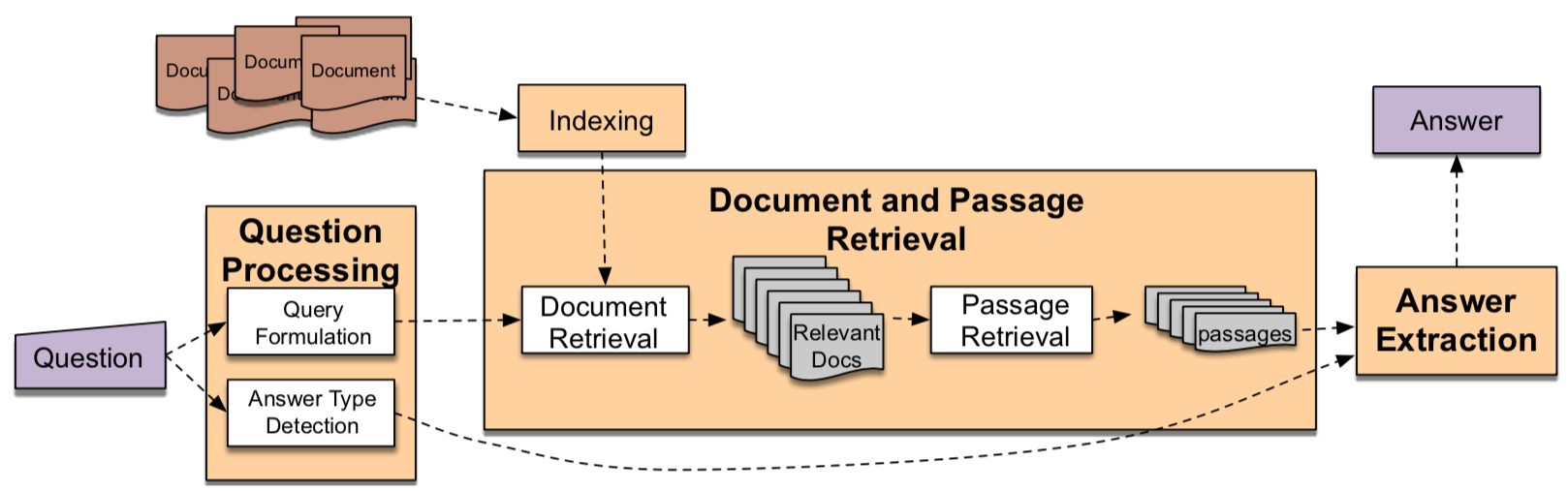}
  \caption{Genaral architecture of text-based QA \citep{Jurafsky:2009}}
  \label{fig:architecture}
\end{figure}

The architecture of Text-based QA, as illustrated in Figure \ref{fig:architecture}, includes three major phases: question processing, document and passage retrieval, and answer extraction. Each of these phases is described below \citep{Jurafsky:2009}.
\begin{enumerate}
\item Question processing: This phase includes two major steps, namely query formulation and answer type detection. In the query formulation step, a query for a given question is generated for retrieving relevant documents by employing an Information Retrieval (IR) engine. The query, which is generated by query reformulation rules, looks like a subset of the intended answer. In the answer type detection step, a classifier is used for classifying questions based on the type of expected answer. Different neural-based or feature-based classifiers can be used in this step.

\item Document and passage retrieval: Generated query in the query formulation step, is passed through an IR engine, and top $n$ retrieved documents are returned. As answer extraction models mostly work on short segments of documents, a passage retrieval model is applied on retrieved documents to receive short segments of text. This is the core component of QA that can find similar passages/sentences to the input question.  
\item Answer extraction: In the final phase of QA, the most relevant answer is retrieved from the given passage. In this step, we need to measure the similarity of the input question and the extracted answer.
\end{enumerate}

As mentioned, estimating the similarity of question and answer sentences is the main important part of text-based QA systems. Since the retrieved sentences are short enough to satisfy users, the output of this step can also be represented to users without any further answer extraction. The similarity of question and answer sentences can be measured by information retrieval or deep learning approaches. In Sections \ref{sec:TextSim_IR} and \ref{sec:TextSim_DL}, we review related works from the information retrieval perspective and the deep learning approaches, respectively.

\section{Question Answer Similarity from Information Retrieval Perspective}
\label{sec:TextSim_IR}
Although lexical-based information retrieval models have been widely used in ad-hoc retrieval, they have less applied to QA tasks because the length of answer sentences is shorter than normal web documents and the vocabulary gap between question and answer sentence in QA is more pronoun than ad-hoc retrieval. It motivated researches to use advanced information retrieval approaches in QA. Some of these approaches are described in this section.
At the end of the section a brief overview of information retrieval approaches is presented in Table \ref{tab:IR_models}.

\textbf{\citet{Yang:2015}}presented the WikiQA dataset for open domain QA. They have evaluated WikiQA and QASent datasets with information retrieval-based models like Word Count (Word Cnt), Weighted Word Count (Wgt Word Cnt), Learning Constrained Latent Representation (LCLR) \citep{Yih:2013}, and Paragraph Vector (PV) \citep{LeAndMikolov:2014}. 
Word Cnt model works by counting the non-stop words in question which also have occurred in the answer sentence. Wgt Word Cnt is the same as Word Cnt, but it also re-weights the counts by Inverse document Frequency (IDF) weight of the question words.
The Idea behind the model proposed by \citet{Yih:2013} is adopting a probabilistic classifier for predicting whether a pair of question and answer are related or not, using semantic model of the question and answer. They used synonym/antonym, hypernym/hyponym and semantic word similarity of each pair of words from question and answer sentences for creating the semantic model. They adopted Learning Constrained Latent Representation (LCLR) \citep{chang:2010} for classifying a pair of question and answer. Details of this classifier is shown in the following equations.

\begin{equation}
\begin{aligned}
& \min _{\theta} \quad \frac{1}{2}\|\theta\|^{2}+C \sum_{i} \xi_{i}^{2}\\
& \text { s.t. } \quad \xi_{i} \geq 1-y_{i} \max _{h} \theta^{T} \phi(x, h)\\
& \underset{h}{\arg \max } \theta^{T} \phi(x, h)
\end{aligned}
\end{equation}

\textbf{\citet{Murdock:2004}} suggested a translation model for QA. In their model, probability of the question ($Q$) given the answer ($A$), denoted as $P(Q|A)$, is computed by the following equations:
\begin{equation}
\begin{aligned}
 p(Q | A)= & \prod_{i=1}^{m}\left[\beta\left(\lambda \sum_{j=1}^{n} p\left(q_{i} | a_{j}\right) p\left(a_{j} | A\right)+(1-\lambda) p\left(q_{i} | C\right)\right)\right. + \\
& (1-\beta)\left(\lambda p\left(q_{i} | D_{A}\right)+(1-\lambda) p\left(q_{i} | C\right)\right) ]\\
\end{aligned}
\end{equation}

\begin{equation}
p\left(q_{i} | a_{j}\right) p\left(a_{j} | A\right)= t_{i} p\left(q_{i} | A\right)+\left(1-t_{i}\right) \sum_{1 \leq j \leq n, a_{j} \neq q_{i}} p\left(q_{i} | a_{j}\right) p\left(a_{j} | A\right)
\end{equation}
where $\lambda$ is the smoothing parameter, $D_A$ is the document containing answer $A$, and $C$ is the collection.
The idea is based on the model proposed by {\citet{Berger:1999}} while different similarity model is used for calculating $P(q_{i}|a_{j})$.

\textbf{\citet{Momtazi:2009}} proposed class-based language models for sentence retrieval in QA. Their class-based language model aims to mitigate the word mismatch problem by finding the relation between words. The Brown word clustering algorithm is adopted for clustering the words in this model and the probability of generating question $Q$, having the answer sentence $S$ is calculated by the following equations:

\begin{equation}
\begin{aligned}
P_{\text { class }}(Q | S)=\prod_{i=1}^{M} P\left(q_{i} | C_{q_{i}}, S\right) P\left(C_{q_{i}} | S\right)
\end{aligned}
\end{equation}
\[P_{\text { class }}\left(C_{q_{i}} | S\right)=\frac{f_{S}\left(C_{q_{i}}\right)}{\sum_{w} f_{S}(w)}\]
where $P (q_{i} | Cq_{i}, S)$ is the probability of term $q_{i}$ having its cluster $(C_{q_{i}})$ and answer sentence model $(S)$, $P (C_{q_{i}} | S)$ is the probability of cluster $C_{q_{i}}$ given the sentence model $S$, $f_{s} (C_{q_{i}})$ is the number of occurrences of all the words in the cluster of term $q_{i}$ in sentence $S$, and $w$ represents the vocabulary words.

\textbf{\citet{momtazi:2011,momtazi:2015}} proposed a trained trigger language model. In their model, the word mismatch problem is mitigated by using the contextual information between words. The Idea behind this model is to find the pair of trigger and target words, whereas appearance of a target word in answer sentence and trigger word in question sentence means as a relation between the related words. They have trained a model for extracting these trigger-target pairs from a corpus, and this model is used for calculating the probability of question word $q_i$, having the answer $S$, denoted as $P(q_i|S)$,  as follows:

\begin{equation}
P_{\text { trigger }}\left(q_{i} | S\right)=\frac{1}{N} \sum_{j=1}^{N} P_{\text { trigger }}\left(q_{i} | s_{j}\right)
\end{equation}

\[P_{\text { trigger }}\left(q_{i} | s_{j}\right)=\frac{f_{C}\left(q_{i}, s_{j}\right)}{\sum_{q_{r}} f_{C}\left(q_{i}, s_{j}\right)}\]

\noindent
where $s_j$  and $q_{i}$ denote the $j^{th}$ term in answer sentence, and $i^{th}$ term in the question sentence, respectively. $f_{c}(q_{i}, s_j)$ is the number of times term $q_{i}$ triggers term $s_{j}$ in the model created upon corpus $C$. $P (q_{i} | S)$ is the probability of $q_{i}$ having the sentence $S$, and $N$ is sentence length.

Having the above probabilities, the probability of question ($Q$), having the answer( $S$), is calculated as follows:

\begin{equation}
\begin{aligned}
P_{\text { trigger }}(Q | S)=\left(\frac{1}{N}\right)^{M} \prod_{i=1}^{M} \sum_{j=1}^{N} \frac{f_{C}\left(q_{i}, s_{j}\right)}{\sum_{q} f_{C}\left(q_{i}, s_{j}\right)}
\end{aligned}
\end{equation}

\noindent
where $M$ is question length.

\textbf{\citet{Yadav:2018}} proposed a model which for each question-answer pair calculates a matching score in three steps. In the first step, IDF weight of each word is calculated by the following equation:

\begin{equation}
\begin{aligned}
i d f\left(q_{i}\right)=\log \frac{N-\operatorname{docfreq} \left(q_{i}\right)+0.5}{\operatorname{docfreq}\left(q_{i}\right)+0.5}
\end{aligned}
\end{equation}
where $N$ is the count of questions, and $docfreq(q_{i})$ is the number of questions which word $q_{i}$ has occurred in. In the second step, one-to-many alignments are performed between terms in question and answer. Cosine similarity between Glove word embedding \citep{Pennington:2014} of each question word $q_{i}$ and each answer word $a_{i}$ is considered as their similarity. Then the top $K^{+}$ most similar words $\{a^{+}_{q^{i},1}, a^{+}_{q^{i},2}, a^{+}_{q^{i},3}, ..., a^{+}_{q^{i},K^{+}} \} $ and $K^{-}$ least similar words $\{a^{-}_{q^{i},1}, a^{-}_{q^{i},2}, a^{-}_{q^{i},3}, ..., a^{-}_{q^{i},K^{+}} \} $ of answer are found. Finally in the third step the similarity score $S(Q, A)$ between each question and answer sentence is calculated by the following equations:

\begin{equation}
\begin{aligned}
s(Q, A)=\sum_{i=1}^{N} i d f\left(q_{i}\right) \cdot \operatorname{align}\left(q_{i}, A\right)
\end{aligned}
\end{equation}
\[\text {align}\left(q_{i}, A\right)=\operatorname{pos}\left(q_{i}, A\right)+\lambda \cdot \operatorname{neg}\left(q_{i}, A\right)\]
\[\operatorname{pos}\left(q_{i}, A\right)=\sum_{k=1}^{K^{+}} \frac{1}{k} \cdot a_{q_{i}, k}^{+}\]
\[\operatorname{neg}\left(q_{i}, A\right)=\sum_{k=1}^{K^{-}} \frac{1}{k} \cdot a_{q_{i}, k}^{-}\]
where $align(q_{i}, A)$ is the alignment score between $q_{i}$ and answer $A$, $\lambda$ is the negative information's weight, $pos(q_{i}, A)$ and $neg(q_{i}, A)$ represent the one-to-many alignment score for the $K^{+}$ most and $K^{-}$ least similar words. They have also proposed two other baselines single-alignment (one-to-one), and one-to-all. In one-to-one approach, just the single alignment score $K^{+} = 1$, the most similar word, is used. In the one-to-all approach, similarity between the question term $q_{i}$ and all the answer terms is considered with the same weight in calculating $align(q_{i}, A)$ which changes the $align(q_{i}, A)$ to the following equation:

\begin{equation}
\begin{aligned}
\operatorname{align}\left(q_{i}, A\right)=\sum_{k=1}^{m} \frac{1}{k} \cdot \operatorname{cosSim}\left(q_{i}, a_{q_{i}, k}^{+}\right)
\end{aligned}
\end{equation}
where $M$ is the count of the words in the answer sentence.

\begin{table}[h]
\caption{Overview of QA Models from Information Retrieval Perspective}
\label{tab:IR_models}
\begin{center}
\begin{small}
\begin{tabular}{| l | c | c |}

\hline\noalign{\smallskip}
	 Model &  Main Idea & Datasets  \\
\noalign{\smallskip}\hline\noalign{\smallskip}

\citet{Yang:2015} Word Cnt		& \bigcell{c}{uses count of co-occurred words in \\ question and answer sentences}   &  WikiQA  \\  \hline
\citet{Yang:2015}  Wgt Word Cnt	& \bigcell{c}{uses  Inverse document Frequency (IDF) for \\ re-weighting the counts in Word Cnt model }   &  WikiQA \\  \hline
\citep{Yih:2013}  LCLR 	& \bigcell{c}{uses probabilistic classifier for classifying \\ question and answer pairs }   & \bigcell{c}{WikiQA \\ TREC-QA}  \\  \hline
\citet{Murdock:2004}	  	 &\bigcell{c}{uses translation model for predicting \\ probability of question given the answer}   &  TREC-QA   \\  \hline
\citet{Momtazi:2009}	 & \bigcell{c}{uses class-based language model for \\ mitigating the word mismatch problem}   &   TREC-QA \\  \hline
\citet{momtazi:2011,momtazi:2015}	& \bigcell{c}{uses a trained trigger language model \\ for mitigating the word mismatch problem by \\ using the contextual information \\ between words}   & TREC-QA   \\  \hline	
\citet{Yadav:2018} & \bigcell{c}{uses one-to-many alignments, and IDF weight \\ for calculating the matching score of \\ question-answer pair}   &    \\  \hline

 \hline
\end{tabular}
\end{small}
\end{center}
\end{table}



\section{Question Answer Similarity from Deep Learning Perspective}
\label{sec:TextSim_DL}
Deep learning models can be divided into three major categories: representation-based, interaction-based, and hybrid \citep{Guo:2016}. Representation-based models construct a fixe$d$-dimensional vector representation for both the question and the candidate answer separately and then perform matching within the latent space. Interaction-based models compute the interaction between each individual term of question and candidate answer sentences where interaction can be identity or syntactic/semantic similarity. Hybrid models combine both interaction and representation models. They consist of a representation component that combines a sequence of words into a fixe$d$-dimensional representation and an interaction component. These components could occur in parallel or serial. In this section, we will review the structure of the proposed deep neural models and specify the type of model according to the mentioned categories. Similar to the previous section, we provide a brief overview of deep learning-based models at the end of this section.


\subsection{Representation-based Models}
\label{sec:Rep-based}


\textbf{\citet{Yu:2014}} proposed a generative neural network-based model for binary classification of each question/answer pair is related or not. This model captures the semantic features of question and answer sentences. Each sample is represented with a triple $(q_{i} , a_{ij} , y_{ij} )$ where $q_{i}  \in Q$ is question, $a_{ij}$ is a candidate answer for question $q_{i}$, and label $y_{ij}$ shows whether $a_{ij}$ is a correct answer for $q_{i}$ or not. For each answer, a related question is generated and then the semantic similarity of generated question and the given question is captured by using the dot product. This similarity is used for predicting whether the candidate answer is a correct answer for the given question or not. The probability of the answer being correct is formulated as: 

\begin{equation}
\begin{aligned}
\mathrm{P}(\mathrm{y}=1 | \mathrm{q}, \mathrm{a})=\sigma\left(\mathrm{q}_{\mathrm{m}}^{\mathrm{T}} \mathrm{M} \mathrm{a}+\mathrm{b}\right)
\end{aligned}
\end{equation}
where  $q^{\prime}=M a$ is the generated question. The model is trained by minimizing the cross-entropy of all labeled data QA pairs as: 

\begin{equation}
\begin{aligned}
\mathcal{L}=-\log \prod_{n} p\left(y_{n} | {q}_{n}, {a}_{n}\right)+\frac{\lambda}{2}\|\theta\|_{F}^{2}
\end{aligned}
\end{equation}
\[=-\sum_{n} y_{n} \log \sigma\left({q}_{n}^{T} f{M} {a}_{n}+b\right)+\left(1-y_{n}\right) \log \left(1-\sigma\left({q}_{n}^{T} {M} {a}_{n}+b\right)\right)+\frac{\lambda}{2}\|\theta\|_{F}^{2}\]
where $\|\theta\|^{2} _{F}$ is the Frobenius norm of $\theta$. 

Each sentence is modeled by the bag of words and bigram approaches. In the bag of words model, a sentence is represented by averaging embeddings of all the words (except stop words) within it. The bigram model has the ability to capture features of bigrams independent of their positions in the sentence. As the architecture of the bigram model is shown in Figure \ref{fig:Yu-2014}, one convolutional layer, and one pooling layer are used for modeling the sentence in the bigram model. Every bigram is projected into a feature value $c_{i}$, which is computed as:

\begin{figure}[bt]
\centering
    \includegraphics[width=0.5\textwidth]{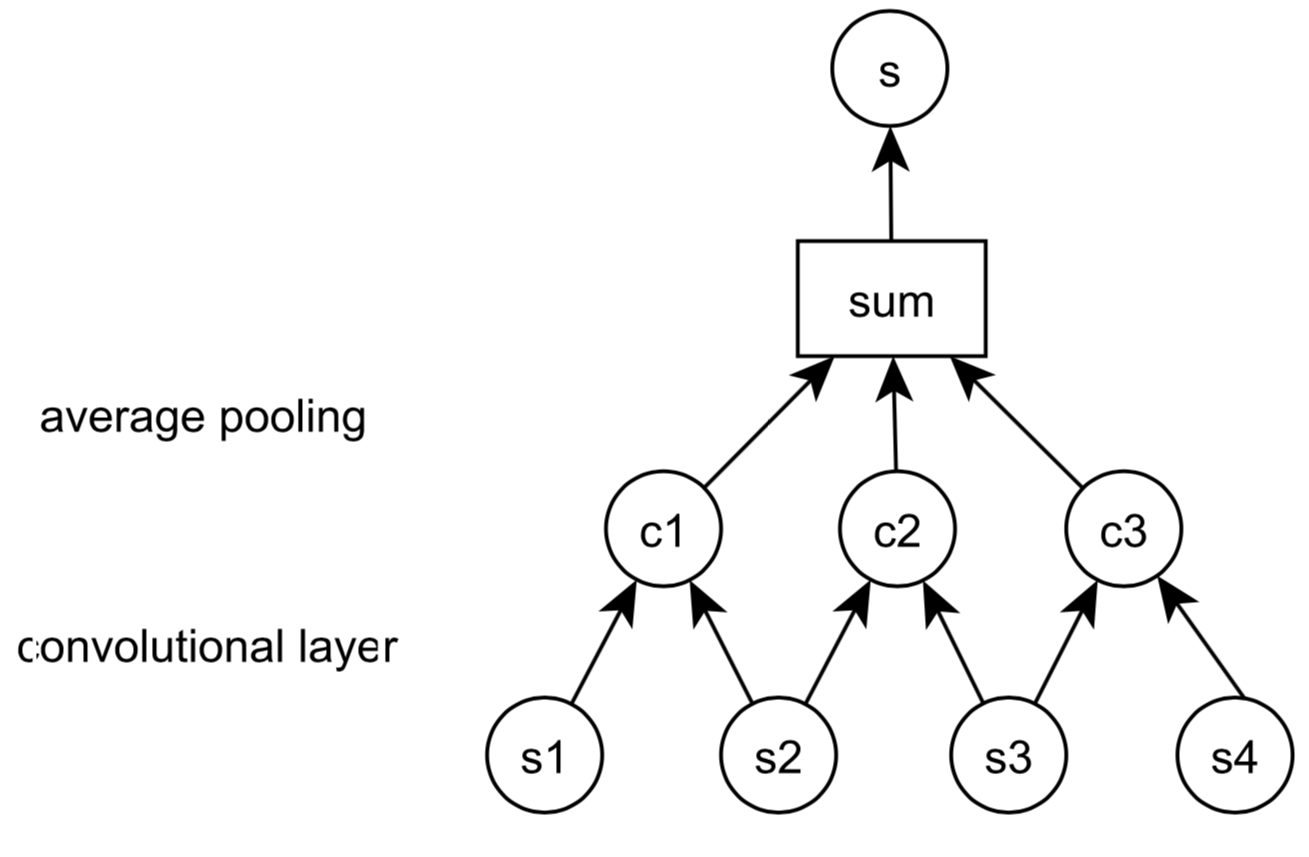}
  \caption{Architecture of \citet{Yu:2014} model}
  \label{fig:Yu-2014}

\end{figure}

\begin{equation}
\begin{aligned}
{c}_{\mathrm{i}}=\tanh \left(\mathrm{T} \cdot \mathrm{s}_{\mathrm{i} :i+1}+\mathrm{b}\right)
\end{aligned}
\end{equation}
where $s$ is the vector representation of the sentence. All bigram features are combined in average pooling layer and finally a full-sentence representation with the same dimensionality as the initial word embeddings are produced by the following equation:

\begin{equation}
\begin{aligned}
{s}=\sum_{i=1}^{|{s}|-1} \tanh \left({T}_{L} {s}_{i}+{T}_{R} {s}_{i+1}+{b}\right)
\end{aligned}
\end{equation}

\begin{figure}[bt]
\centering
    \includegraphics[width=0.7\textwidth]{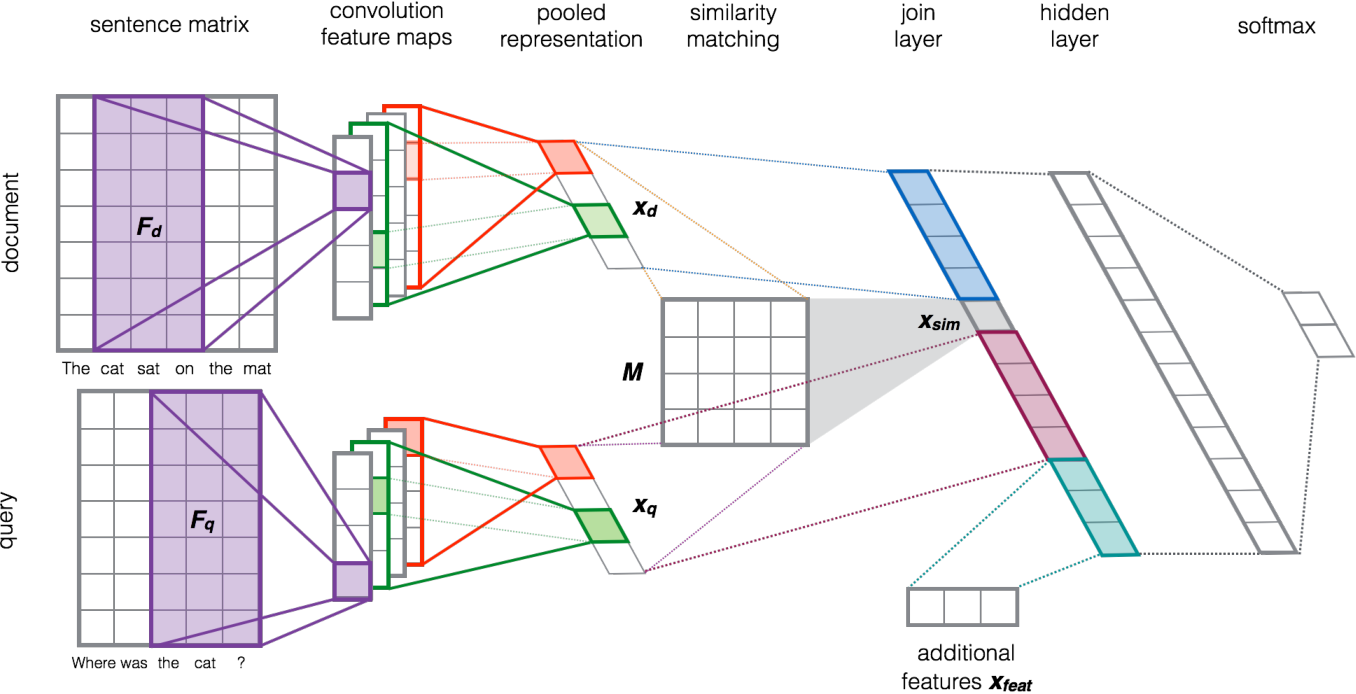}
  \caption{Architecture of \citet{Severyn:2015} model}
  \label{fig:Severyn2015}

\end{figure}

\textbf{\citet{Severyn:2015}} proposed a framework for answer sentence selection. They divided their task into two main subtasks: (1) mapping the original space of words to a feature space encoding, and (2) learning a similarity function between pairs of objects. They used a Convolutional Neural Network (CNN) architecture for learning to map input text, either query or document, to a vector space model. For the second part, they used the idea of noisy channel approach for finding a transformation of the document to be as close as possible to the query:
$Sim (x_{q}, x_{d}) = x_{q}^{T} M x_{d}$.
To this end, they used a neural network architecture to train the similarity matrix $M$. According to Figure \ref{fig:Severyn2015}, the vector representation of the query and the document that is derived from the first CNN model are jointly fed to the second CNN to train and build the similarity matrix.

\begin{figure}[bt]
\centering
    \includegraphics[width=0.75\textwidth]{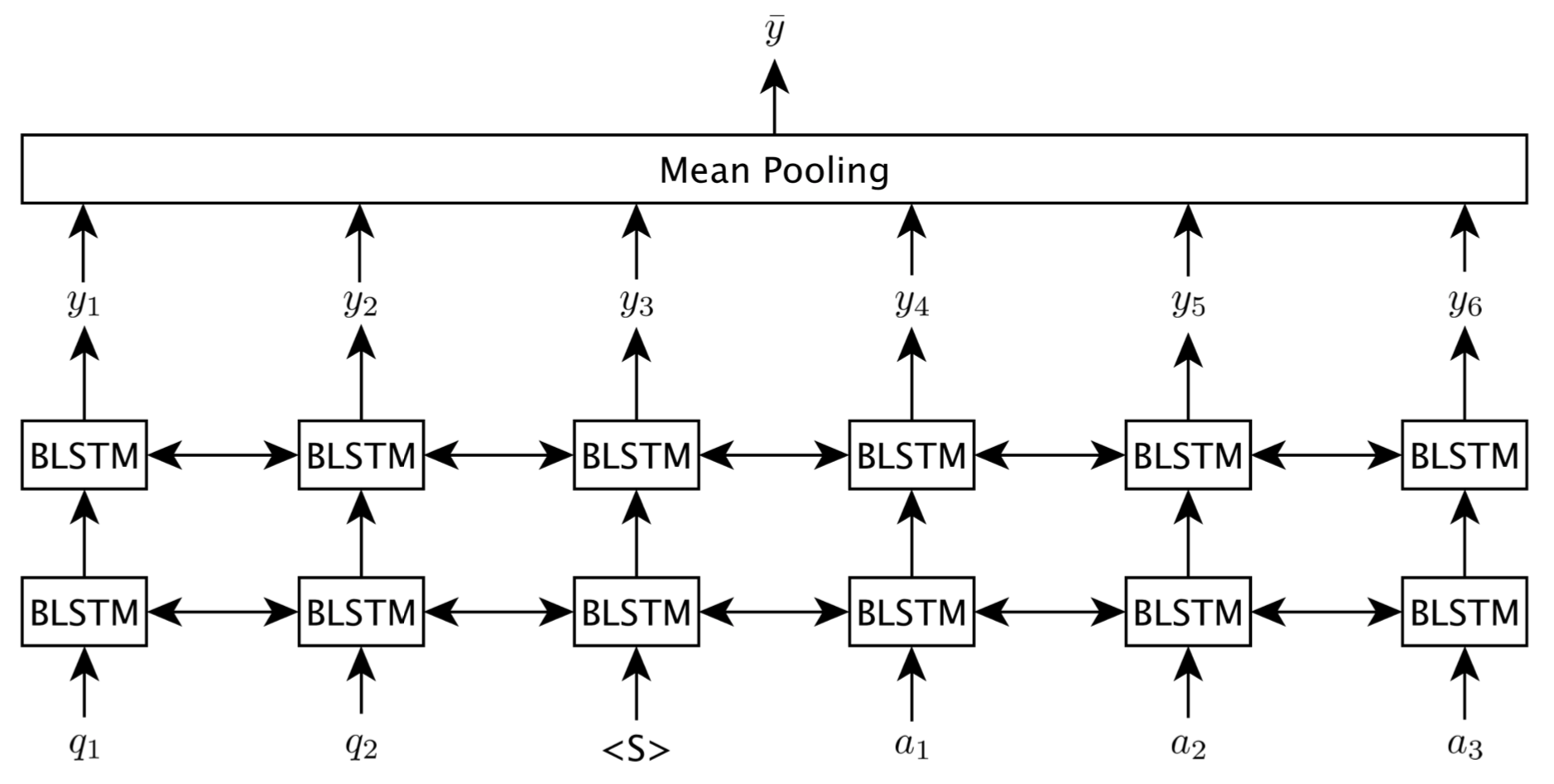}
  \caption{Architecture of \citet{wang:2015} model}
  \label{fig:wang:2015}
\end{figure}

\textbf{\citet{wang:2015}} used a multilayer stacked Bidirectional Long Short-term Memory (BiLSTM) for answer sentence selection task. As represented in Figure \ref{fig:wang:2015}, a sequence of word2vec representation of question and answer sentence terms are fed to this model. Symbol, $<S>$, is placed between question $q$ and answer $a$ for distinguishing the question and answer. Among different Recurrent Neural Network (RNN) architectures, stacked BiLSTM is chosen as first bidirectional RNN extracts the contextual information of question and answer pair from both directions, in other words, it uses the future information, second stacked BiLSTM provides better results due to its ability in extracting higher levels of abstraction, and third LSTM is a more complicated RNN block which mitigates the gradient vanishing problem of standard RNNs. The final output of each time step indicates whether the given answer is a correct answer for the question or not.

In this model, the stacked BiLSTM relevance model is combined by Gradient Boosted Regression Tree (GBDT) method for exact matching the proper nouns and cardinal numbers in question and answer sentences.

\begin{figure}[bt]
\centering
    \includegraphics[width=0.55\textwidth]{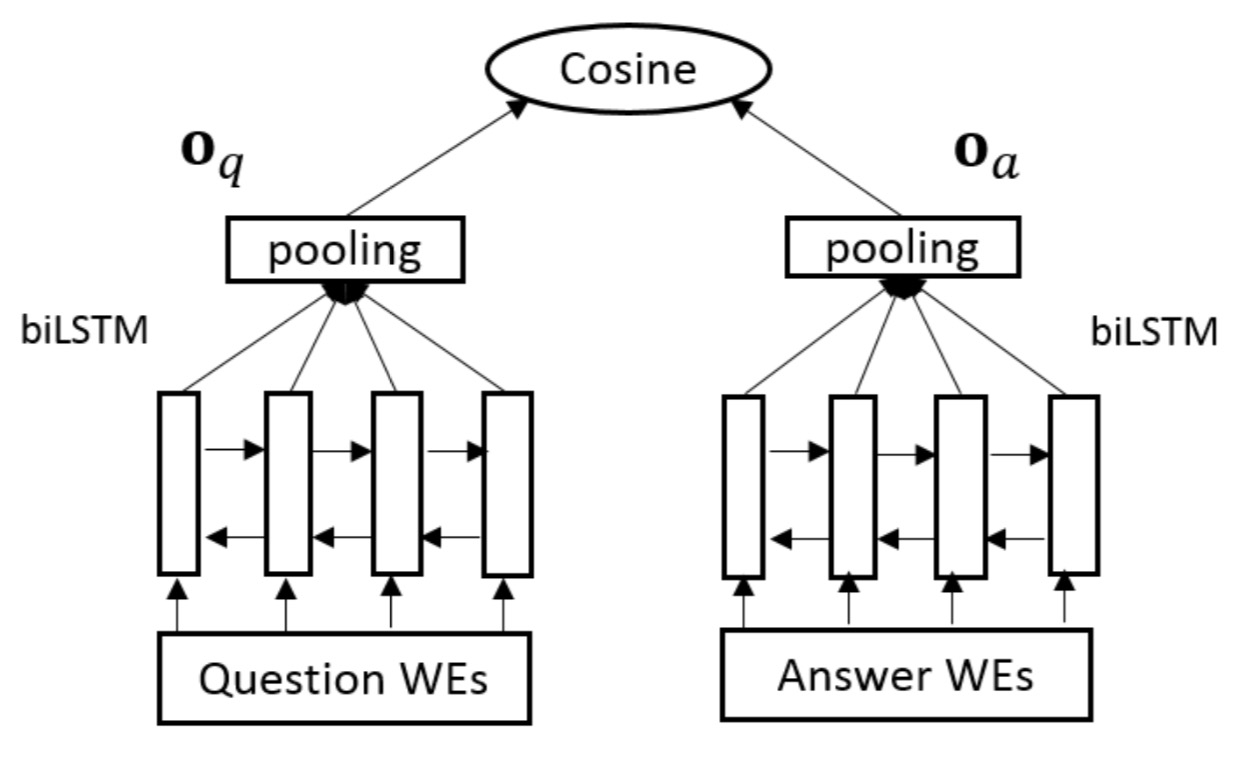}
  \caption{Architecture of QA-LSTM basic model \citep{Tan:2016} }
  \label{fig:Basic-QA-LSTM}
\end{figure}

\textbf{\citet{Tan:2016}} proposed a basic model called QA-LSTM, shown in Figure \ref{fig:Basic-QA-LSTM}, for sentence matching. According to this figure, in the basic model, word embeddings of question and answer sentences are fed into a BiLSTM network. A fixed-size representation is obtained for each sentence in three different ways: (1) concatenating the last output of both directions, (2) average pooling and max pooling over all the outputs of the BiLSTM, and finally, (3) using the cosine similarity, semantic matching between question and answer sentences are scored. LSTM is a powerful architecture in capturing long-range dependencies, but it suffers from not paying attention to local $n$-grams. Although convolutional structures pay more attention to local $n$-grams, they do not consider long-range dependencies. Therefore each of the CNN and RNN blocks has its own pros and cons. Three different variants of the basic QA-LSTM are proposed in this work which one of them belongs to the hybrid models and is described in section \ref{sec:hybrid-models}. In the following, two other variants of QA-LSTM, belonging to the representation-based models, are described.

\begin{enumerate}

\item Convolutional pooling LSTM: As is shown in Figure \ref{fig:Conv-pooling-LSTM}, the pooling layer is replaced with a convolutional layer for capturing richer local information and on top of this layer an output layer is placed for generating a representation of the input sentence. Representation of the input sentence is generated by the following equations:

\begin{equation}
\begin{aligned}
C = tanh ( W_{cp} Z )   ,   [ O_{j} ] = max_{1<l<L} [C_{j,l}]
\end{aligned}
\end{equation}
where $Z \in R^{k|h| \times L}$and $m$-th column is generated by concatenation of the $k$ hidden vectors of BiLSTM centralized in the $m$-th token of the sequence, $L$ is the length of the sequence, and $W_{cp}$ is the network parameter.

\begin{figure}[bt]
\centering
    \includegraphics[width=0.55\textwidth]{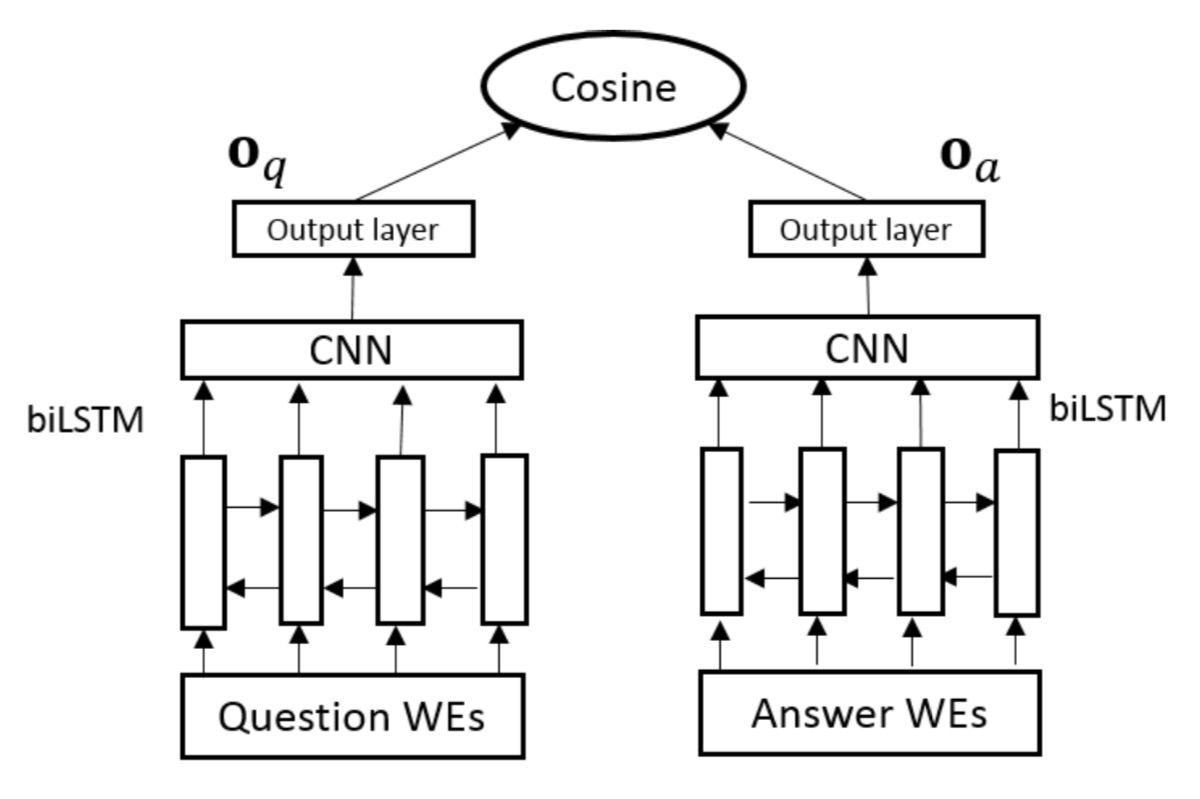}
  \caption{Architecture of Conv-pooling LSTM model \citep{Tan:2016} }
  \label{fig:Conv-pooling-LSTM}
\end{figure}

\item Convolutional-based LSTM: Architecture of this model is shown in Figure \ref{fig:Conv-based-LSTM}. In this model, word embeddings are first fed to a CNN for retrieving the local $n$-gram interactions at the lower level. The output of the convolution is then fed into the BiLSTM network for capturing long-range dependencies. Max pooling is used over the output of the BiLSTM for producing the sentence representation. The output of the convolution layer, named $X$, is obtained by the following equation:

\begin{equation}
\begin{aligned}
X = tanh (W_{cb} D)
\end{aligned}
\end{equation}
where $D \in R ^{kE \times L}$ is input of this model and column $l$ of $D$ is concatenation of $k$ word vectors of size $E$ centered at the $l$-th word.

\begin{figure}[bt]
\centering
    \includegraphics[width=0.6\textwidth]{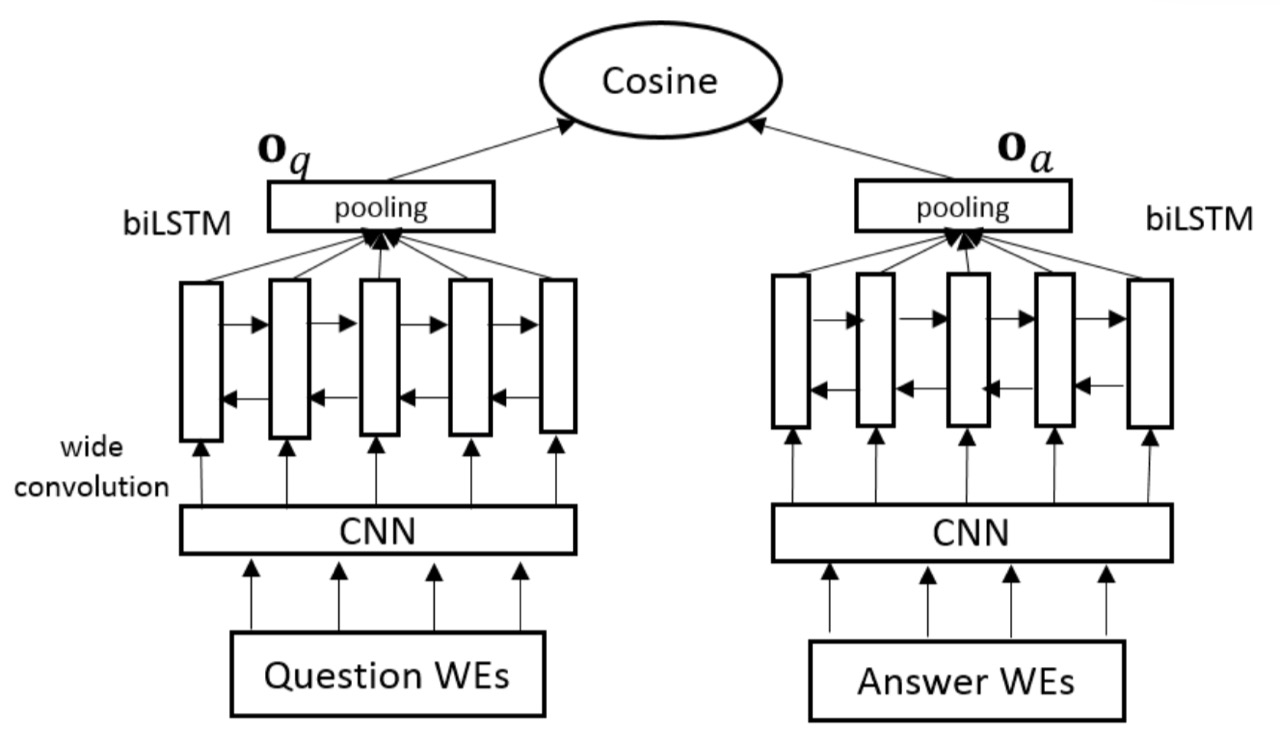}
  \caption{Architecture of Conv-based LSTM model \citep{Tan:2016} }
  \label{fig:Conv-based-LSTM}
\end{figure}
\end{enumerate}


\textbf{\citet{Yin:2016}} proposed a Basic CNN (BCNN) model and three Attention-based CNN (ABCNN) models for text matching. ABCNN models belong to the hybrid category and are described in section \ref{sec:hybrid-models}. In the following, the architecture of BCNN is described.

\textbf{Basic CNN (BCNN):} Architecture of this model is shown in Figure \ref{fig:BCNN}. This model is based on the Siamese architecture {\citep{Bromley:1993}}. The model provides a representation of each sentence using convolutional, $w-ap$ pooling and $all-ap$ pooling layers, and then compares these two representations with logistic regression. Different layers in BCNN are as follows:

\begin{figure}[bt]
\centering
    \includegraphics[width=9cm]{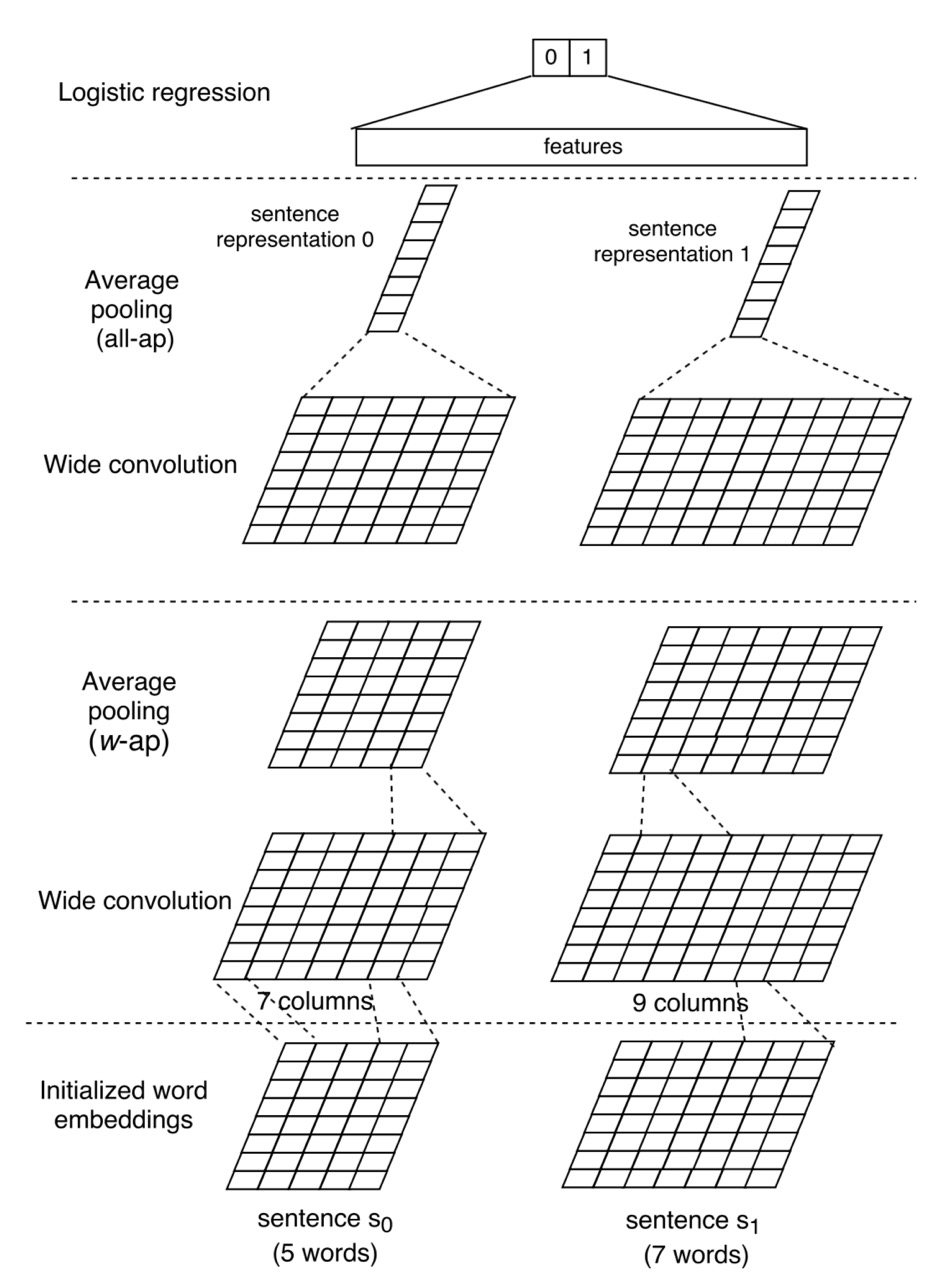}
  \caption{Architecture of BCNN model \citep{Yin:2016}}
  \label{fig:BCNN}
\end{figure}

\begin{enumerate}

\item Input layer: Each sentence is passed to the model with a $d_{0}\times s$ matrix, where $d_{0}$ is the dimension of word2vec {\citep{Mikolov:2013}}  embedding of each word and $s$ is the maximum length of the two sentences (the shorter sentence is padded to the larger sentence length).

\item Convolution layer: Embedding of words within a sentence with window size of $w$ are concatenated and represented as $c_{i} \in R^{w . d_{0}}$ where $0<i<s+w$ ($s$ is length of the sentence). Then each $c_{i}$ is converted to $p_{i}$ by the following equation:

\begin{equation}
\begin{aligned}
P_{i} = tanh (W.c_{i} + b)
\end{aligned}
\end{equation}
let $W \in R ^{d_{1} \times w.d_{0}}$ be the convolution weights, and $b \in R^{d_{1}}$ be the bias.

\item Average pooling layer: This model utilizes two types of average pooling, namely $all-ap$ and $w-ap$, for extracting the robust features from convolution. $All-ap$ pooling is used in the last convolution layer and calculates the average of each column. $W-ap$ pooling is used in the middle convolution layers and calculates the average of each $w$ consecutive columns.

\item Output layer: In the output layer logistic regression is applied to final representations in order to classify the question and answer sentences as related or not.

\end{enumerate}


\textbf{\citet{Tay:2017}} proposed Holographic-dual LSTM (HD-LSTM), a binary classifier model for QA task. As is shown in Figure \ref{fig:HD-LSTM}, HD-LSTM consists of four major parts. In the representation layer, two multi-layered LSTMs denoted as Q-LSTM and A-LSTM are used for learning the representations of the question and answer. A holographic composition is used for measuring the similarity of the outputs of Q-LSTM and A-LSTM. Finally, a fully connected hidden layer is used for performing the binary classification of the QA pair as correct or incorrect. Each part of HD-LSTM is described in the following:

\begin{enumerate}
\item Learning QA Representations: Instead of learning word embeddings, pre-trained weights of SkipGram embeddings {\citep{Mikolov:2013}} denoted as $W$ are used in this layer. Embeddings of both the question and the answer sequences are fed into Q-LSTM and A-LSTM. Representation of question and answer is generated in the last hidden output of Q-LSTM and A-LSTM.

\begin{figure}[bt]
\centering
    \includegraphics[width=0.98\textwidth]{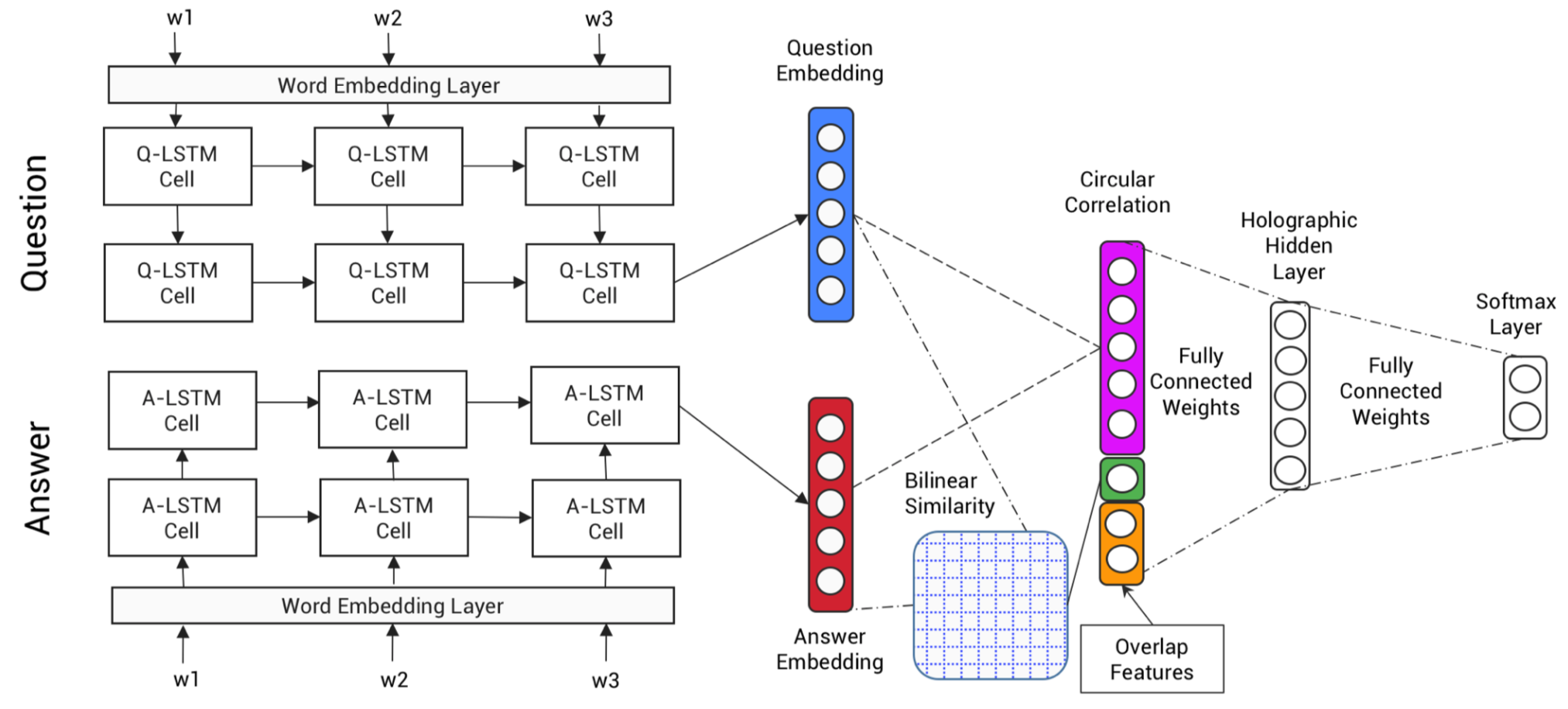}
  \caption{Architecture of HD-LSTM model \citep{Tay:2017}}
  \label{fig:HD-LSTM}
\end{figure}

\item Holographic Matching of QA pairs: Embeddings of the question and answer which learned in the previous layer are passed into the holographic layer and circular correlation of vectors is used for modeling the similarity of them. The similarity of question and answer is modeled by the following equation:

\begin{equation}
\begin{aligned}
[q \star a]_{k}=\sum_{i=0}^{d-1} q_{i} a_{(k+i) \bmod d} \\
q \star a=\mathcal{F}^{-1}(\overline{\mathcal{F}(q)} \odot \mathcal{F}(a))
\end{aligned}
\end{equation}
where $F$ is Fast Fourier transform, $q$ is question, $a$ is answer, and $d$ is the dimension of embeddings. In the above equation, question and answer embeddings must have the same dimension.

\item Holographic Hidden Layer: This is a fully connected dense layer. Input and output of this layer are $[ [q \star a], Sim (q, a), X_{f e a t} ]$ and $h_{out}$, respectively. $(X_{f e a t})$ is word overlap feature, and $Sim (q, a)$ is bilinear similarity function between $q$ and $a$ which is defined as: 
\begin{equation}
\begin{aligned}
\operatorname{Sim}(q, a)=\vec{q}^{T} M \vec{a}
\end{aligned}
\end{equation}

where $M \in R^{n \times n}$ is a similarity matrix between $q \in R^{n}$ and $a \in R^{n}$.
Concatenation of $Sim (q, a)$ with $[q \star a]$ makes the model perform worse. So, in order to mitigate this weakness $X_{f e a t}$ is concatenated to make the model work better. 

\item SoftMax layer: A softMax layer with the following equation is used at last: 

\begin{equation}
\begin{aligned}
\mathrm{P}=\text { SoftMax }\left(\mathrm{W}_{\mathrm{f}}. \mathrm{h}_{\mathrm{out}}+\mathrm{b}_{\mathrm{f}}\right) 
\end{aligned}
\end{equation}
where $W_{f}$ and $b_{f}$ are network parameters.
\end{enumerate}


\subsection{Interaction-based Models}


\textbf{\citet{Yang:2016}} proposed aNMM-1 and aNMM-2 neural matching models. ANMM-1 works in three major steps as follows:
\begin{enumerate}
\item Building QA matching matrix: Each cell in this matrix represents the similarity of the corresponding question and answer words. The similarity is calculated by the dot product of the normalized word embeddings.

\item Learning semantic matching: Various length of answer sentences results in variable size for the QA matrix. To fix this problem, value shared weights method is used. In value shared weights method, each node is weighted based on its value where the value of a node represents the similarity between two words. Input to the hidden layer for each question term is defined as follows:

\begin{equation}
\begin{aligned}
h_{j}=\delta\left(\sum_{k=0}^{K} w_{k} \cdot x_{j k}\right)
\end{aligned}
\end{equation}

where $j$ is the index of the question term, $w_{k}$ is the model parameter, and $x_{j k}$ is the sum of all matching signals within the range $k$ (the range of possible matching signals is divided to $k$ equal bins, and each matching score is assigned to one bin).
\item Question attention network: An attention layer with question word embedding weights is applied to hidden states $h_{j}$. Finally match score is computed by the following equation:

\begin{equation}
\begin{aligned}
y=\sum_{j=1}^{M} g_{j} \cdot h_{j}=\sum_{j=1}^{M} \frac{\exp \left({v} \cdot {q}_{j}\right)}{\sum_{l=1}^{L} \exp \left({v} \cdot {q}_{l}\right)} \cdot \delta\left(\sum_{k=0}^{K} w_{k} \cdot x_{j k}\right)
\end{aligned}
\end{equation}
where $v$ is the model's parameter and dot product of the question word embedding and $v$ are fed to the softMax function.

\end{enumerate}

\begin{figure}[bt]
\centering
    \includegraphics[width=0.95\textwidth]{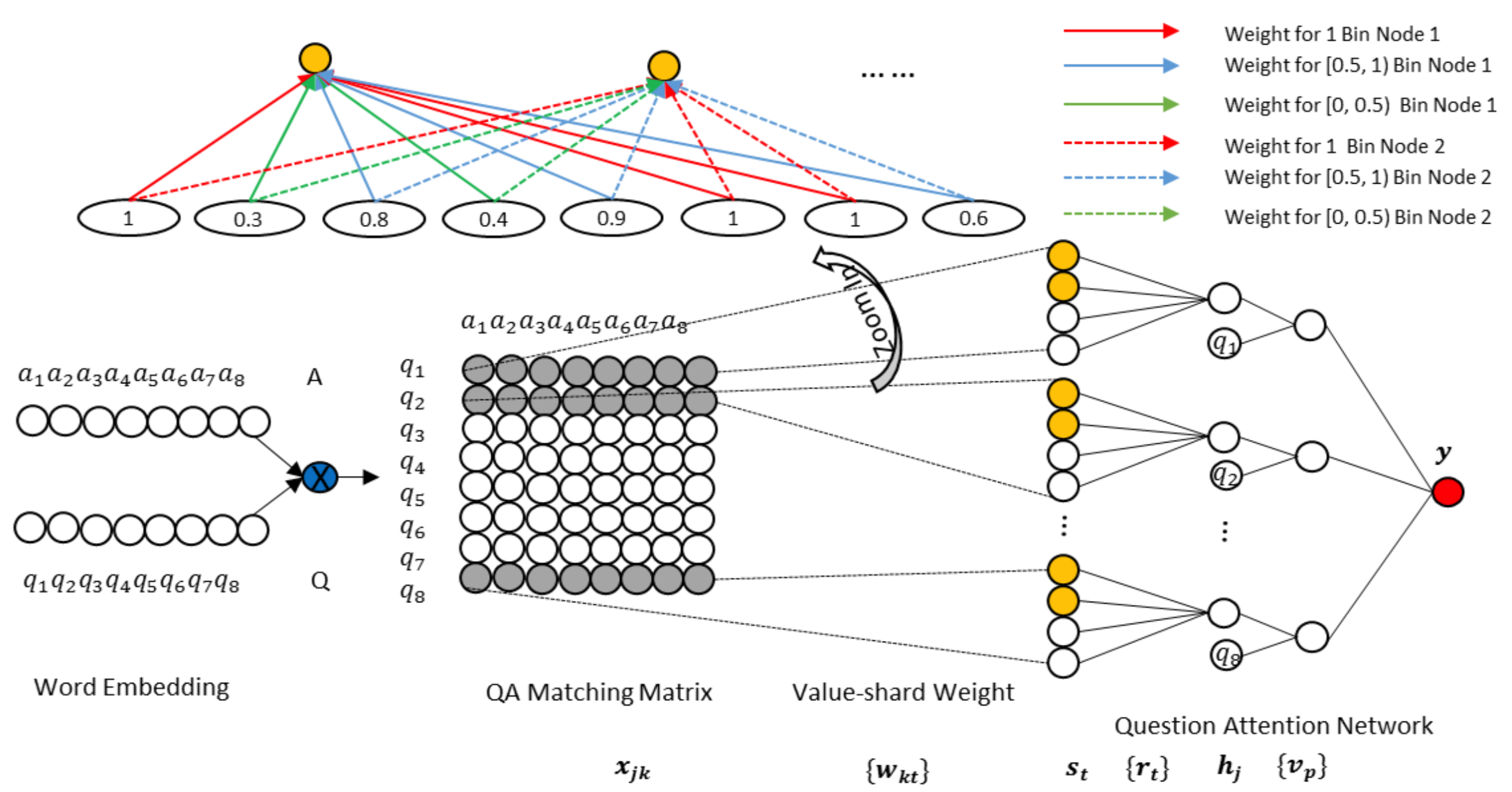}
  \caption{Architecture of ANMM-2 model \citep{Yang:2016} }
  \label{fig:ANMM2}
\end{figure}

In aNNM-2, more than one value-shared weights are used for each question answer matching vector then in the first hidden layer, there are multiple intermediate nodes. As architecture of aNMM-2 is shown in Figure \ref{fig:ANMM2}, the final output of the model $y$ is defined as: 

\begin{equation}
\begin{aligned}
y=\sum_{j=1}^{M} \tau\left({v} \cdot {q}_{j}\right) \cdot \delta\left(\sum_{t=0}^{T} r_{t} \cdot \delta\left(\sum_{k=0}^{K} w_{k t} x_{j k}\right)\right)
\end{aligned}
\end{equation}
where $T$ is the number of nodes in hidden layer 1, $r_{t}$ is the model parameter from hidden layer 1 to hidden layer 2, and $\tau(v, q_{j})$ is calculated as follows: 

\begin{equation}
\frac{\exp (v . q j)}{\sum_{l=1}^{L} \exp (v . q l)}
\end{equation}

\begin{figure}
\centering
    \includegraphics[width=0.65\textwidth]{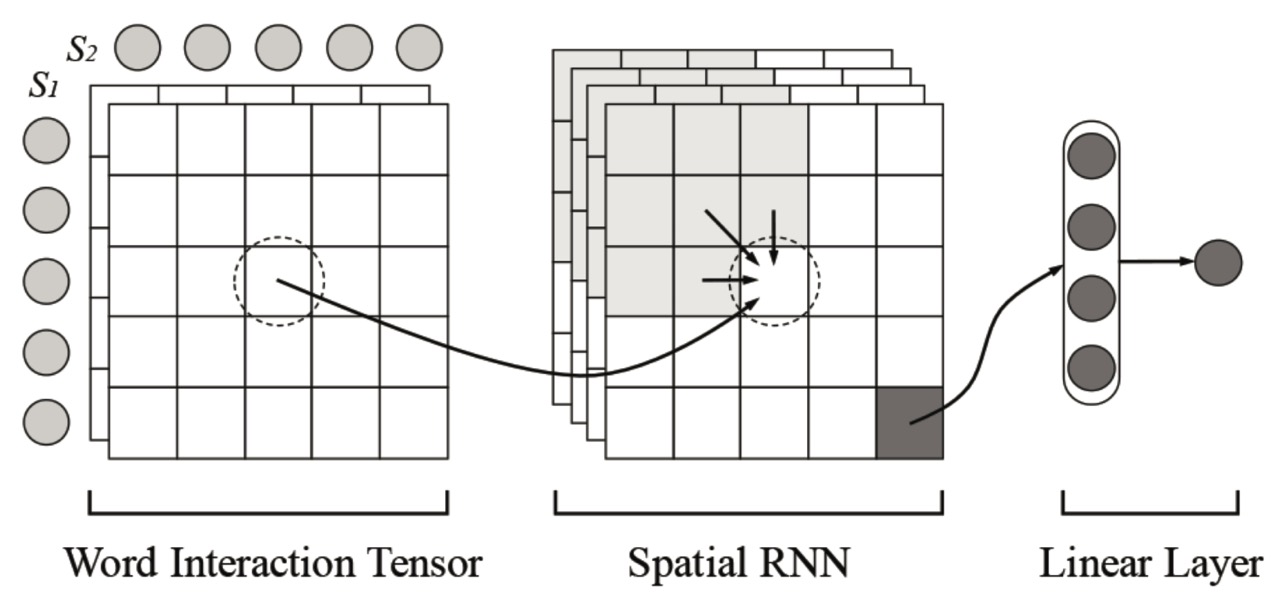}
  \caption{\citet{Wan:2016b} (Match-SRNN) model}
  \label{fig:Match-SRNN}
\end{figure}

\textbf{\citet{Wan:2016b}} proposed a recursive semantic matching model called Match-SRNN. According to Figure \ref{fig:Match-SRNN}, which shows the architecture of Match-SRNN, Match-SRNN works in three major steps: In the first step word-level interactions are modeled. In the second step a special case of interaction between two prefixes of two different sentence ($S1[ 1:i ]$ and $S2[ 1:j ]$ ) is modeled as a function of the interaction between $S1[1:i-1]$ and $S2[1:j]$, $S1[1:i]$ and $S2[1:j-1]$, $S1[1:i-1]$ and $S2[1:j-1]$, and interaction between words $w_{i}$ and $v_{j}$. Then the equation for this special interaction is:

\begin{equation}
\begin{aligned}
h_{i j}=f\left(h_{i-1, j}, h_{i, j-1}, h_{i-1, j-1}, s\left(w_{i,}, v_{j}\right)\right)
\end{aligned}
\end{equation}
where $h_{i j}$ is the interaction between $S1[1:i]$ and $S2[1:j]$. This recursive way of modeling the interaction helps to capture the long-term dependencies between two sentences. In the third step, a linear function is used for measuring the matching score of two given sentences. More details about these steps are below.

\begin{enumerate}

\item Neural tensor network: The interaction between two words $w_{i}$ and $v_{j}$ is captured by a neural tensor network according to the following equation:

\begin{equation}
\begin{aligned}
\vec{s}_{i j}=F\left(u\left(w_{i}\right)^{T} T^{[1 : c]} u\left(v_{j}\right)+W \left[ \begin{array}{c}{u\left(w_{i}\right)} \\ {u\left(v_{j}\right)}\end{array}\right]+\vec{b}\right)
\end{aligned}
\end{equation}
where $s_{ij}$ is a vector representation of the similarity between $w_{i}$ and $v_{j}$ words, $T_{i}$ is one slice of the tensor parameters, $W$ and $b$ are parameters and $F(Z) = max (0, Z)$.

\item Spatial RNN: In this layer, GRU is used as an RNN because of its easy implementation for implementing a Spatial-GRU which models the $h_{i j}$. Figure \ref{fig:Spatial-GRU} shows the architecture of the 1D-GRU and Spatial-GRU which is used in this work.

\begin{figure}[bt]
\centering
    \includegraphics[width=0.75\textwidth]{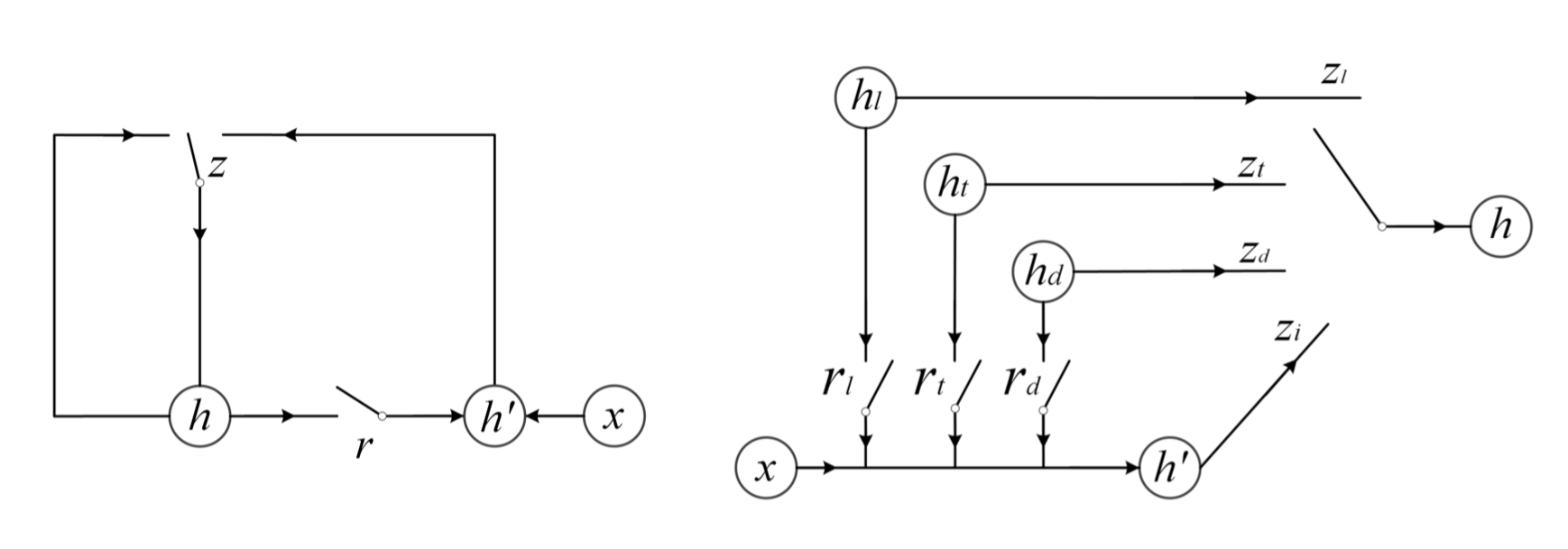}
  \caption{Architecture of Spatial-GRU (right) and GRU (left)}
  \label{fig:Spatial-GRU}
\end{figure}

According to the right part in Figure \ref{fig:Spatial-GRU} Spatial-GRU has four updating gates, and three reset gates. Function $f$ in the Spatial-GRU is computed as follow:

\begin{equation}
\begin{aligned}
\vec{q}^{T}=\left[\vec{h}_{i-1, j}^{T}, \vec{h}_{i, j-1}^{T}, \vec{h}_{i-1, j-1}^{T}, \vec{s}_{i j}^{T}\right]^{T}
\end{aligned}
\end{equation}
\[\vec{r}_{l}=\sigma\left(W^{\left(r_{l}\right)} \vec{q}+\vec{b}^{\left(r_{l}\right)}\right), \overrightarrow{r_{t}}=\sigma\left(W^{\left(r_{t}\right)} \vec{q}+\vec{b}^{\left(r_{t}\right)}\right)\]
\[\vec{r}_{d}=\sigma\left(W^{\left(r_{d}\right)} \vec{q}+\vec{b}^{\left(r_{d}\right)}\right), \vec{r}^{T}=\left[\vec{r}_{l}^{T}, \vec{r}_{t}^{T}, \vec{r}_{d}^{T}\right]^{T}\]
\[\vec{z}_{i}^{\prime}=W^{\left(z_{i}\right)} \vec{q}+\vec{b}^{\left(z_{i}\right)}, \overrightarrow{z_{l}^{\prime}}=W^{\left(z_{l}\right)} \vec{q}+\vec{b}^{\left(z_{l}\right)}\]
\[\vec{z}_{t}^{\prime}=W^{\left(z_{t}\right)} \vec{q}+\vec{b}^{\left(z_{t}\right)}, \vec{z}_{d}^{\prime}=W^{\left(z_{d}\right)} \vec{q}+\vec{b}^{\left(z_{d}\right)}\]
\[\left[\vec{z}_{i}, \vec{z}_{l}, \vec{z}_{t}, \vec{z}_{d}\right]=\operatorname{SoftmaxByRow} \left(\left[\vec{z}_{i}^{\prime}, \vec{z}_{l}^{\prime}, \vec{z}_{t}^{\prime}, \vec{z}_{d}^{\prime}\right]\right)\]
\[\vec{h}_{i j}^{\prime}=\phi\left(W \vec{s}_{i j}+U\left(\vec{r} \odot\left[\vec{h}_{i, j-1}^{T}, \vec{h}_{i-1, j}^{T}, \vec{h}_{i-1, j-1}^{T}\right]^{T}\right)+\vec{b}\right)\]
\[\vec{h}_{i j}=\vec{z}_{l} \odot \vec{h}_{i, j-1}+\vec{z}_{t} \odot \vec{h}_{i-1, j}+\vec{z}_{d} \odot \vec{h}_{i-1, j-1}+\vec{z}_{i} \odot \vec{h}_{i j}^{\prime}\]

\item Linear Scoring Function: Final matching score of two given sentences is computed by the equation: $M (S_{1}, S_{2}) = W^{(s)}h_{mn} + b^{(s)}$ where $h_{mn}$ is the global interaction between two sentences, and $W^{(s)}$ and $b^{(s)}$ are network parameters.

\end{enumerate}

\begin{figure}[bt]
\centering
    \includegraphics[width=0.98\textwidth]{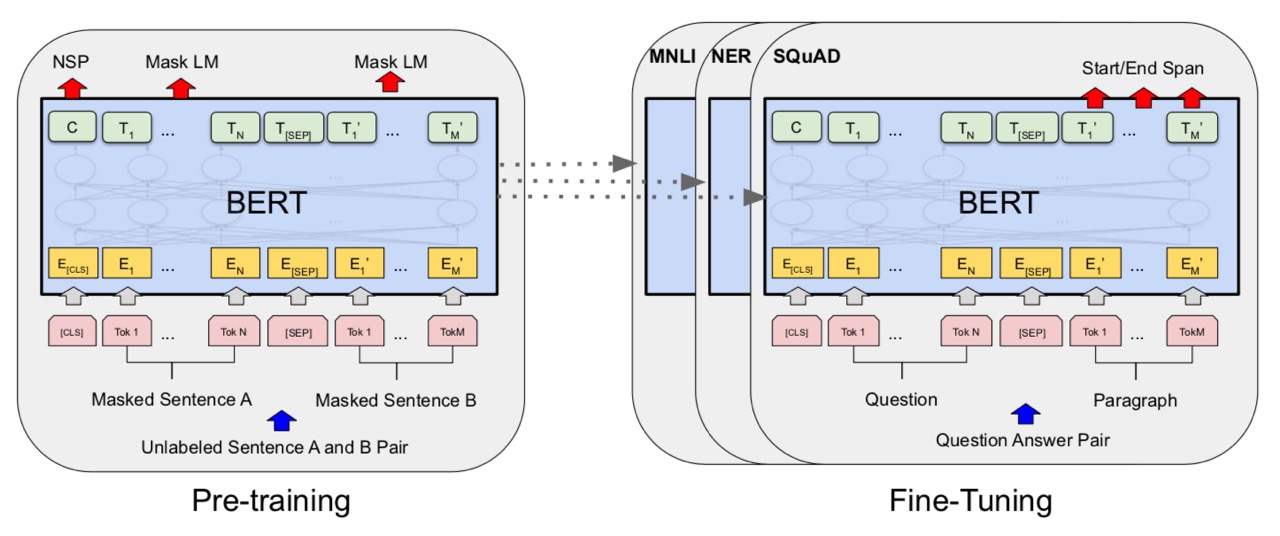}
      \caption{Architecture of BERT in pre-training and fine-tuning \citep{Devlin:2019}}
  \label{fig:bert_arch}
\end{figure}
\begin{figure}[bt]
\centering
    \includegraphics[width=0.98\textwidth]{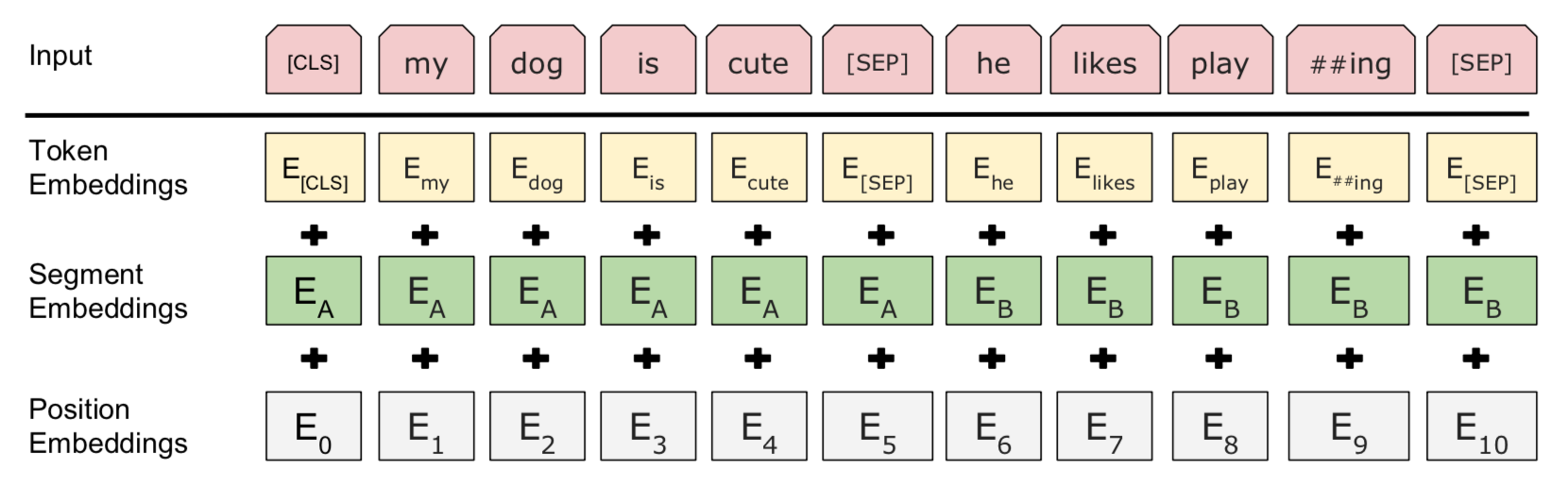}
      \caption{Building input representation of BERT \citep{Devlin:2019}}
  \label{fig:Bert_input_rep}
\end{figure}

\textbf{\citet{Devlin:2019}} proposed Bidirectional Encoder Representations from Transformers (BERT) model which is a language modeling neural network. BERT has a multi-layered bidirectional architecture, in which each layer is a transformer encoder. The transformer was proposed by \citet{Vaswani:2017} and has an encoder-decoder architecture. The same decoder segment of the transformer model is used in BERT.  BERT is used in a wide range of NLP downstream tasks, including QA, natural language inference, and text classification for capturing the textual dependency among given sequences. BERT is pre-trained on large corpora by two different approaches, namely masked language model and next sentence prediction, and then fine-tuned on each specific downstream task based on the application. The architecture of BERT including both pre-training and fine-tuning steps is shown in Figure \ref{fig:bert_arch}. Input of BERT is a sequence of input representation of words. Input representation of each word is built by summing the word embedding, segment embedding, and position embedding as shown in Figure \ref{fig:Bert_input_rep}. In the QA domain, a $(CLS)$ token is used in the first position of the sequence, then the question and the candidate answer followed by a $(SEP)$ token are placed in the sequence, respectively. The output of BERT is an encoded representation for each token. BERT is also used in question answering task and As BERT uses cross-match attention between question and answer sentences it is considered as an interaction-based model.

\begin{figure}[bt]
\centering
    \includegraphics[width=0.98\textwidth]{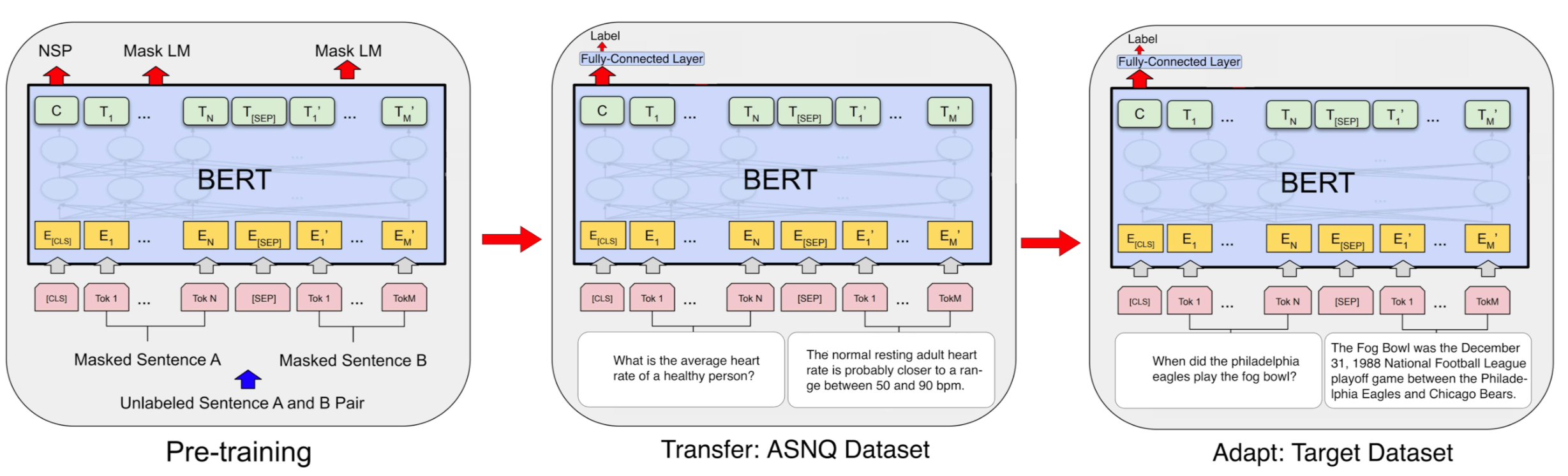}
      \caption{Architecture of the TANDA model \citep{Garg:2019}}
  \label{fig:TANDA}
\end{figure}

\textbf{\citet{Garg:2019}} proposed the TANDA model, which utilizes BERT and RoBERTa pre-trained language models for modeling the dependency between two sequences of sentences for Answer Sentence Selection (AS2). 
The small size of data for fine-tuning BERT may lead to an unstable, and noisy model. For mitigating this problem, they have used two distinct fine-tuning steps for AS2. In the first fine-tuning step which is performed on a large corpus for the AS2 task, BERT is transferred to an AS2 model instead of being a language model only. Then the model is adapted to a specific domain of question types by fine-tuning the model on the target dataset. The architecture of the TANDA model is shown in Figure \ref{fig:TANDA}. A pair of a question and an answer is attached with a $[SEP]$ token and passed to the BERT model. The encoded representation of $[CLS]$ token is passed to a fully-connected layer followed by a sigmoid function for predicting the matching score of the given question and answer pair.

\subsection{Hybrid Models}
\label{sec:hybrid-models}
\begin{figure}[bt]
\centering
    \includegraphics[width=6cm]{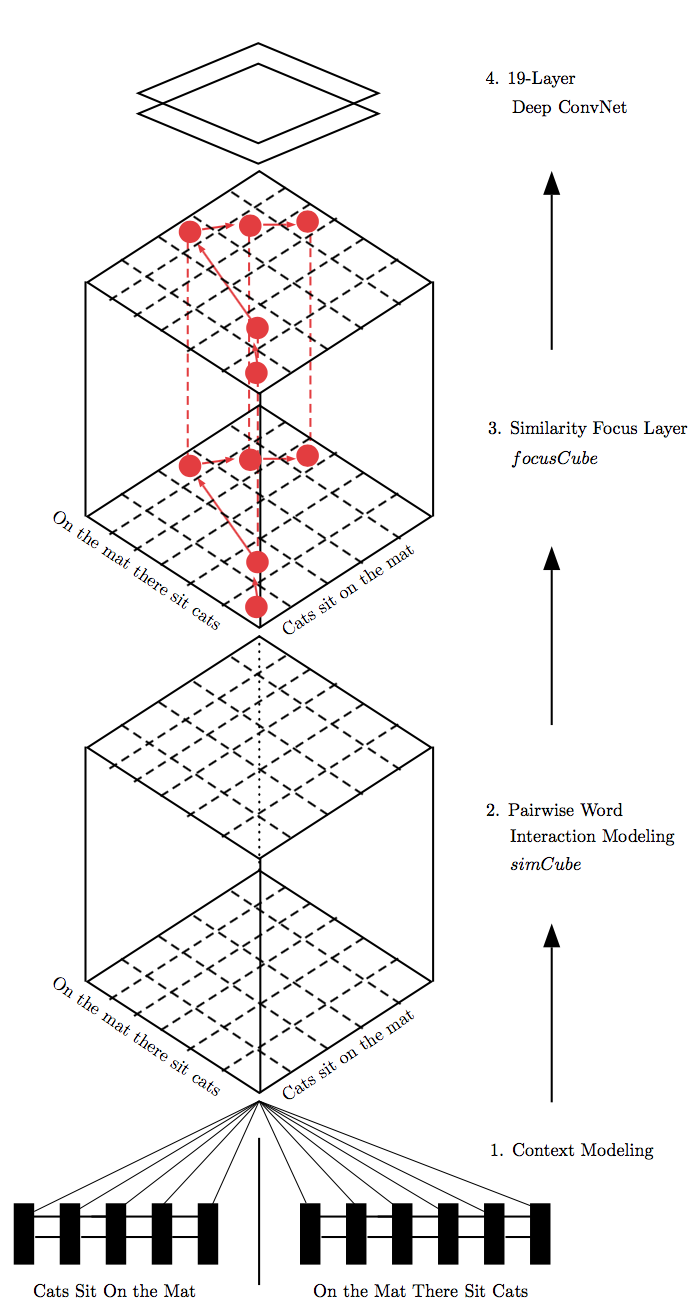}
  \caption{Architecture of \citet{HeAndLin:2016} model}
  \label{fig:HeAndLin:2016}
\end{figure}

 \textbf{\citet{HeAndLin:2016}} proposed a model for QA task which consists of four major components. Architecture of the model is shown in Figure \ref{fig:HeAndLin:2016}. Different components of this model are described below: 

\begin{enumerate}
\item Context modeling: This is the first component and uses a BiLSTM for modeling context of each word.
\item Pairwise word interaction modeling: This component compares two hidden states of BiLSTM with Cosine, L2 Euclidean, and dot product distance measures:

\begin{equation}
CoU (h_{1},h_{2}) = \{Cos (h_{1},h_{2}), L2Euclid (h_{1},h_{2})), DotProduct (h_{1},h_{2})) \} 
 \end{equation}

Output of this component is a cube with size $R^{13 . |sent1| . |sent2|}$ ,where $|sent1|$ and $|sent2|$ are the size of the first and the second sentences, respectively. For each pair of words 12 different similarity distances and one extra padding are considered.

\item Similarity focus: In this layer word interactions are assigned weights by maximizing the weight of the important word interactions. The output of this component is a cube named FocusCube and words identified as important have more weight in this cube.
\item Similarity classification: In this layer, CNN is used for finding the patterns of strong pairwise word interactions. Question and answer sentences in FocusCube are fed to this layer and a similarity score is computed.

\end{enumerate}
  
\textbf{\citet{Wan:2016a}} proposed MV-LSTM for matching two sentences using representation of different positions of sentences. As shown in Figure \ref{fig:MV-LSTM}, representation of different positions in each sentence is created and multiple tensors are created by calculating the interaction between different positions of these two sentences with different similarity metrics. Then a $k$-max pooling layer and a multi-layered LSTM are used for modeling the matching score of two given sentences. Architecture of MV-LSTM is explained in more details in following three steps:

\begin{figure}[bt]
\centering
    \includegraphics[width=0.75\textwidth]{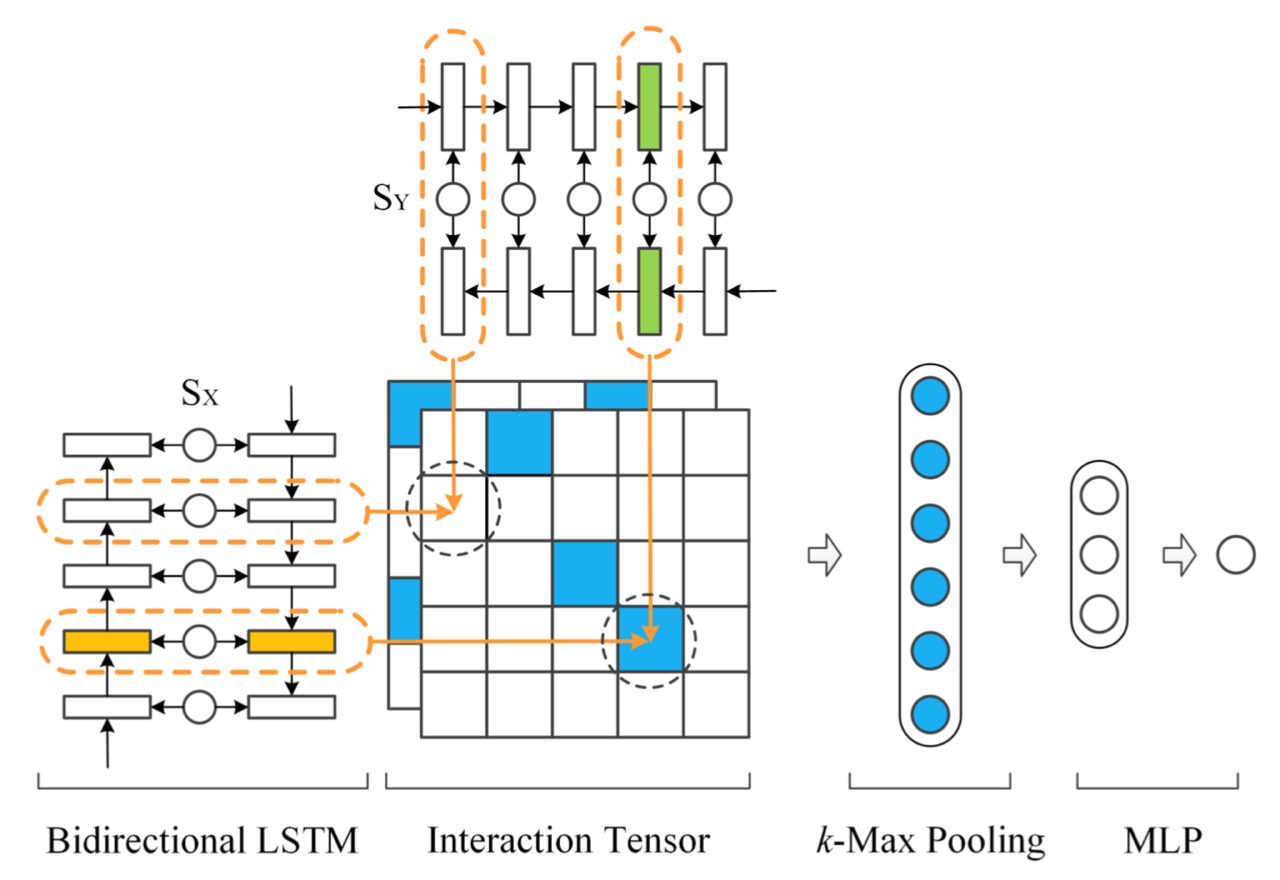}
  \caption{Architecture of MV-LSTM model \citep{Wan:2016a} }
  \label{fig:MV-LSTM}
\end{figure}

\begin{enumerate}

\item Positional Sentence Representation: Positional sentence representation or representation of sentence at one position is obtained by BiLSTM . BiLSTM is used for capturing the long and short-term dependencies in one sentence. An LSTM layer similar to the implementation used in {\citep{Graves:2013}} is used here. Given a sentence $S = (x_{0}, x_{1}, ..., x_{T})$, LSTM represents position $t$ of the sentence $h_{t}$ as follows:

\begin{equation}
\begin{aligned}
i_{t}=\sigma\left(W_{x i} x_{t}+W_{h i} h_{t-1}+b_{i}\right)
\end{aligned}
\end{equation}
\[f_{t}=\sigma\left(W_{x f} x_{t}+W_{h f} h_{t-1}+b_{f}\right)\]
\[c_{t}=f_{t} c_{t-1}+i_{t} \tanh \left(W_{x c} x_{t}+W_{h c} h_{t-1}+b_{c}\right)\]
\[o_{t}=\sigma\left(W_{x o} x_{t}+W_{h o} h_{t-1}+b_{o}\right)\]
\[h_{t}=o_{t} \tanh \left(c_{t}\right)\]

Utilizing the BiLSTM layer, two different representations $\overrightarrow{h}_{t}$ and $\overleftarrow{h}_{t}$ are generated for each position and final representation of each position is considered as the concatenation of these two representations $[ \overrightarrow{h}_{t}. \overleftarrow{h}_{t}]$.

\item Interactions between two sentences: Cosine, bilinear, and tensor similarity functions are used in this step for modeling the interaction between two positions. Bilinear function which captures more complicated interactions compared to cosine is as follows:
\begin{equation}
\begin{aligned}
S (u, v) = u^{t} M v + b
\end{aligned}
\end{equation}
where $b$ is bias and $M$ is a matrix for reweighting $u$ and $v$ in different dimensions. Tensor function models the interaction between two vectors more powerfully. It uses the following equation for modeling the interaction:

\begin{equation}
\begin{aligned}
s(u, v)=f\left(u^{T} M^{[1 : c]} v+W_{u v} \left[ \begin{array}{l}{u} \\ {v}\end{array}\right]+b\right)
\end{aligned}
\end{equation}

where $W_{uv}$ and $b$ are parameters, $M_{i}$ is one slice of the tensor parameter, and $f$ is rectifier function. Output of the cosine and bilinear similarities are interaction matrices while the output of the tensor layer is an interaction tensor.

\item Interaction aggregation: The third step uses the interaction between different positional sentence representations in order to measure the matching score of two given sentences. $K$-max pooling is applied to extract a vector $q$ which includes the top $k$ values of a matrix, or the top $k$ values of each slice of the tensor. A new representation $r$ is obtained by feeding the output of $k$-max pooling into a fully connected hidden layer. And finally, the matching score $s$ is obtained by the following function:

\begin{equation}
\begin{aligned}
r = f (W_{r}q + b_{r}), s = W_{s}r + b_{s}
\end{aligned}
\end{equation}
where $W_{r}$ and $W_{s}$ are model parameters and $b_{s}$ and $b_{r}$ are biases.
\end{enumerate}


\begin{figure}[bt]
\centering
    \includegraphics[width=0.6\textwidth]{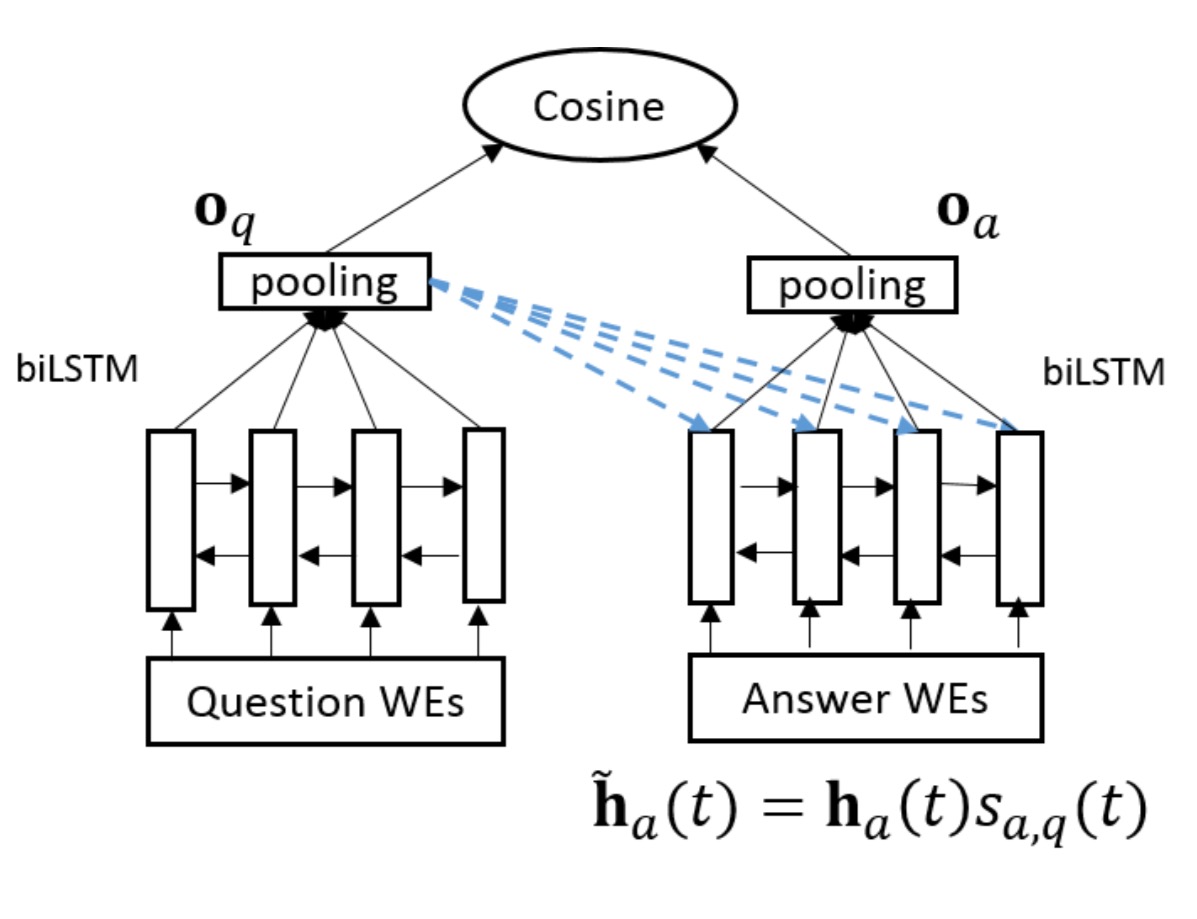}
  \caption{Architecture of Attentive-LSTM model \citep{Tan:2016} }
  \label{fig:Attentive-LSTM}
\end{figure}

\textbf{\citet{Tan:2016}} proposed attentive LSTM, a variant of QA-LSTM model, for mitigating some problems of two other variants of QA-LSTM: convolutional-based LSTM and convolutional pooling LSTM. These two previous models, which are described in section \ref{sec:Rep-based}, suffer from a common issue that happens when the answer is very long and contains a lot of not related words to the question sentence. Attention mechanism by considering the question in constructing the answer sentence's representation can solve this issue. In this model, the attention mechanism works by learning weights for hidden vectors of BiLSTM. As shown in Figure \ref{fig:Attentive-LSTM}, output of the BiLSTM is multiplied by a softMax weight, which is obtained from the question representation. The model gives more weight to each word of the answer, based on the information from question representation. Finally representation of the answer sentence is obtained by the following equations:

\begin{equation}
\begin{aligned}
{m}_{a, q}(t)={W}_{a m} {h}_{a}(t)+{W}_{q m} {o}_{q}
\end{aligned}
\end{equation}
\[s_{a, q}(t) \propto \exp \left({w}_{m s}^{T} \tanh \left({m}_{a, q}(t)\right)\right)\]
\[\widetilde{{h}}_{a}(t)={h}_{a}(t) s_{a, q}(t)\]
where $h_{a}(t)$ is the output vector of answer BiLSTM at time step $t$, $o_{q}$ is question representation, $W_{am}$, $W_{qm}$, and $W_{ms}$ are attention parameters, and $\widetilde{h}_{a}(t)$ indicates the attention-based representation of $h_{a}(t)$.


\textbf{\citet{Wang:2016b}} proposed four inner attention-based RNN models. These models try to mitigate the attention bias problem which traditional attention-based RNN models suffer from. In the following, first a traditional attention-based RNN model and then four variants of inner attention-based RNN (IARNN) models are described.

\begin{enumerate}
\item Traditional attention based RNN models (OARNN): In OARNN first of all an RNN block is used for encoding sentences, and then attention weights from question embedding are used in generating answer sentence's representation. This type of attention mechanism, which is done after learning embeddings, biases toward the later hidden states, because they contain more information than the nearer ones about the sentence. Architecture of OARNN is shown in Figure \ref{fig:OARNN}. This model is named OARNN (stands for outer attention-based RNN) as it adds the attention layer after RNN block. Last hidden layer or average of all hidden states are considered as representation of the question sentence, where the representation of answer is obtained by using attention weights from question representation. 

\begin{figure}[bt]
\centering
    \includegraphics[width=0.95\textwidth]{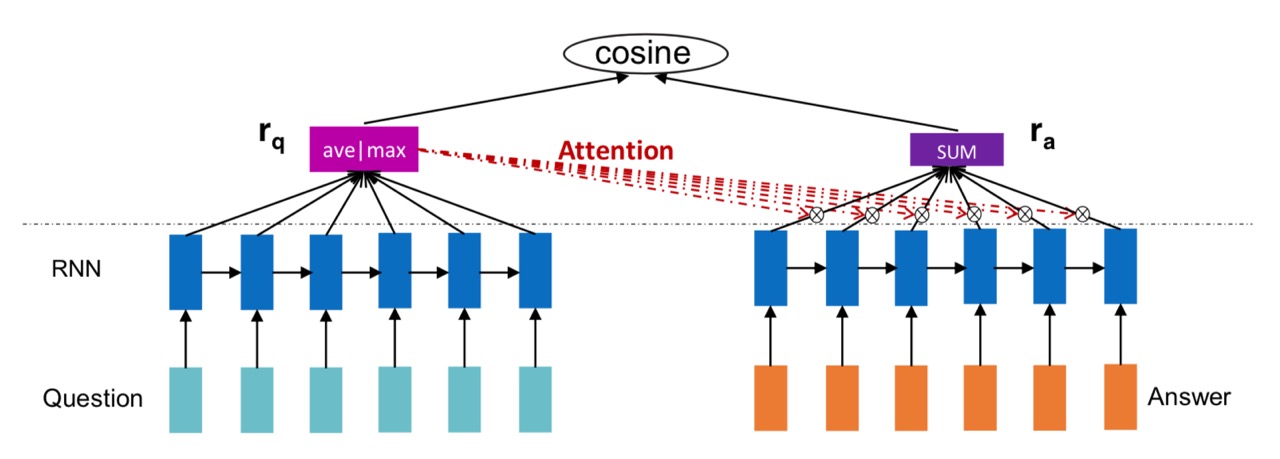}
  \caption{Architecture of OARNN model \citep{Wang:2016b} }
  \label{fig:OARNN}
\end{figure}

\item Inner attention-based RNNs (IARNN): IARNN models are proposed to mitigate the bias problems of OARNN in generating the representation of the answer sentence. In these models, the attention weights are added before that RNN blocks generate hidden layers. The architecture of four different IARNN models is described in the following.

\begin{enumerate}
\item {IARNN-WORD:} Architecture of this model is shown in Figure \ref{fig:IARNN-word}. Representation of each word is generated using the question attention weights, then the whole sentence's representation is obtained by using the RNN model. GRU is chosen among RNN blocks because it has fewer parameters and trains fast. Representation of the sentence is generated by a weighted average of the hidden states $h_{t}$. 

\begin{figure}[h]
\centering
    \includegraphics[width=0.65\textwidth]{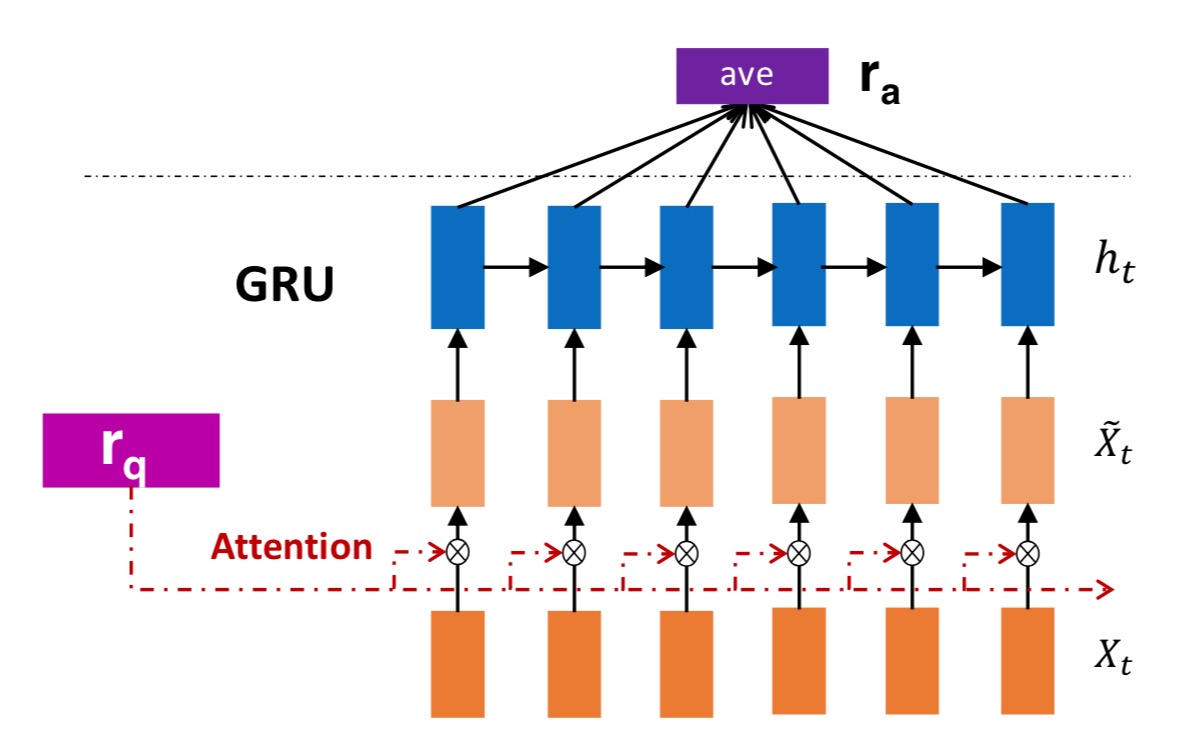}
  \caption{Architecture of IARNN-word model \citep{Wang:2016b}}
  \label{fig:IARNN-word}
\end{figure}

\item {IARNN-Context:} Due to the inability of IARNN-WORD model in capturing the multiple related words, in IARNN-Context, contextual information of answer sentence is fed into the attention weights. Architecture of this model is shown in the Figure \ref{fig:IARNN-Context}.

\begin{figure}[h]
\centering
    \includegraphics[width=0.65\textwidth]{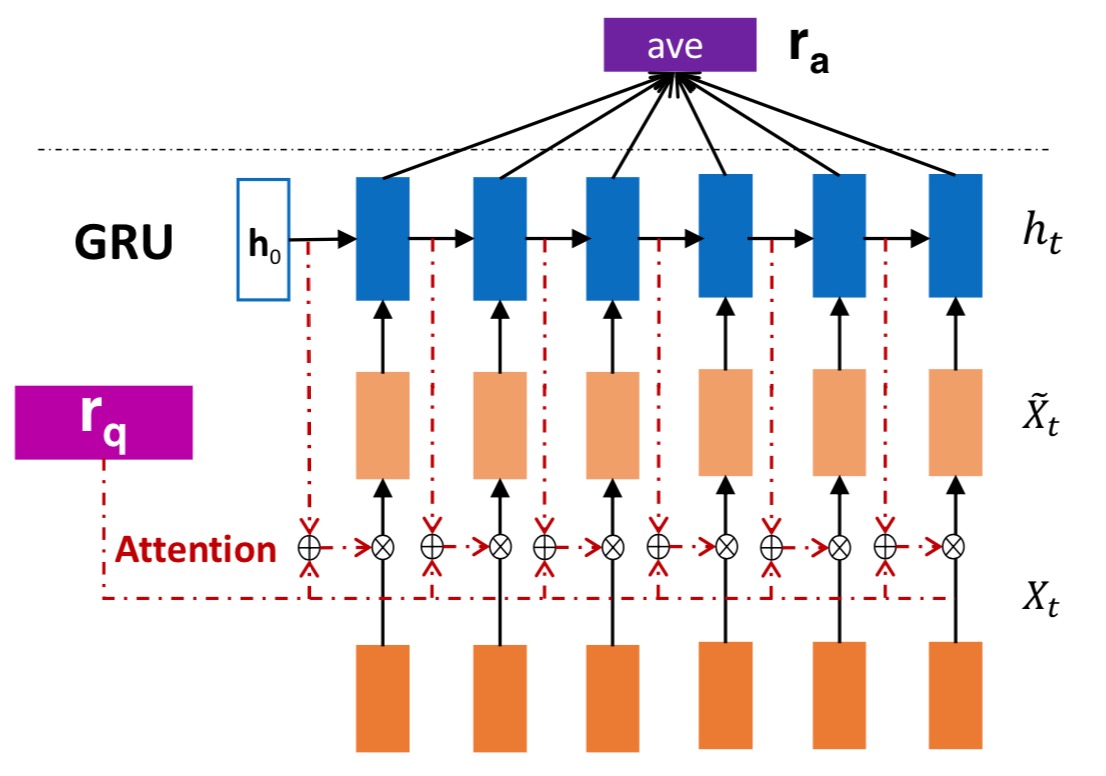}
  \caption{Architecture of IARNN-Context model \citep{Wang:2016b} }
  \label{fig:IARNN-Context}
\end{figure}

\item {IABRNN-GATE:} As GRU gates control the flow of the information in hidden stages, attention information is fed to these gates. Architecture of this model is shown in the Figure \ref{fig:IABRNN-GATE}.

\begin{figure}[bt]
\centering
    \includegraphics[width=0.5\textwidth]{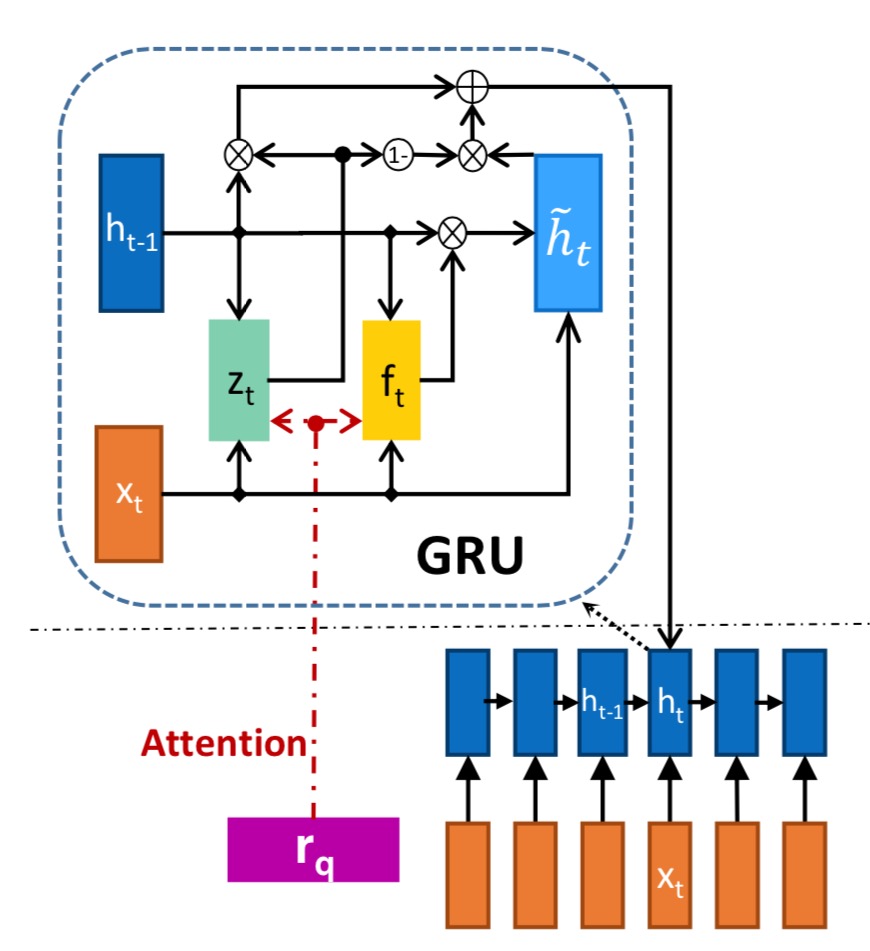}
  \caption{Architecture of IABRNN-GATE model  \citep{Wang:2016b}}
  \label{fig:IABRNN-GATE}
\end{figure}

\item {IARNN-OCCAM:} This model is named after the Occam's Razor which says: "Between the whole words set, the fewest number of words which can represent the sentence must be chosen". Based on the type of question, a different number of relevant words to the question are required for answering the question. For example "what" and "where" questions need a smaller number of relevant words than "why" and "how" questions in answer sentence. This issue is handled in IARNN-OCCAM by using a regulation value. Therefore more sparsity should be imposed on the summation of the attention in "what" and "where" questions and a smaller number should be assigned to the regulation value in "why" and "how" questions.
This regulation model just could be used in IARNN-context and IARNN-word models.

\end{enumerate}
\end{enumerate}


\textbf{\citet{Yin:2016}} proposed four different attention-based variants of BCNN (ABCNN) for text-matching task. In the following, the architecture of these ABCNN models is described.

\begin{itemize}
\item Attention-based CNN (ABCNN): Three different attention-based models are proposed in this work. In ABCNN-1, which is shown in Figure \ref{fig:ABCNN-1}, an attention matrix A is generated by comparing each unit of two feature maps. Let $S_{1}$ and $S_{0}$ be feature maps representing a sentence. Each row in the matrix $A$ shows the attention distribution of the corresponding unit in $S_{0}$ respect to $S_{1}$, and each column of $A$ represents the attention distribution of the corresponding unit in $S_{1}$ respect to $S_{0}$. Then matrix $A$ is transformed into two attention feature map matrices with the same dimension of the representation feature map. According to Figure \ref{fig:ABCNN-1} representation feature map and attention feature map are fed to the convolution layer as order-3 tensors. Given representation of two feature maps of sentences $i={0,1}$ $(F_{i,r} \in R^{d \times s} )$, each cell in attention matrix $(A \in R^{s \times s})$ is computed as follows:

\begin{equation}
\begin{aligned}
{A}_{i, j}=\operatorname{match-score}\left({F}_{0, r}[ :, i], {F}_{1, r}[ :, j]\right)
\end{aligned}
\end{equation}
where $\frac{1}{(1+|\mathrm{x}-\mathrm{y}|)}$ is match-score function for inputs $x$ and $y$.
Attention matrix $A$ is converted to two given feature maps ( $F_{0 , a}$ and $F_{1 , a}$ ) by the following equations where $W_{0}$, $W_{1} \in R^{d \times s}$ are model parameters to be learned.

\begin{equation}
\begin{aligned}
{F}_{0, a}={W}_{0} \cdot {A}^{\top}, \quad {F}_{1, a}={W}_{1} \cdot {A}
\end{aligned}
\end{equation}

A higher-level representation for the corresponding sentence is generated by passing these matrices to the convolution layer.

\begin{figure}[bt]
\centering
    \includegraphics[width=0.95\textwidth]{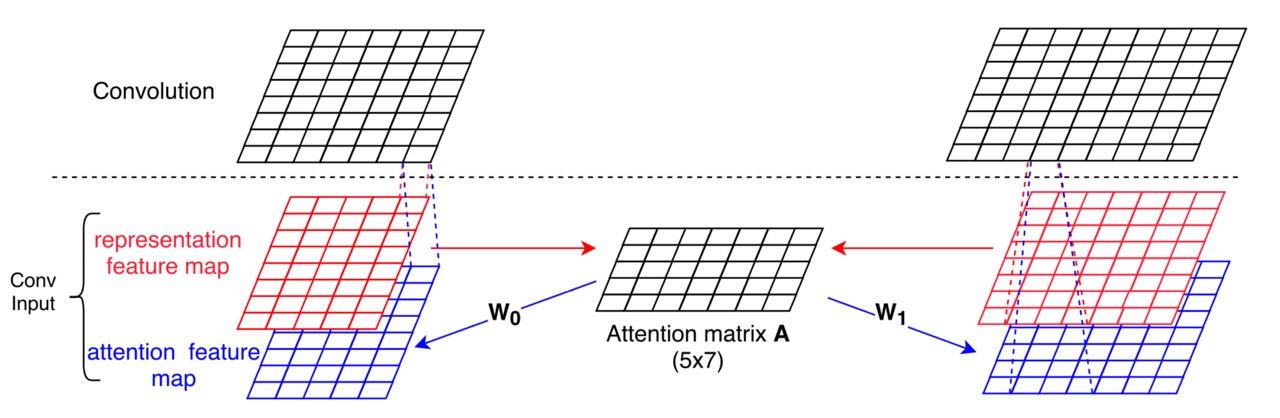}
  \caption{ Architecture of ABCNN-1 model \citep{Yin:2016}}
  \label{fig:ABCNN-1}
\end{figure}

\item ABCNN-2: In this architecture (shown in Figure \ref{fig:ABCNN-2}) attention mechanism is applied to the output of convolutional layers. Each cell in the attention matrix $A$ is calculated by comparing corresponding units from convolution outputs. Each row in the convolution output matrix represents one unit of the given sentence. The attention weight of each unit is computed by summing all the attention values of that unit. Attention weight of unit $j$ in sentence $i$ is shown with $a_{ i , j}$ and computed by:

\begin{equation}
\begin{aligned}
a_{0, j}=\sum A[ j,: ], a_{1, j}=\sum A[ :, j ]
\end{aligned}
\end{equation}
then the new feature map $F^{p}_{i , r} \in R^{d \times s_{i}}$ is calculated with $w-ap$ pooling as follows:

\begin{equation}
\begin{aligned}
{F}_{i, r}^{p}[ :, j]=\sum_{k=j : j+w} a_{i, k} {F}_{i, r}^{c}[ :, k], \quad j=1 \ldots s_{i}
\end{aligned}
\end{equation}

\item ABCNN-3: As it is shown in Figure \ref{fig:ABCNN-3}, ABCNN-3 combines two previous models by employing the attention mechanism before and after the convolution layer. The output of the convolution layer has a larger granularity than its input. That means if the input of the convolution layer has a word-level granularity, then its output has phrase-level granularity. Therefore in the ABCNN-1 model, attention mechanism is employed on a smaller level of granularity than the ABCNN-2 model, and in the ABCNN-3, it is employed on two different levels of granularity.

\begin{figure}[h]
\centering
    \includegraphics[width=0.95\textwidth]{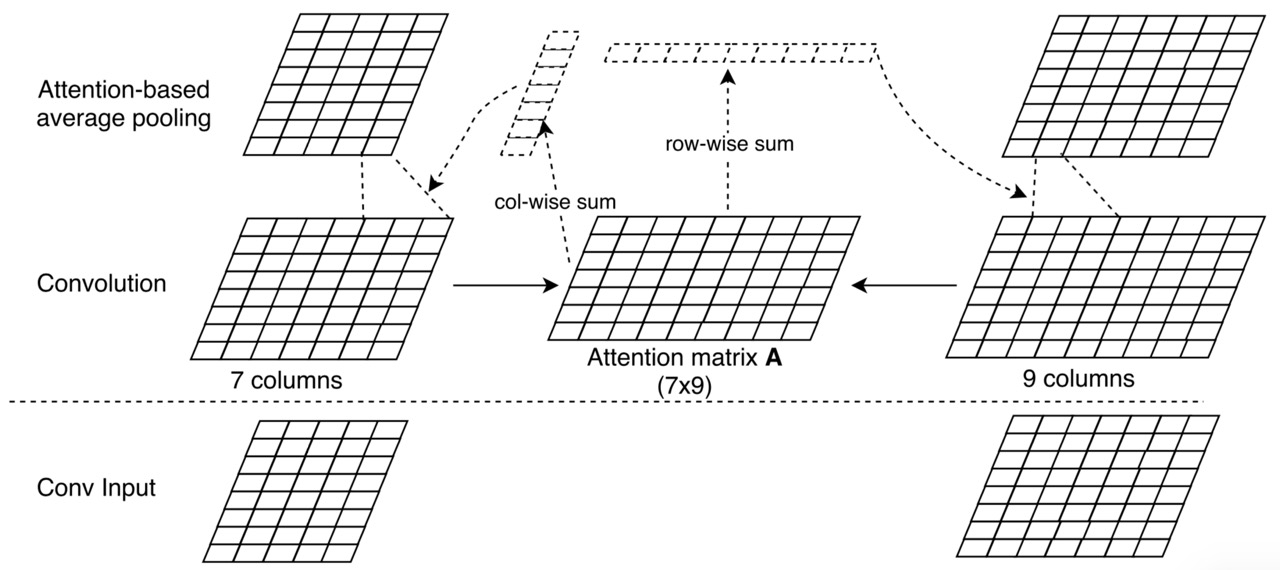}
  \caption{ Architecture of ABCNN-2 model \citep{Yin:2016} }
  \label{fig:ABCNN-2}
\end{figure}

\begin{figure}[h]
\centering
    \includegraphics[width=0.95\textwidth]{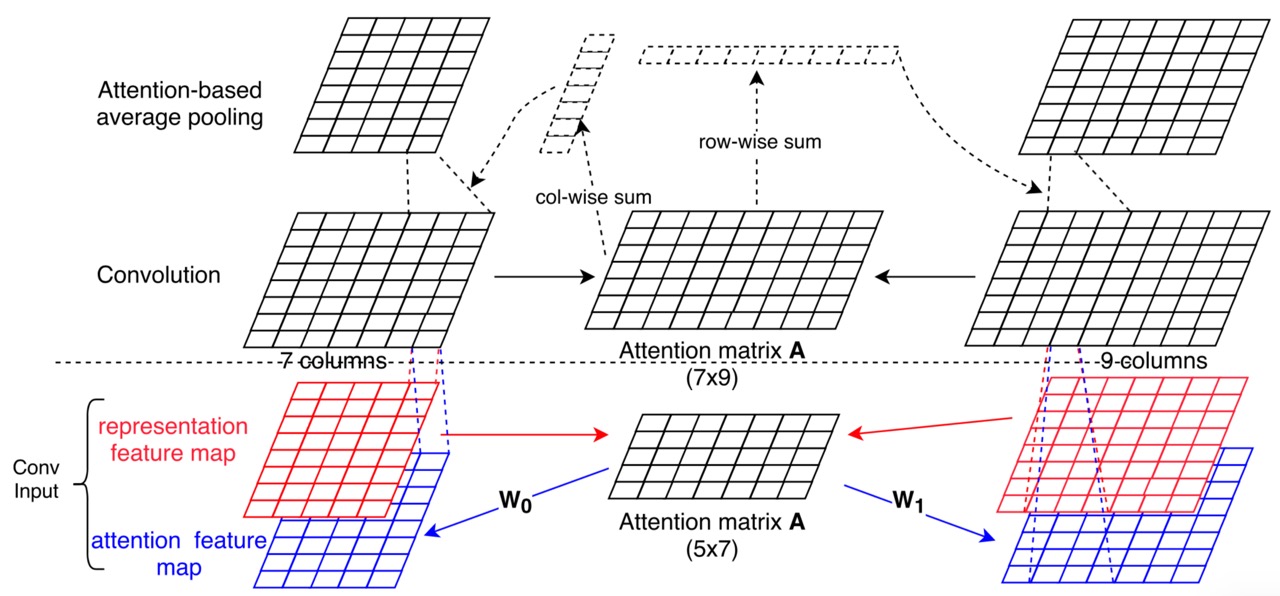}
  \caption{Architecture of ABCNN-3 model \citep{Yin:2016} }
  \label{fig:ABCNN-3}
\end{figure}

\end{itemize}

\begin{figure}[bt]
\centering
    \includegraphics[width=0.75\textwidth]{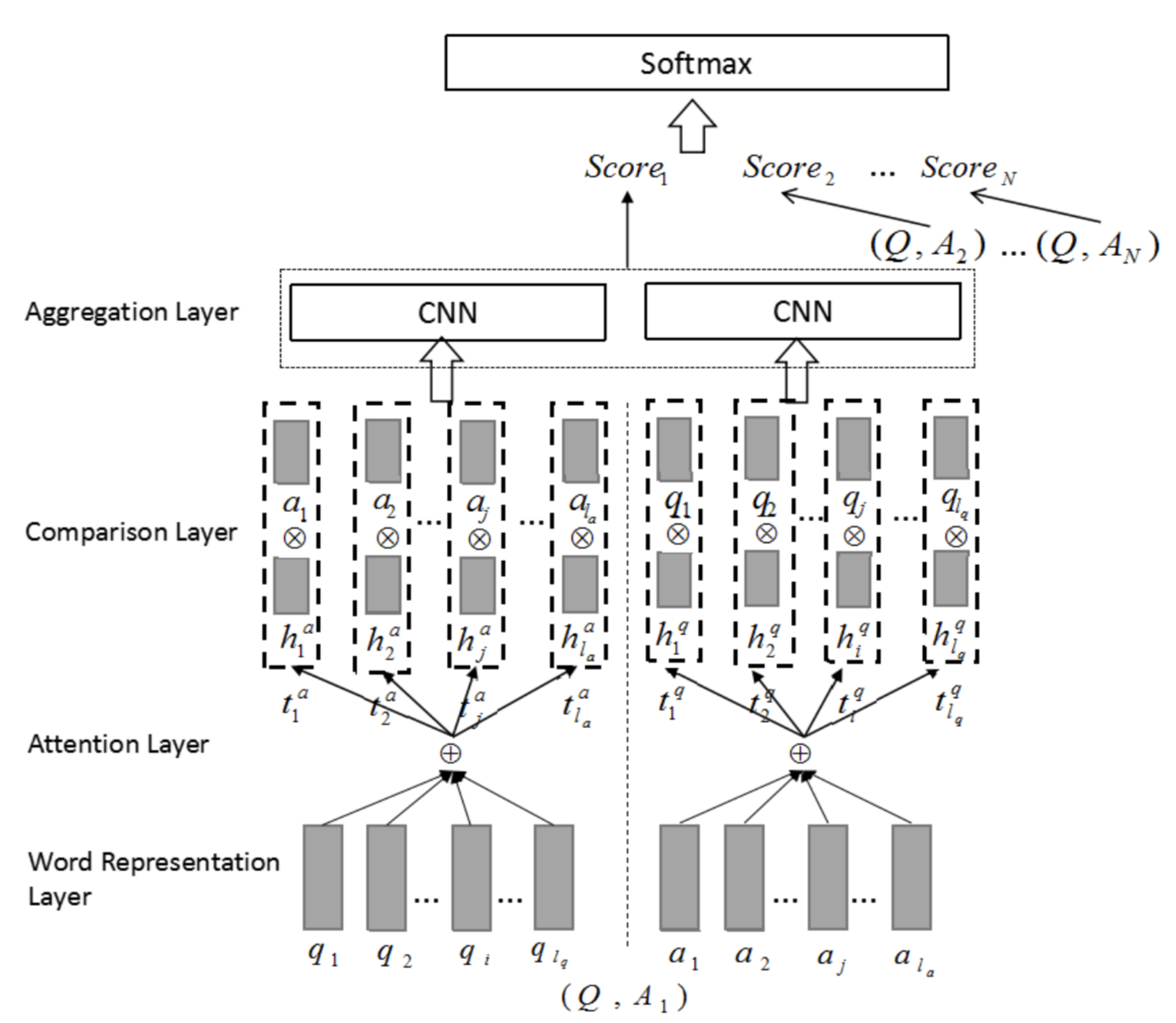}
  \caption{Architecture of \citet{Bian:2017} model }
  \label{fig:Bian-2017}
\end{figure}

\textbf{\citet{Bian:2017}} proposed a model which estimates the relevance score $P (y | Q, A)$ between question $(Q)$ and answer $(A)$. As we see in Figure \ref{fig:Bian-2017}, this model consists of four major layers. First, word representation of each sentence is passed to an attention layer and then the output of the attention layer is compared by sentence representation. The output of comparison layers is passed to a CNN layer for aggregating and then the matching score of two given sentences is obtained in this layer. These layers are described below.

\begin{enumerate}
\item Word representation layer: Word representations of question $q = (q_{1}, ..., q_{l_{q} }))$ and answer $a = (a_{1}, ..., a_{l_{a}}))$ are fed to the attention layer.
\item Attention layer: The aim of applying the attention layer is finding the relevance between local text substructure of question and answer pairs. $h_{j}^{a}$ and $h_{i}^{q}$ are obtained in this layer by the following equations:

\begin{equation}
\begin{aligned}
w_{i j}^{a}=\frac{\exp \left(e_{i j}\right)}{\sum_{k=1}^{\ell_{q}} \exp \left(e_{k j}\right)}, \quad w_{i j}^{q}=\frac{\exp \left(e_{i j}\right)}{\sum_{k=1}^{\ell_{a}} \exp \left(e_{i k}\right)}
\end{aligned}
\end{equation}

\[ {h}_{j}^{a}=\sum_{i=1}^{\ell_{q}} w_{i j}^{a} {q}_{i}, \quad {h}_{i}^{q}=\sum_{j=1}^{\ell_{a}} w_{i j}^{q} {a}_{j} \]

where $e_{i j} = q_{i} . a_{j}$ and $w$ indicate attention weight. This attention model has two problems: First, only a small number of interactions between two sentences are related and the semantic relation is being ambiguous by considering irrelevant interactions. It is proper to consider just relevant interactions. Second, if one token from answer sentence doesn't have any semantic matching with all the words from the question sentence, it is better to omit that token. For tackling the mentioned problems, two filtering approaches, which are called $k$-max attention and $k$-threshold attention, are proposed.
Implementation of these two filtering models in computing $h_{j a}$ is described below.

\begin{itemize}

\item $K$-max attention: This filtering model helps to discard irrelevant fragments, by sorting $w$ in decreasing order and preserving the top $k$ weights and setting other weights to zero.

\item $K$- threshold attention: This filtering just preserves attention weights which are larger than $K$. This filtering works by omitting the units with no semantic matching in another sentence.

\end{itemize}

\item Comparison: Each sentence and weighted version of the other sentence which is obtained in attention layer are compared in this layer. for example $a_{j}$ is compared with $h_{j a}$ by using comparison function $f$ as follows:

\begin{equation}
\begin{aligned}
{t}_{j}^{a}=f\left({a}_{j}, {h}_{j}^{a}\right)={a}_{j} \otimes{h}_{j}^{a}
\end{aligned}
\end{equation}
where $t_{j a}$ represents the comparison result.

\item Aggregation: Comparison vectors of the previous layer for each sentence are aggregated using a one-layer CNN, and finally, the relevance score between question and answer sentences is computed by the following equation:

\begin{equation}
\begin{aligned}
{r}_{{a}}=\operatorname{CNN}\left(\left[{t}_{1}^{a}, \ldots, {t}_{l_{a}}^{a}\right]\right), \quad {r}_{{q}}=\operatorname{CNN}\left(\left[{t}_{1}^{q}, \ldots, {t}_{l_{q}}^{q}\right]\right)\\
\end{aligned}
\end{equation}
\[ \text { Score }=\left[{r}_{{a}}, {r}_{{q}}\right]^{T} {W} \]

\end{enumerate}

\begin{figure}[bt]
\centering
    \includegraphics[width=0.75\textwidth]{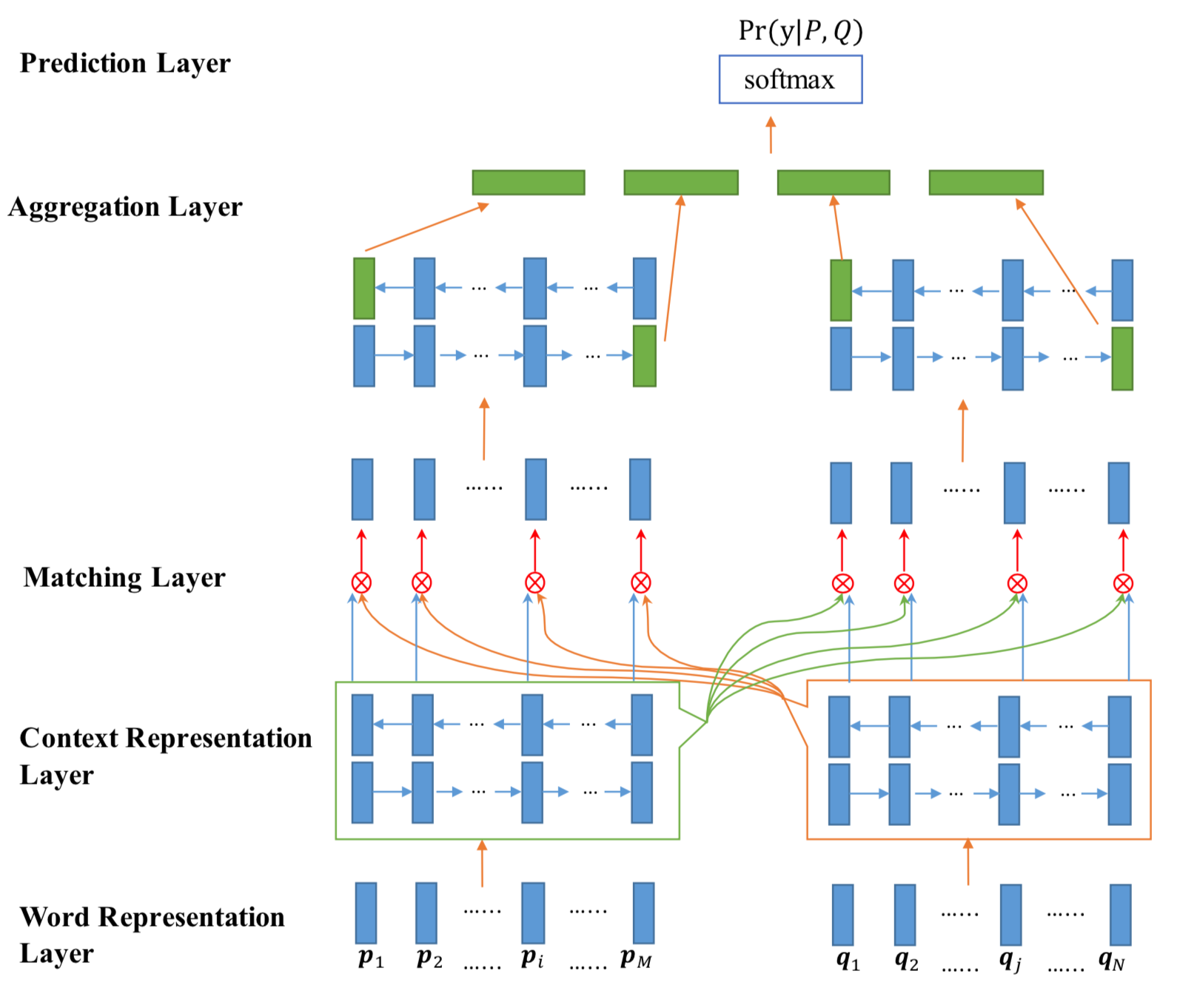}
  \caption{Architecture of BiMPM model \citep{Wang:2017}}
  \label{fig:BiMPM-1}
\end{figure}

\textbf{\citet{Wang:2017}} proposed Bilateral Multi-Perspective Matching Model (BiMPM), a paraphrase-based method for the QA task. In BiMPM, each sample is represented with $(P, Q, y)$, where $P = (p_{1}, p_{2}, ..., p_{M})$ is answer sentence with length $M$,  $Q = (q_{1}, q_{2}, ... , q_{N})$ is question sentence with length $N$, and $y = {0,1}$ is a label indicating whether the answer is related to the question or not. $y=1$ means $P$ is a relevant answer for question $Q$ and $y=0$ means $P$ is not a relevant answer for question $Q$. Figure \ref{fig:BiMPM-1} shows the architecture of BiMPM. BiMPM consists of five major layers which are described in the following.

\begin{enumerate}
\item Word representation layer: Each word is represented with a $d$-dimensional vector constructed by a word embedding and a character-composed embedding. Word embeddings are obtained from pre-trained GloVe {\citep{Pennington:2014}} or word2vec {\citep{Mikolov:2013}} embeddings. Character-composed embeddings are generated by feeding characters of words into an LSTM {\citep{SeppHochreiter:1997}}.

\item Context representation layer: A BiLSTM is used in order to combine contextual information of a sentence with its representation. 

\item Matching layer: In this layer, each contextual embedding of one sentence is compared with all the contextual representations of the other sentence using a multi-perspective matching operation. Also, question and answer sentences are matched in two directions. Multi-perspective cosine matching function is defined as:

\begin{equation}
\begin{aligned}
 M = f_m(v_1, v_2; W) \\
\end{aligned}
\end{equation}
where $v_1$ and $v_2$ are $d$-dimensional vectors, $W \in R^{l \times d}$ is trainable parameter and each perspective is controlled by one row of $W$, and $M$ is a $l$-dimensional vector. Each $M_k$ is calculated as follows: 

\begin{equation}
\begin{aligned}
M_k = cosine (w_k \circ v_1, w_k \circ v_2) \\
\end{aligned}
\end{equation}
let $w_k$ be $k$-th row of $W$ and $\circ$ be element-wise multiplication.
 
Four different matching strategies are proposed based on different $f_m$ functions. These matching strategies are shown in Figure \ref{fig:BIMPM-matchings} and described just for one direction in the following.
 
\begin{figure}[bt]
\centering
     \includegraphics[width=0.75\textwidth]{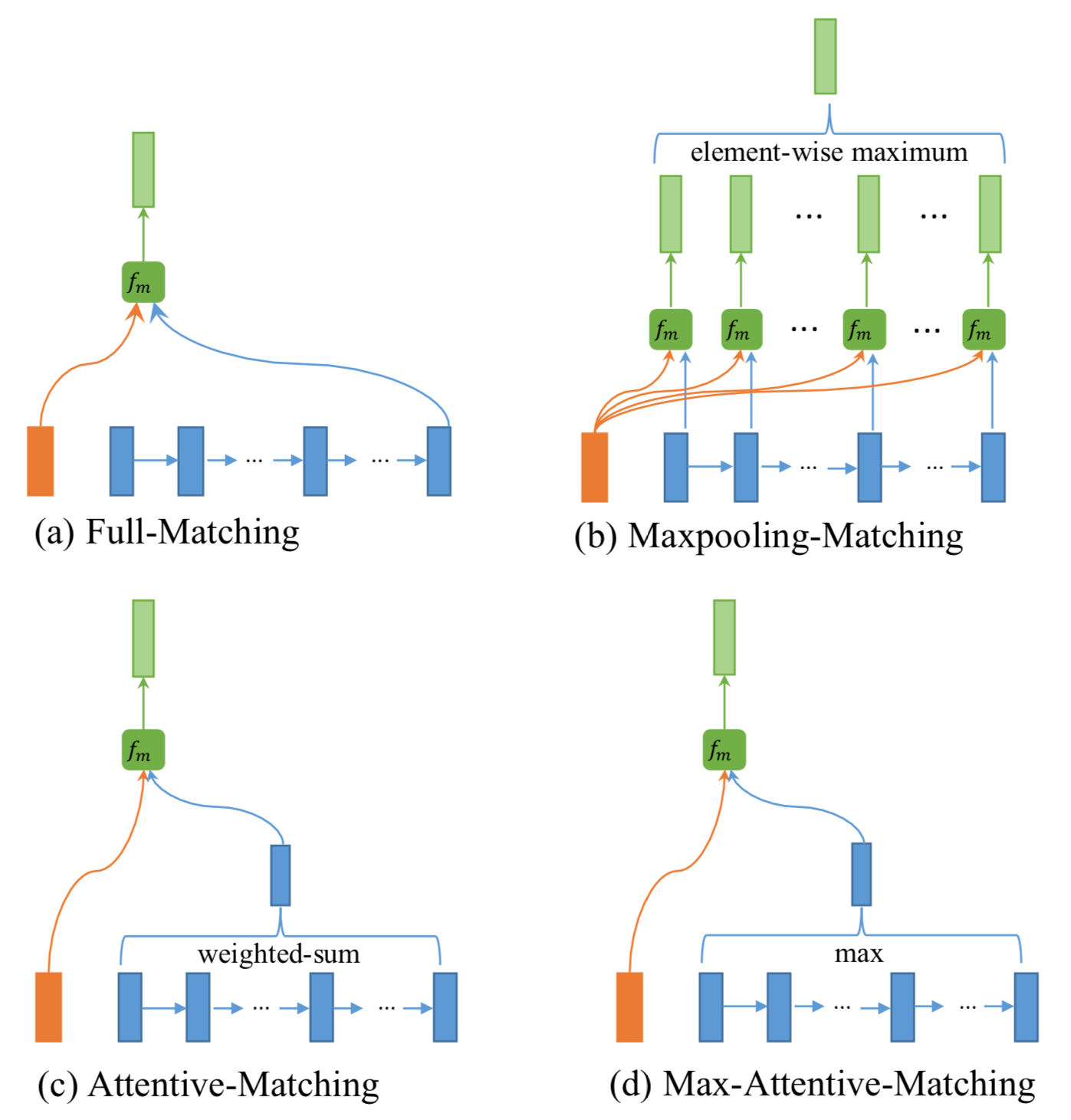}
  \caption{Architecture of BiMPM-matchings \citep{Wang:2017}}
  \label{fig:BIMPM-matchings}
\end{figure}

\begin{itemize}
\item Full-Matching: Each forward time step representation of the answer sentence $\overrightarrow{h_{i}^{p}}$ is compared with every forward time step representations of the question sentence $\overrightarrow{h_{N}^{q}}$. This strategy is shown in Figure \ref{fig:BIMPM-matchings} (a).

\item Maxpooling-Matching: Maximum similarity for each forward time step representation of the answer sentence with all the time steps of the forward representation of the question sentence is returned. This strategy is shown in Figure \ref{fig:BIMPM-matchings} (b).

\item Attentive-Matching: Cosine similarity between each forward time step representation of the answer sentence $\overrightarrow{h_{i}^{p}}$ and each forward time step representation of the question sentence $\overrightarrow{h_{i}^{q}}$ is considered as attention weight. This strategy is shown in Figure \ref{fig:BIMPM-matchings} (c). Then the new representation of the question sentence $(Q)$ called $h_{i}^{mean}$ is generated by calculating the weighted average of its forward time steps representations by using attention weights. And finally, the matching vector for each time step representation of the answer sentence is calculated with its corresponding attentive vector $h_{i}^{mean}$.

\item Max-Attentive-Matching: This strategy is different from an attentive-matching strategy just in generating attentive vector ($h_{i}^{mean}$). Attentive vector here is the contextual embedding with the highest similarity. This strategy is shown in Figure \ref{fig:BIMPM-matchings} (d). Finally, for each direction, all of these strategies are applied for each time-step and eight generated vectors are concatenated and considered as the matching in that direction.
 
 \end{itemize}
 
\item Aggregation layer: Two sequences of matching vectors of both sentences, obtained from the matching layer, are fed into a BiLSTM. Then a fixed-length matching vector is obtained by concatenating the last four output vectors of two BiLSTMs.

\item Prediction layer: In this layer, $P(y| P, Q)$ is predicted using a two-layered feed-forward neural network followed by a softMax layer. The fixed-length matching-vector is fed to this layer.
\end{enumerate}


\textbf{\citet{WangAndJiang:2017}} proposed a compare-aggregate model for matching two sentences. This model for each pair of $Q$ and $A$ predicts a label $y$ which shows whether the candidate answer $A$ is a correct answer for question $Q$ or not. According to the architecture of this model, which is shown in the Figure \ref{fig:Wang-and-jiang-2017}, this model consists four following major layer:

\begin{figure}[bt]
\centering
    \includegraphics[width=0.9\textwidth]{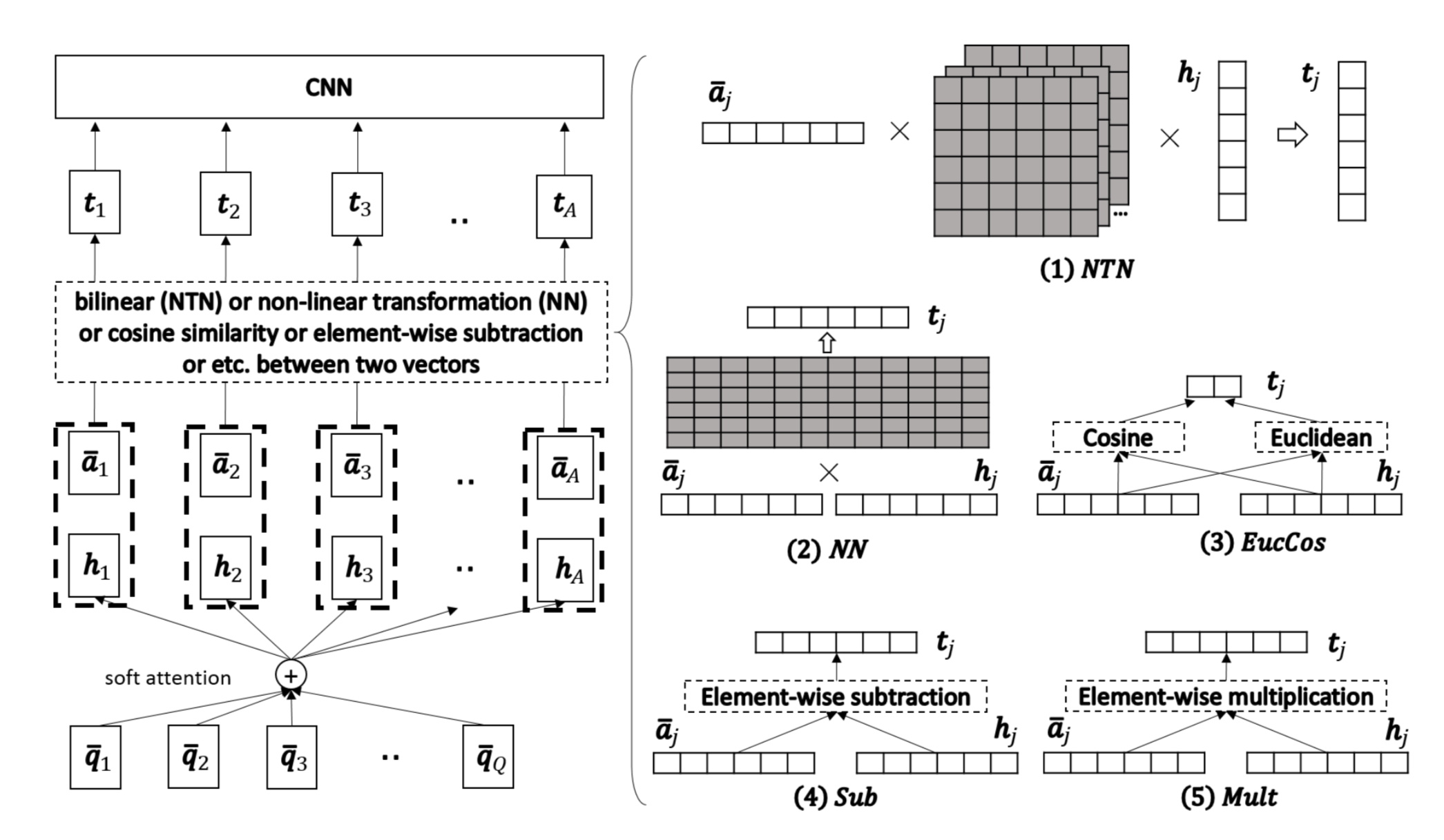}
  \caption{Architecture of \citet{WangAndJiang:2017} (compare-aggregate) model }
  \label{fig:Wang-and-jiang-2017}
\end{figure}

\begin{enumerate}
\item Preprocessing layer: This layer constructs an embedding for each word which represents the word and its contextual information. $Q$ and $A$ are inputs of this layer. A version of LSTM/GRU, which uses only the input gates, is applied for generating $\overline{Q} \in R^{ l \times Q}$ and $\overline{A} \in R^{ l \times A}$ matrices. 

\begin{equation}
\begin{aligned}
\overline{{Q}}=\sigma\left({W}^{\mathrm{i}} {Q}+{b}^{{i}} \otimes {e}_{Q}\right) \odot \tanh \left({W}^{\mathrm{u}} {Q}+{b}^{\mathrm{u}} \otimes {e}_{Q}\right)\\
\overline{{A}}=\sigma\left({W}^{\mathrm{i}} {A}+{b}^{\mathrm{i}} \otimes {e}_{A}\right) \odot \tanh \left({W}^{\mathrm{u}} {A}+{b}^{\mathrm{u}} \otimes {e}_{A}\right)
\end{aligned}
\end{equation}

where $W^{i} , W^{u} \in R^{ l \times d}$ and $b^{i}$, $b^{u} \in R^{l}$ are parameters, and $( . \otimes {e}_{X})$ generates a matrix by repeating the vector on the left for $X$ times. 

\item Attention layer: This layer is applied to the output of the previous layer. Attention-weighted vector $H \in R^{l \times A}$ is obtained by the following equations. The $j^{th}$ column of $H$ indicates the part of $Q$ that best matches the $j^{th}$ word in $A$.

\begin{equation}
\begin{aligned}
& {G}=\operatorname{softmax}\left(\left({W}^{\mathrm{g}} \overline{{Q}}+{b}^{{g}} \otimes {e}_{Q}\right)^{\mathrm{T}} {A}\right)\\
& {H}=\overline{{Q}} {G}
\end{aligned}
\end{equation}

where $W^{g} \in R^{l \times l}$ and $b^{g} \in R$ are parameters, and $G \in R^{Q \times A}$ is the attention weight matrix. 

\item Comparison layer: Embedding of each word $a_{j}$ in the answer is matched with the corresponding attention weight $h_{j}$ and the comparison result is indicated with vector $t_{j}$. In this work, six different comparison functions are introduced.

\begin{itemize}
\item Neural Net (NN):

\begin{equation}
\begin{aligned}
 {t}_{j}=f\left(\overline{{a}}_{j}, {h}_{j}\right)=\operatorname{ReLU}\left({W} \left[ \begin{array}{l}{{a}_{j}} \\ {{h}_{j}}\end{array}\right]+{b}\right) \\
\end{aligned}
\end{equation}

\item Neural Tensor Net (NTN):

\begin{equation}
\begin{aligned}
{t}_{j}=f\left(\overline{{a}}_{j}, {h}_{j}\right)=\operatorname{ReLU}\left({a}_{j}^{\mathrm{T}} {T}^{[1 \ldots l]} {h}_{j}+{b}\right) \\
\end{aligned}
\end{equation}

\item Euclidean distance or cosine similarity (EucCos):

\begin{equation}
\begin{aligned}
 {t}_{j}=f\left(\overline{{a}}_{j}, {h}_{j}\right)=\left[ \begin{array}{l}{\left\|{a}_{j}-{h}_{j}\right\|_{2}} \\ {\cos \left(\overline{{a}}_{j}, {h}_{j}\right]}\end{array}\right]   \\
\end{aligned}
\end{equation}

\item Subtraction (Sub):

\begin{equation}
\begin{aligned}
{t}_{j}=f\left(\overline{{a}}_{j}, {h}_{j}\right)=\left(\overline{{a}}_{j}-{h}_{j}\right) \odot\left(\overline{{a}}_{j}-{h}_{j}\right)  \\
\end{aligned}
\end{equation}

\item Multiplication (Mult):

\begin{equation}
\begin{aligned}
{t}_{j}=f\left(\overline{{a}}_{j}, {h}_{j}\right)=\overline{{a}}_{j} \odot {h}_{j}   \\
\end{aligned}
\end{equation}

\item Submult + NN:

\begin{equation}
\begin{aligned}
{t}_{j}=f\left(\overline{{a}}_{j}, {h}_{j}\right)=\operatorname{ReLU}\left({W}\left[ \begin{array}{c}{\left(\overline{{a}}_{j}-{h}_{j}\right) \odot\left(\overline{{a}}_{j}-{h}_{j}\right)} \\ {\overline{{a}}_{j} \odot {h}_{j}}\end{array}\right]+{b}\right) \\
\end{aligned}
\end{equation}

\end{itemize}

Among these comparison functions, NN and NTN do not capture the similarity well. EucCos may ignore some important information. Sub and Mult are similar to the Euclidean distance and Cosine similarity, and the last model is the combination of the Sub, Mult, and NN.

\item Aggregation: A one-layer CNN is used for combining $t_{j}$ vectors. The output of the aggregation is $r \in R^{n l}$ which is used in the final classifier.

\begin{equation}
\begin{aligned}
{r}=\mathrm{CNN}\left(\left[{t}_{1}, \ldots, {t}_{A}\right]\right)  \\
\end{aligned}
\end{equation}

\end{enumerate}

\begin{figure}[bt]
\centering
    \includegraphics[width=0.75\textwidth]{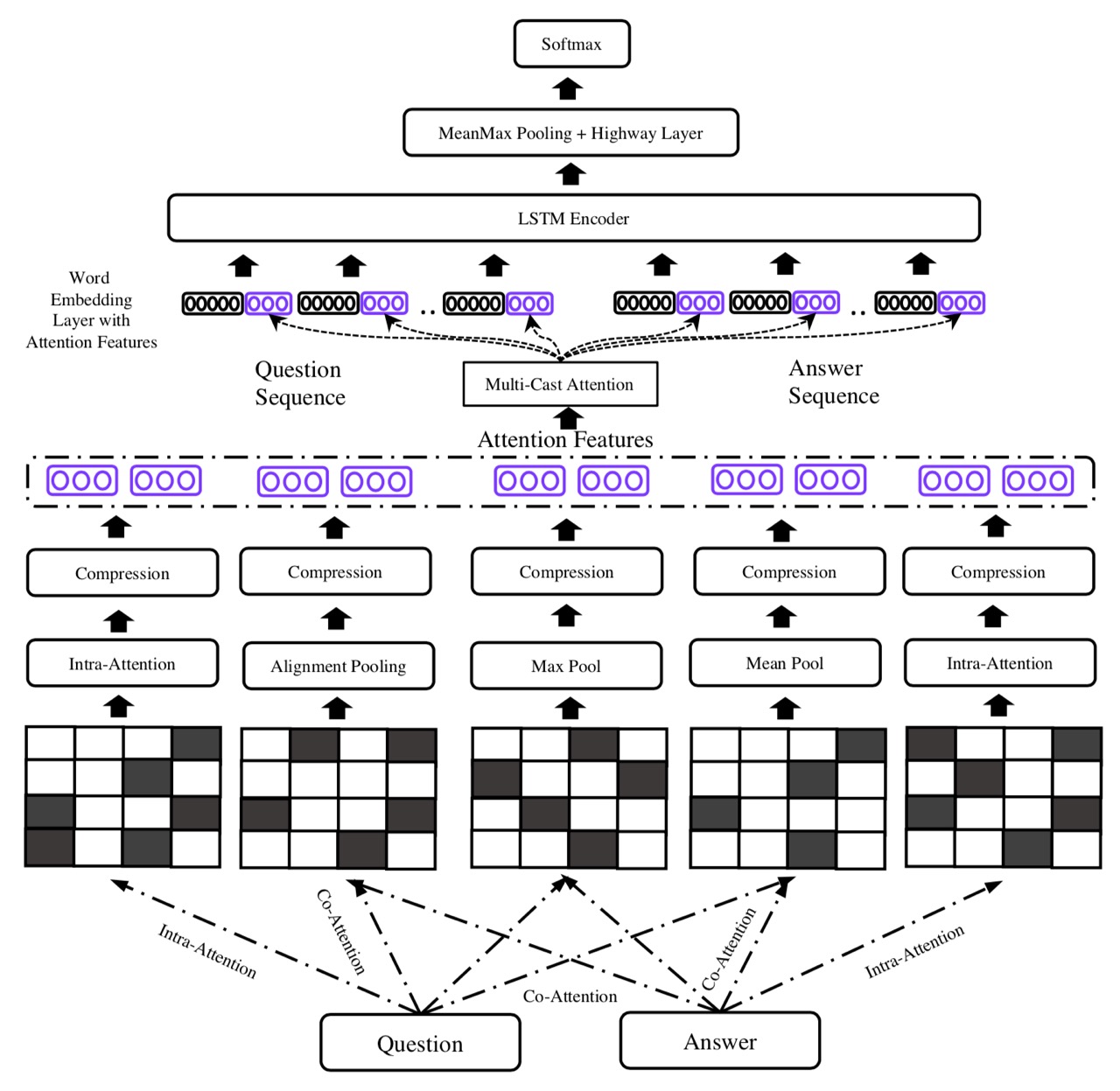}
  \caption{Architecture of MCAN model \citep{Tay:2018}}
  \label{fig:Tay-2018}
\end{figure}

\textbf{\citet{Tay:2018}} proposed Multi-Cast Attention Network (MCAN) for retrieval-based QA. Inputs of MCAN are two sentences: question $q$ and document $d$ sentences. As is shown in Figure \ref{fig:Tay-2018}, MCAN has five major layers. These layers are described in the following.

\begin{enumerate}

\item Input Encoder: Input sentences are fed to this network as one-hot encoded vectors and word embeddings are generated by passing through an embedding layer. Highway encoders like RNNs control the flow of the information by using a gating mechanism. A highway encoder layer is used for detecting important and not important words in a given sentence. A single highway network is formulated as: 

\begin{equation}
\begin{aligned}
y=H\left(x, W_{H}\right) \cdot T\left(x, W_{T}\right)+\left(1-T\left(x, W_{T}\right)\right) \cdot x
\end{aligned}
\end{equation}

where $H(.)$ and $T(.)$ are one-layer affine transforms with ReLU and sigmoid activation functions, and $W_{H}$, $W_{T} \in R^{r \times d}$.

\item Co-Attention: In this layer, a similarity matrix which denotes the similarity between each pair of words across both sentences is learned by the following formulations:

\begin{equation}
\begin{aligned}
 s_{i j}=F\left(q_{i}\right)^{T} F\left(d_{j}\right) \\
\end{aligned}
\end{equation}
\[ s_{i j}=q_{i}^{T} M d_{j} \]
\[ s_{i j}=F\left[q_{i};d_{j}] \right] \]

where $F$ could be a multi-layered perceptron.

\begin{enumerate}

\item {Extractive Pooling:} Max-pooling and mean-pooling are two variants of this type. Formulation of these two poolings are as below:

\begin{equation}
\begin{aligned}
q^{\prime}=\operatorname{Soft}\left(\max _{c o l}(s)\right)^{\top} q \text { and } d^{\prime}=\operatorname{Soft}\left(\max _{r o w}(s)\right)^{\top} d\\
q^{\prime}=\operatorname{Soft}(\operatorname{mean}(s))^{\top} q \text { and } d^{\prime}=\operatorname{Soft}\left(\operatorname{mean}_{r o w}(s)\right)^{\top} d
\end{aligned}
\end{equation}

where Soft is softMax function, and $d^{\prime}$ and $q^{\prime}$ are co-attentive representations of the document and question. Performance of these poolings varies on different datasets, but in general, max-pooling pays attention to words based on their maximum influence, and mean-pooling pays attention to words based on their total influence on the words of the other sentence.

\item {Alignment pooling:} Word pairs from two sentences are realigned in this pooling strategy. Co-attentive representations are learned as below:

\begin{equation}
\begin{aligned}
d_{i}^{\prime} :=\sum_{j=1}^{\ell_{q}} \frac{\exp \left(s_{i j}\right)}{\sum_{k=1}^{\ell_{q}} \exp \left(s_{i k}\right)} q_{j} \text { and } q_{j}^{\prime} :=\sum_{i=1}^{\ell_{d}} \frac{\exp \left(s_{i j}\right)}{\sum_{k=1}^{\ell_{d}} \exp \left(s_{k j}\right)} d_{i}
\end{aligned}
\end{equation}

let $d_{i}^{\prime}$ be the sub-phrase of $q$ which is aligned to $d_{i}$. 

\item {Intra-attention:} Intra-attention attempts to represent long-term dependencies in one sentence. Representation of each sentence is learned regardless of the other sentence. So, it is applied on both the document and the question separately. Co-attentive representations are learned as below:

\[  x_{i}^{\prime} :=\sum_{j=1}^{\ell} \frac{\exp \left(s_{i j}\right)}{\sum_{k=1}^{\ell} \exp \left(s_{i k}\right)} x_{j} \]

where $x_{i}^{\prime}$ is Intra-attention representation of $x_{i}$. 
\end{enumerate}

\item Multi-Cast Attention: This model utilizes all of the mentioned pooling functions. The following values are calculated for the output of each co-attention function.

\begin{equation}
\begin{aligned}
f_{c}=F_{c}([\overline{x} ; x])  , f_{m}=F_{c}(\overline{x} \odot x) ,  f_{s}=F_{c}(\overline{x}-x)\\
\end{aligned}
\end{equation}

where $\overline{x}$ denotes the co-attention representation of $x$, and $F_{C}$ is a compression function. In the above formulations, $\overline{x} $ and $x$ are compared by three different operators for modeling the difference between $\overline{x}$ and $x$ from different perspectives. Difference between $\overline{x}$ and $x$ is an $n$-dimensional vector which is compressed by a compression function to a scalar. Three different compression functions: sum, fully-connected layer, and Factorization Machines (FM) are used. As is shown in Figure \ref{fig:Tay-2018}, given a document question pair, co-attention with three different poolings (1) mean-pooling, (2) Max-pooling, (3) alignment-pooling are applied on pair of question and document and (4) Intra-attention is applied on document and question separately. 12 scalars are generated for each word and concatenated with word embedding. Then each word $w_i$ is represented as $w_{i} = [w_{i}; z_{i}]$ where $z \in R^{12}$ is output of multi-cast layer.

\item LSTM encoder: Casted representation of words of a sentence that are generated by multi-cast attention are fed to an LSTM encoder, and a meanMax pooling is applied to hidden states of the LSTM. Casted representations of words help the LSTM network with its knowledge about each sentence and between question and document, in extracting long-term dependencies.

\begin{equation}
\begin{aligned}
 \mathrm{H}_{\mathrm{i}}=\mathrm{LSTM}(\mathrm{u}, \mathrm{i}) ,  \forall \mathrm{i} \in[1,2, \ldots, 1] \\
\end{aligned}
\end{equation}
\[   \mathrm{H}=\mathrm{MeanMax}\left[\mathrm{h}_{1} \ldots \mathrm{h}_{l}\right] \]

\item Prediction layer and optimization: Given representation of the document and the question, prediction is computed by using two-layer highway network and a softMax layer as follows: 

\begin{equation}
\begin{aligned}
y_{o u t}=H_{2}\left(H_{1}\left(\left[x_{q} ; x_{d} ; x_{q} \odot x_{d} ; x_{q}-x_{d}\right]\right)\right)\\
y_{\text {pred}}=\operatorname{softmax}\left(W_{F} \cdot y_{o u t}+b_{F}\right)
\end{aligned}
\end{equation}
where $H_{1}$, and $H_{2}$ are highway network layers with ReLU activation and $W_{F} \in R^{ h \times 2}$ , $b_{F} \in R^{2}$.

\end{enumerate}

\begin{figure}[bt]
\centering
    \includegraphics[scale=0.32]{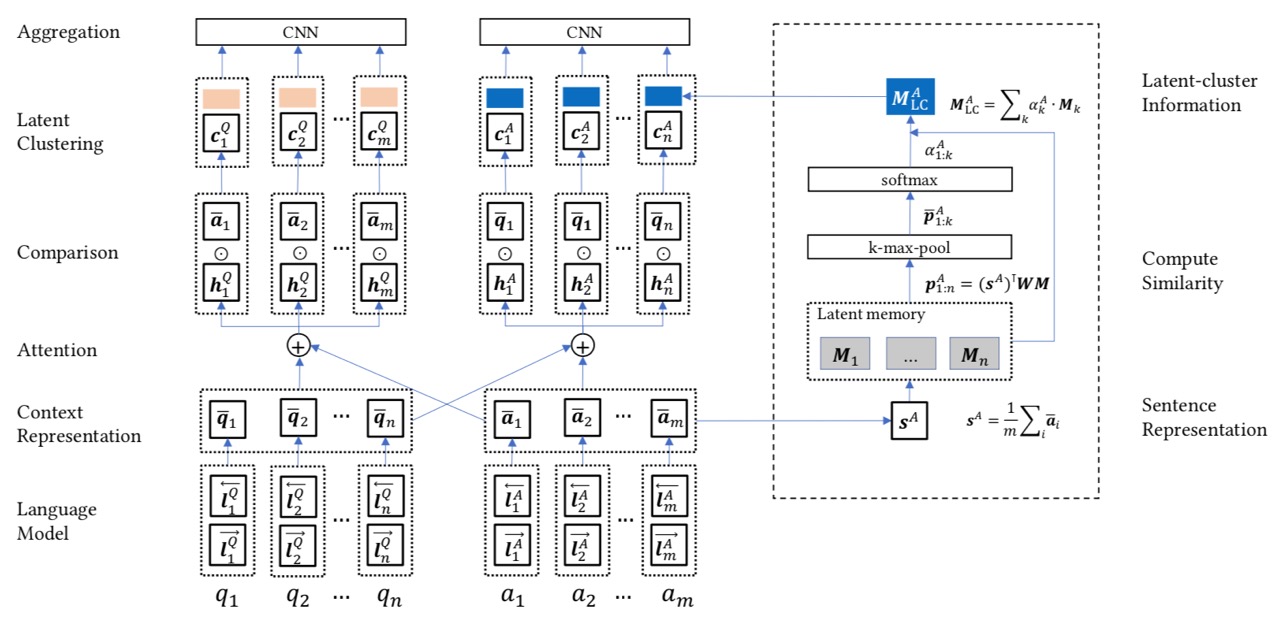}
      \caption{Architecture of CompClip model \citep{Yoon:2019}}
        \label{fig:Yoon-2019}
\end{figure}

\textbf{\citet{Yoon:2019}} proposed CompClip for predicting matching score $(y)$ of given pair of question ($Q={q_{1}, ..., q_{n}}$) and answer ($A={a_{1}, ..., a_{n}}$). For improving the performance of CompClip, they have applied the transfer learning technique by training it on question-answering NLI (QNLI) corpus \citep{wang:2018}. They have also used pointwise learning to rank approach for training this model. The most prominent feature of their work is using the ELMo language model for achieving more meaningful contextual information of question and answer sentences. The architecture of CompClip consists of six layers, as is illustrated in Figure \ref{fig:Yoon-2019}. These layers are described below.

\begin{enumerate}

\item Language model: Instead of using a word embedding layer, the Elmo language model \citep{peters:2018} is used for extracting contextual information of the given sentence in a more efficient way. After applying the ELMo language model, a new representation of question and answer is denoted as $L^{Q}$ and $L^{A}$, respectively.

\item Context representation: This part of model learns the weight $W$ for extracting contextual information of given sentence and generating its contextual representation by following equations:

\begin{equation}
\begin{aligned}
\overline{\mathbf{Q}}=\sigma\left(\mathbf{W}^{i} \mathbf{Q}\right) \odot \tanh \left(\mathbf{W}^{u} \mathbf{Q}\right) \\
\overline{\mathbf{A}}=\sigma\left(\mathbf{W}^{i} \mathbf{A}\right) \odot \tanh \left(\mathbf{W}^{u} \mathbf{A}\right)
\end{aligned}
\end{equation}
after applying Elmo language model, $Q$ and $A$ are replaced by $L^{Q}$ and $L^{A}$, respectively.

\item Attention: Attentional representation of question $H^{Q}$ and answer $H^{A}$ sentences are generated utilizing dynamic-clip attention \citep{Bian:2017} as follows:

\begin{equation}
\begin{aligned}
\mathbf{H}^{Q}=\overline{\mathbf{Q}} \cdot \operatorname{softmax}\left(\left(\mathbf{W}^{q} \overline{\mathbf{Q}}\right)^{\top} \overline{\mathbf{A}}\right)\\
\mathbf{H}^{A}=\overline{\mathbf{A}} \cdot \operatorname{softmax}\left(\left(\mathbf{W}^{a} \overline{\mathbf{A}}\right)^{\top} \overline{\mathbf{Q}}\right)
\end{aligned}
\end{equation}

\item Comparison: Each term form question and answer sentences are compared by element-wise multiplication of question and answer representations with $H^{A}$ and $H^{Q}$, respectively.

\begin{equation}
\begin{aligned}
\mathbf{C}^{Q}=\overline{\mathbf{A}} \odot \mathbf{H}^{Q}, \left(\mathbf{C}^{Q} \in {R}^{l \times A}\right)\\
\mathbf{C}^{A}=\overline{\mathbf{Q}} \odot \mathbf{H}^{A}, \left(\mathbf{C}^{A} \in {R}^{l \times Q}\right)
\end{aligned}
\end{equation}

\item Aggregation layer: For aggregating outputs of comparison layer, a CNN with $n$-types of filters is employed. Aim of this layer is computing the matching score ($score$) between question and answer as follows:

\begin{equation}
\begin{aligned}
\mathbf{R}^{Q}=\mathrm{CNN}\left(\mathbf{C}^{Q}\right), \mathbf{R}^{A}=\mathrm{CNN}\left(\mathbf{C}^{A}\right)
\end{aligned}
\end{equation}
\[ \text { score }=\sigma\left(\left[\mathbf{R}^{Q} ; \mathbf{R}^{A}\right]^{\top} \mathbf{W}\right) \]

\item Latent clustering: In order to improve performance of model, latent clustering information of corpus is used for obtaining cluster information of question and answer sentences. Latent clustering information of sentence $s$ is generated using the following equations: 

\begin{equation}
\begin{aligned}
\mathbf{p}_{1: n}=\mathbf{s}^{\top} \mathbf{W} \mathbf{M}_{1: n}
\end{aligned}
\end{equation}
\[\overline{\mathbf{p}}_{1: k}=k-\max -\operatorname{pool}\left(\mathbf{p}_{1: n}\right)\]
\[\alpha_{1: k}=\operatorname{softmax}\left(\overline{\mathbf{p}}_{1: k}\right)\]
\[\mathbf{M}_{\mathrm{LC}}=\Sigma_{k} \bar{\alpha}_{k} \mathbf{M}_{k}\]

where ${M}_{1: n} \in {R}^{d^{\prime} \times n}$ is latent memory and ${W} \in {R}^{d \times d^{\prime}}$ is parameter of the model. Latent clustering function $f$ is applied on context representation of question and answer sentences and cluster information of question and answer, $M_{LC}^Q$ and $M_{LC}^A$ vectors,  are generated, respectively. $M_{LC}^Q$ and $M_{LC}^A$ are concatenated with $C^{Q}$ and $C^{A}$ which results in generating $C^{Q}_{new}$ and $C^{A}_{new}$ representations, respectively. $C^{Q}_{new}$ and $C^{A}_{new}$ could be considered as input of aggregation layer.

\begin{equation}
\begin{aligned}
\mathbf{M}_{\mathrm{LC}}^{Q}=f\left(\left(\Sigma_{i} \bar{q}_{i}\right) / n\right), \bar{q}_{i} \subset \overline{\mathbf{Q}}_{1: n}
\end{aligned}
\end{equation}
\[\mathbf{M}_{\mathrm{LC}}^{A}=f\left(\left(\Sigma_{i} \bar{a}_{i}\right) / m\right), \bar{a}_{i} \subset \overline{\mathbf{A}}_{1: m}\]
\[\mathbf{C}_{\text {new }}^{Q}=\left[\mathbf{C}^{Q} ; \mathbf{M}_{\mathrm{LC}}^{Q}\right], \mathbf{C}_{\text {new }}^{A}=\left[\mathbf{C}^{A} ; \mathbf{M}_{\mathrm{LC}}^{A}\right] \]

\end{enumerate}


\textbf{\citet{Yang:2019}} presented the RE2 model, which is a text matching model. RE2 is a simple and fast model for extracting more effective features from the two given sequences. The name RE2 comes from augmented features used in this model: residual vectors, embedding vectors, and encoded vectors. RE2 includes two distinct components with shared parameters for processing each of the sequences as well as one prediction layer on top of these two components. The component for processing the given text includes N similar blocks. The architecture of RE2 is shown in Figure \ref{fig:RE2}. Each component of this model is described in the following:

\begin{figure}[bt]
\centering
  \includegraphics[width=0.5\textwidth]{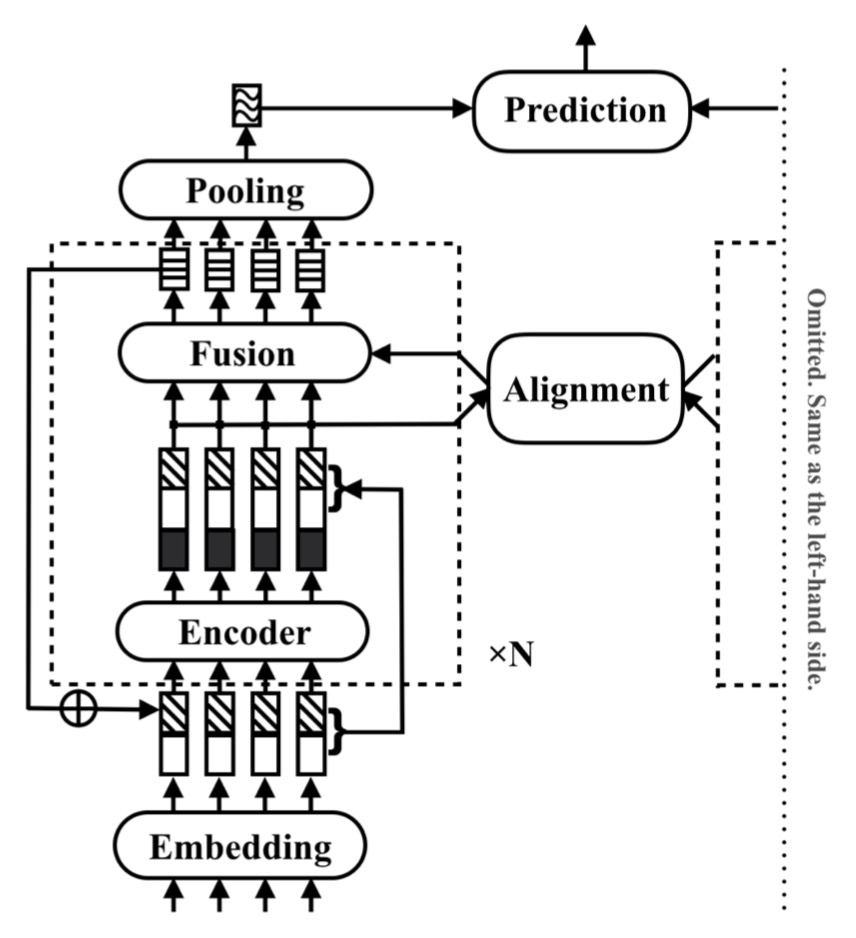}
  \caption{Architecture of the RE2 model\citep{Yang:2019}}
  \label{fig:RE2}
\end{figure}

\begin{enumerate}

\item Augmented Residual Connections: Blocks are connected to each other by residual connections. Input to the first block is word embeddings. The input of other blocks are constructed as follows:

\begin{equation}
\begin{aligned}
x_{i}^{(n)}=\left[x_{i}^{(1)} ; o_{i}^{(n-1)}+o_{i}^{(n-2)}\right]
\end{aligned}
\end{equation}
where $x_{i}^{(n)}$ is the $i$-th token of the input of the $n$-th block for $n>2$, and $o_{i}^{(n)}$ is the $i$-th token of the output of the $n$-th block.
As we see, pure embedding vectors, aligned features which are fed through the previous block, and extracted features from the encoder are passed to the Alignment layer.

\item Alignment Layer: For aligning two text sequences $a$ and $b$, a simple attention mechanism is used for modeling the attentional representation of two given sequences $a^{\prime}$ and $b^{\prime}$.

\begin{equation}
\begin{aligned}
e_{i j}=F\left(a_{i}\right)^{T} F\left(b_{j}\right)\\
\end{aligned}
\end{equation}
\[  a_{i}^{\prime}=\sum_{j=1}^{l_{b}} \frac{\exp \left(e_{i j}\right)}{\sum_{k=1}^{l_{b}} \exp \left(e_{i k}\right)} b_{j} \]
\[ b_{j}^{\prime}=\sum_{i=1}^{l_{a}} \frac{\exp \left(e_{i j}\right)}{\sum_{k=1}^{l_{a}} \exp \left(e_{k j}\right)} a_{i} \]
where F is a feed-forward neural network.

\item Fusion Layer: This layer compares the aligned and simple representations from three perspectives and merges the results.

\begin{equation}
\begin{aligned}
\bar{a}_{i}^{1}=G_{1}\left(\left[a_{i} ; a_{i}^{\prime}\right]\right)\\
 \end{aligned}
\end{equation}
\[  \bar{a}_{i}^{2}=G_{2}\left(\left[a_{i} ; a_{i}-a_{i}^{\prime}\right]\right) \]
\[\bar{a}_{i}^{3}=G_{3}\left(\left[a_{i} ; a_{i} \circ a_{i}^{\prime}\right]\right)  \]
\[ \bar{a}_{i}=G\left(\left[\bar{a}_{i}^{1} ; \bar{a}_{i}^{2} ; \bar{a}_{i}^{3}\right]\right) \]
where $G_{1}$, $G_{2}$, $G_{3}$, and $G$ are single-layer feed-forward networks.

\item Prediction Layer: Output of the two parallel components after deploying a pooling layer is used for creating the input of the multi-layer feed-forward network for predicting the score $y$ as follows: 

\begin{equation}
\begin{aligned}
\hat{\mathbf{y}}=H\left(\left[v_{1} ; v_{2}\right]\right)
 \end{aligned}
\end{equation}
where $v_{1}$ and $v_{2}$ are outputs of two text matching components and $H$ is a multi-layer feed-forward network.

\end{enumerate}

Considering all the described models in this section, a brief overview of deep learning-based models is presented in Table \ref{tab:deep_learning_models}.

\begin{center}
\begin{small}
\begin{longtable}{| l | l | p{5cm} | c |}
\caption{Overview of Deep Learning-based Models (I: Interaction-based, R: Representation-based, H: Hybrid)}
\label{tab:deep_learning_models}\\
\hline\noalign{\smallskip}
	 Model & Name and Type  & Main Idea & Datasets  \\
\noalign{\smallskip}\hline\noalign{\smallskip}
 {\citet{Yu:2014} }           &  R & uses count of co-occurring words as an additional feature, captures more complex semantic features by CNN   & TREC-QA   \\  \hline	 
 {\citet{wang:2015}}        &  R & uses GBDT method for exact matching of the proper nouns and cardinal numbers, models contextual information by using BiLSTM  &  TREC-QA \\  \hline
 {\citet{Severyn:2015}}  &  R & uses CNN for effectively learning the representation of a given sentence, has capability of including any additional similarity features  &  TREC-QA \\  \hline	   	 
 {\citep{Tan:2016}} 	   &  \bigcell{l}{R: QA-LSTM \\ R: Conv-pooling LSTM \\ R: Conv-based LSTM \\ H: Attentive-LSTM} &  uses different combination of BiLSTM and CNN by attention mechanism for representing answer according to question & \bigcell{c}{TREC-QA \\ InsuranceQA}   \\  \hline
 {\citep{Yin:2016}} 	   & \bigcell{l}{R: BCNN \\ H: ABCNN \\ H: ABCNN-2 \\ H: ABCNN-3} & uses attention in CNN with different approaches & WikiQA   \\  \hline
 {\citep{Yang:2016}} 	   & \bigcell{l}{ I: aNMM-1 \\ I: aNMM-2 }& uses attention for estimating the  question term importance   & TREC-QA \\  \hline
 {\citep{Wan:2016b}}	   & I: Match-SRNN  & uses a recursive method for modeling the interaction-matrix  & Yahoo!  \\  \hline
{\citep{Wan:2016a}}	   & H: MV-LSTM & uses BiLSTM for generating a rich model of context, matches different positions of two sentences & Yahoo!   \\  \hline
 {\citep{Wang:2016b}}  & \bigcell{l}{ H: OARNN \\ H: IARNN-WORD \\ H: IARNN-Context \\ H: IARNN-OCCAM \\ H: IABRNN-GATE} & introduces an attention mechanism for mitigating the bias problem, incorporates attention to different positions  of an RNN & \bigcell{c}{TREC-QA \\ WikiQA \\ InsuranceQA } \\  \hline
 {\citep{Tay:2017}}  &  R: HD-LSTM & uses circular correlation for modeling the relationship of question and answer sentences &  \bigcell{c}{ TREC-QA \\ Yahoo!}  \\  \hline
 {\citep{Bian:2017}}	 & H & incorporates dynamic-clip attention for avoiding noise in attentional representation &\bigcell{c}{ TREC-QA \\ WikiQA}  \\  \hline	   
 {\citep{Wang:2017}}	   & H: BiMPM  & encodes question and answer using BiLSTM and matches them in two directions using different matching strategies & \bigcell{c}{TREC-QA \\ WikiQA}  \\  \hline
 {\citep{WangAndJiang:2017}} & H & uses six different comparison functions  & \bigcell{c}{WikiQA \\ InsuranceQA \\ MovieQA} \\  \hline	   
{\citep{Tay:2018}}  & H: MCAN & attaches casted attention to word-level representation for hinting the LSTM & TREC-QA  \\  \hline
{\citep{Yoon:2019}}	   & H: CompClip   & uses pre-trained ELMo language model, transfer learning, and latent clustering & \bigcell{c}{TREC-QA \\ WikiQA}  \\  \hline	  
{\citep{Yang:2019}}	   &  H: RE2 &  a simple and fast text-matching model for extracting  augmented features &  WikiQA \\  \hline	    	   	   	    
 {\citep{Garg:2019}}	   & I: TANDA  & uses pre-trained language models: BERT and RoBERTa, adapts pre-trained model to QA by additional fine-tuning step & \bigcell{c}{ TREC-QA \\ WikiQA}  \\  \hline 	   
 
 \hline
\end{longtable}
\end{small}
\end{center}


\section{Datasets for QA}
\label{sec:datasets}
In this section, we describe five datasets that have been widely used for evaluating the QA tasks.

\begin{enumerate}
\item WikiQA is an open domain QA dataset \citep{Yang:2015}. This dataset is collected from Bing query logs. Question-like queries that are issued by at least 5 different users and have clicked to Wikipedia pages are selected as questions in this dataset and the sentences of the summary section of the corresponding Wikipedia page are considered as candidate answers to the related question. The candidate answers are labeled as correct or incorrect with crowdsourcing and then the correct answer is selected. This dataset consists of 3047 questions and 1473 answers, more statistics about this dataset is presented in Table \ref{tab:WikiQA-Data}. A noted feature of WikiQA is that not all the questions in this dataset have the correct answer which makes it possible to use this dataset in answer triggering component. The answer triggering component's task is finding whether the question has an answer or not.

\item TREC-QA is collected from the Text REtrieval Conference (TREC) 8-13 QA dataset \citep{Wang:2007}. Questions of TREC 8-12 are used as training dataset and questions of TREC 13 are used as development and test dataset. Statistics of TREC-QA dataset are presented in Table \ref{tab:TRECQA-Data}. TREC-QA contains two training datasets: TRAIN, and TRAIN-ALL. TRAIN dataset includes the first 94 questions of TREC 8-12 and its candidate answers are judged manually, while in the TRAIN-ALL dataset, correct answers are recognized by matching the answers with predefined patterns of the answer regular expression. 

\begin{table}[h]
\begin{center}
\caption{Statistics of the WikiQA Dataset}
\label{tab:WikiQA-Data}
\begin{tabular}{ l l l l l }

\hline\noalign{\smallskip}
& Train &  Validation & Test & Total \\
\noalign{\smallskip}\hline\noalign{\smallskip}

\# of questions & 2118 & 296 & 633 & 3047 \\
\# of sentences & 20360 & 2733 & 6165 & 29258 \\
\# of answers & 1040 & 140 & 293 & 1473 \\
Average length of questions & 7.16 & 7.23 & 7.26 & 7.18 \\
Average length of sentences & 25.29 & 24.59 & 24.95 & 25.15 \\
\hline
\# of questions w/o answers & 1245 & 170 & 390 & 1805 \\
\hline
\end{tabular}
\end{center}
\end{table}


\begin{table}[h]
\begin{center}
\caption{Statistics of the TREC-QA Dataset}
\label{tab:TRECQA-Data}

\begin{tabular}{ l l l l l  }
\hline
 &  Train-all & Train & Validation & Test  \\
\hline
\# Questions     & 1229        & 94            & 82            &100     \\
\# QA pairs       & 53417      & 4718        &1148          &  1517   \\
 \% correct        & 12.00\%   &7.40\%     & 19.30\%    & 18.70\%     \\
 \#Answers/Q   & 43.46         &50.19      & 14.00    & 15.17     \\
 judgement      & automatic  & manual & manual  & manual\\ 
 \hline
\end{tabular}
\end{center}
\end{table}

\item MovieQA is gathered from diverse data sources \citep{Tapaswi:2016} which is a unique feature of this dataset. It contains 14944 questions where each question is associated with five answers, including one correct answer and four deceiving answers. Statistics of this dataset is shown in Table \ref{tab:MovieQA-Data}.

\begin{table}[bt]
\begin{center}
\caption{Statistics of the MovieQA Dataset}
\label{tab:MovieQA-Data}
\begin{tabular}{l l l l l  }
\hline\noalign{\smallskip}
& Train &  Validation & Test & Total \\
\noalign{\smallskip}\hline\noalign{\smallskip}

Movies with Plots and Subtitles & & & &\\
\hline
\# of Movies & 269 & 56 & 83 & 408 \\
\# of QA & 9848 & 1958 & 3138 & 14944\\
Q \# words & 9.3 & 9.3 & 9.5 & 9.3 $\pm$ 3.5 \\
CA. \# words & 5.7 & 5.4 & 5.4 &5.6 $\pm$ 4.1 \\
WA. \# words & 5.2 & 5.0 & 5.1 & 5.1 $\pm$ 3.9\\
\hline
\end{tabular}
\end{center}
\end{table}

\item InsuranceQA is a close domain dataset for QA in the insurance domain \citep{Feng:2015}. Question/answer pairs of this dataset are collected from the internet. This dataset includes Train, Development, Test1, and Test2 parts. More detailed statistics about InsuranceQA is presented in Table \ref{tab:InsuranceQA-Data}.

\begin{table}[bt]
\begin{center}
\caption{Statistics of the InsuranceQA Dataset}
\label{tab:InsuranceQA-Data}
\begin{tabular}{l  l  l  l l}

\hline\noalign{\smallskip}
                                     &Train     &  Dev       & Test1      & Test2  \\
\noalign{\smallskip}\hline\noalign{\smallskip}
Questions                     & 12887   &1000   & 1800         & 1800\\
Answers                       &18540     &1454   & 2616      & 2593\\
Question Word Count  &92095     &7158    & 12893     &12905\\
                                           
\hline
\end{tabular}
\end{center}
\end{table}

\item Yahoo! Dataset is collected from Yahoo! Answers QA system. Yahoo! includes 142,627 question/answer pairs. But in the literature \citep{Wan:2016b, Wan:2016a} a subset of this dataset is selected as positive pair and for each question in this subset, four other negative pairs are constructed. The (question, answer) pairs in which the question and its answer length are between 5 to 50 words are selected as positive pairs (including 60564 pairs). Four negative answers for each question are selected by querying the whole answers set by the correct answer and selecting the four answers randomly among 1000 top retrieved answers.
\end{enumerate}


\section{Evaluation}
For evaluation of QA, three metrics are used: Mean Reciprocal Rank (MRR), Mean Average Precision (MAP), and accuracy. MRR indicates the ability of the system to answer a question. Reciprocal Rank (RR) for one question is inverse of the rank of the first correct answer, if there exists any correct answer, or zero if there exists no correct answer. RR of each question is computed and then the average of RRs is considered as MRR.

\begin{equation}
\mathrm{MRR}=\frac{1}{|Q|} \sum_{i=1}^{|Q|} \frac{1}{\operatorname{rank}_{i}}
\end{equation}
\[\mathrm{rank}_{i}= \left\{\begin{array}{c}{\frac{1}{\text {the rank of the first correct answer}}} \\ {\text { (if a correct answer exists) }} \\ {\text { 0. }} \\ {\text { (if no carrect answer) }}\end{array}\right.\]

The second metric is the MAP. MAP is the mean of the average of the precisions at each rank when a correct answer is detected \citep{Diefenbach:2018}. Precision@k is the number of correct answers among the first $k$ retrieved answers divided by $k$. Precision indicates how many of the answers are correct and is calculated as follows:

\begin{equation}
\mathrm{MAP}=\frac{1}{Q} \sum_{q=1}^{Q} {AP(q)}
\end{equation}
\[\mathrm{AP(q)}=\frac{1}{k} * \sum_{k=1}^{K} {p @ k}\]
\[\mathrm{Precision@k}=\frac{\text { number of correct answers among first } k \text { results}}{k}\]
where $k$ is the rank of correctly retrieved answers.

Accuracy is used for evaluating datasets whose questions have just one correct answer or one label. Accuracy indicates how many of the questions are answered correctly.
\begin{equation}
\mathrm{Accuracy}=\frac{\text {number of correct answered questions}}{\text{number of questions}}
\end{equation}


\section{Available Results from the Literature}

As mentioned in Section \ref{sec:datasets}, WikiQA \citep{Yang:2015}, TREC-QA \citep{Wang:2007}, MovieQA \citep{Tapaswi:2016}, InsuranceQA \citep{Feng:2015}, and Yahoo! \citep{Wan:2016b, Wan:2016a} are the main datasets that have been used for evaluating text-based QA systems. In this section, we report the results of different models on the mentioned datasets to have a comparison of the quality of the state-of-the-art models. It has to be mentioned that considering a large number of models reviewed in this paper, it is not possible to reimplement all of them to have a comprehensive comparison and we only report the results based upon their availability.

\begin{center}
\begin{small}
\begin{longtable}{| l | c | c | c |}
\caption{Results of Different Models on the WikiQA Dataset.}
\label{tab:results-WikiQA}\\
\hline\noalign{\smallskip}
Model & Setting & MAP & MRR \\
\noalign{\smallskip}\hline\noalign{\smallskip}
\hline
 \hiderowcolors
\citet{Yih:2013} & LCLR & 0.5993  & 0.6086  \\ \hline
\citet{LeAndMikolov:2014} & PV & 0.5110  & 0.5160  \\ \hline

\multirow{4}{*}{\citet{Yang:2015}} & Word Count & 0.5707 & 0.6266 \\ 
 & Wgt Word Count & 0.5961 & 0.6515 \\ 
 & PV-Cnt & 0.5976 & 0.6058 \\
 & CNN-Cnt & 0.6520 & 0.6652 \\ \hline
 \citet{He:2016} &   &  0.6930 &  0.7090 \\ \hline
\citet{Miao:2016} & &  0.689 & 0.707 \\ \hline
\citet{Wang:2016d} & & 0.706 & 0.723 \\ \hline
\citet{Rao:2016} & & 0.701 &  0.718 \\ \hline
 \hline
  \textbf{R: \citet{Yu:2014}} & CNN & 0.6190 & 0.6281 \\ \hline
\multirow{ 2}{*}{\textbf{R: \citet{Yin:2016}}} & BCNN, one-conv & 0.6629 & 0.6813 \\
 & BCNN, two-conv & 0.6593 & 0.6738 \\ \hline
\hline

\multirow{4}{*}
{\textbf{I: \citet{Garg:2019}}} & TANDA (BERT-b)          & 0.893   & 0.903 \\
  					    & TANDA (BERT-L)          & 0.903   & 0.912 \\ 
   					    & TANDA (RoBERTa-B)   & 0.889   & 0.901 \\
    					    & TANDA (RoBERTa-L)    & \textbf{0.920}   & \textbf{0.933} \\ \hline
\hline
\multirow{7}{*}{\textbf{H: \citet{Wang:2016b}}} & IARNN-word & 0.7098 & 0.7234 \\
 & IARNN-Occam(word) & 0.7121 & 0.7318 \\
 & IARNN-context & 0.7182 & 0.7339 \\
 & IARNN-Occam(context) & 0.7341 & 0.7418 \\
 &  IABRNN-GATE & 0.7258 &  0.7394\\
 &  GRU & 0.6581 & 0.6691 \\
 & OARNN & 0.6881 & 0.7013 \\ \hline
\multirow{6}{*}{\textbf{H: \citet{Yin:2016}}} & ABCNN-1, one-conv & 0.6810 & 0.6979  \\
 & ABCNN-1, two-conv & 0.6855 & 0.7023 \\
 & ABCNN-2, one-conv & 0.6885 & 0.7054 \\
 & ABCNN-2, two-conv & 0.6879  & 0.7068  \\
 & ABCNN-3, one-conv & 0.6914 & 0.7127 \\
 & ABCNN-3, two-conv & 0.6921 &0.7108  \\ \hline
 \textbf{H: \citet{HeAndLin:2016}} &  & 0.7090 & 0.7234 \\ \hline
\multirow{3}{*}{\textbf{H: \citet{Bian:2017}}} & listwise & 0.746  & 0.759 \\
 & with $k$-max & 0.754 & 0.764 \\
 & with $k$-threshold & 0.753  & 0.764 \\ \hline
\textbf{H: \citet{Wang:2017}} & BiMPM & 0.718 & 0.731 \\ \hline
\multirow{6}{*}{\textbf{H: \citet{WangAndJiang:2017}}} & NN & 0.7102 & 0.7224 \\
 &  NTN & 0.7349 & 0.7456  \\
 & EucCos &  0.6740 & 0.6882  \\
 & Sub & 0.7019 & 0.7151 \\
 & Mult & 0.7433 & 0.7545 \\
 & SUBMULT+NN & 0.7332  & 0.7477 \\ \hline
\multirow{4}{*}{\textbf{H: \citet{Yoon:2019}}} & Comp-Clip & 0.714 & 0.732  \\
 &  Comp-Clip + LM & 0.746 & 0.762  \\
 & Comp-Clip + LM + LC &  0.764 & 0.784  \\
 & Comp-Clip + LM + LC +TL & 0.834  & 0.848 \\ \hline
 \textbf{H: \citet{Yang:2019}} & RE2 & 0.7452 & 0.7618 \\ \hline
\end{longtable}
\end{small}
\end{center}
\vspace{-3mm}


Table \ref{tab:results-WikiQA} reports the results of different representation-based (\textbf{R}), interaction-based (\textbf{I}), and hybrid (\textbf{H}) methods on the WikiQA dataset and compares it with other baseline models. 
As can be seen TANDA \citep{Garg:2019} achieved the best results over all methods on the WikiQA dataset in MRR and MAP metrics. Although this interaction-based technique is the best state-of-the-art model in the field, comparing the rest of models, we can see that the general performance of hybrid models is better than interaction-based models and the next best results are all from the hybrid models.

Table \ref{tab:results-TREC-QA} reports the results of different models which are described in Section 6 as well as their baseline methods on the TREC-QA dataset. According to this table, the TANDA \citep{Garg:2019} model achieved the best result in the MAP.

\vspace{5mm}
\begin{center}
\begin{small}
\begin{longtable}{| l | c | c | c |}
\caption{Results of Different Models on TREC-QA Dataset.}
\label{tab:results-TREC-QA}\\

\hline\noalign{\smallskip}
Model & Setting & MAP & MRR \\
\noalign{\smallskip}\hline\noalign{\smallskip}
\hline
 \hiderowcolors
\citet{Cui:2005} & & 0.4271 & 0.5259 \\ \hline
\citet{Wang:2007} &  & 0.6029  & 0.6852  \\  \hline
\citet{HeilmanAndSmith:2010} &  & 0.6091 & 0.6917  \\  \hline
\citet{WangAndManning:2010} &  & 0.5951 & 0.6951 \\ \hline
\citet{Yao:2013} &  & 0.6307 & 0.7477  \\ \hline
\citet{Feng:2015} & Architecture-II & 0.711 & 0.800\\ \hline
\citet{Rao:2016} & & 0.801 & 0.877 \\ \hline
\hline 
\multirow{3}{*}{\textbf{R:{\citet{Yih:2013}}}} & LR & 0.6818 & 0.7616  \\ 
 & BDT & 0.6940 & 0.7894  \\
& LCLR & 0.7092  & 0.7700  \\ \hline

\multirow{8}{*}{\textbf{R:{\citet{Yu:2014}}}} & TRAIN bigram + count  & 0.7058 & 0.7800 \\ 
& TRAIN-ALL bigram + count  & 0.7113  & 0.7846 \\
& TRAIN unigram + count & 0.6889 & 0.7727 \\
& TRAIN-ALL unigram + count  & 0.6934 & 0.7677 \\
& TRAIN unigram  & 0.5387 & 0.6284 \\
& TRAIN-ALL unigram & 0.5470 & 0.6329 \\
& TRAIN bigram  & 0.5476 & 0.5476 \\
& TRAIN-ALL bigram  & 0.5693 & 0.6613 \\ \hline

\multirow{ 2}{*}{\textbf{R:{\citet{Yang:2015}} }}& Word Count & 0.5707 & 0.6266 \\ 
& Wgt Word Count & 0.5961 & 0.6515 \\ \hline

\multirow{2}{*}{\textbf{R:{\citet{Severyn:2015}}}} & TRAIN & 0.7329 & 0.7962 \\ 
& TRAIN-ALL & 0.7459 & 0.8078 \\ \hline

\multirow{ 5}{*}{\textbf{R:{\citet{wang:2015}}}} & BM25 & 0.6370 & 0.7076 \\
& Single-Layer LSTM & 0.5302 & 0.5956 \\
& Single-Layer BiLSTM& 0.5636 & 0.6304 \\
& Three-Layer BiLSTM & 0.5928 & 0.6721 \\
&  Three-Layer BiLSTM + BM25 & 0.7134 & 0.7913 \\ \hline

\multirow{ 6}{*}{\textbf{R:{\citet{Tan:2016}}}} & QA-CNN & 0.714 & 0.807 \\ 
& QA-LSTM (max-pooling) & 0.733 & 0.819  \\
& Conv-pooling LSTM & 0.742 & 0.819 \\
& Conv-based LSTM & 0.737 & 0.827  \\
& HD-LSTM TRAIN  & 0.7520 & 0.8146 \\
& HD-LSTM TRAIN-ALL  & 0.7499 & 0.8153 \\ \hline \hline

{\textbf{I:{\citet{Yang:2016}}}} & aNMM, TRAIN-ALL  & 0.7495 & 0.8109 \\ \hline 
 
 \multirow{4}{*}
{\textbf{I: \citet{Garg:2019}}} & TANDA (BERT-b)          & 0.912   & 0.951 \\
  					    & TANDA (BERT-L)          & 0.912   & 0.967 \\ 
   					    & TANDA (RoBERTa-B)   & 0.914   & 0.952 \\
    					    & TANDA (RoBERTa-L)    & \textbf{0.943}   & \textbf{0.974} \\ \hline \hline
 
 {\textbf{H:{\citet{HeAndLin:2016}}}} &  & 0.7588 & 0.8219 \\ \hline 
{\textbf{H:{\citet{Tan:2016}}}} & Attentive LSTM  & 0.753 & 0.830 \\ \hline

\multirow{ 7}{*}{\textbf{H:{\citet{Wang:2016b}}}} & IARNN-word & 0.7098 & 0.7757 \\
& IARNN-Occam(word) & 0.7162 & 0.7916 \\
& IARNN-context & 0.7232 & 0.8069 \\
& IARNN-Occam(context) & 0.7272 & 0.8191\\
& IABRNN-GATE & 0.7369 & 0.8208 \\
& GRU & 0.6487 & 0.6991 \\
& OARNN & 0.6887 & 0.7491 \\ \hline
 
\multirow{3}{*}{\textbf{H:{\citet{Bian:2017}}}} & listwise & 0.810 & 0.889 \\
 & with $k$-max & 0.817 & 0.895\\
 & with $k$-threshold & 0.821 & 0.899 \\ \hline
  
{\textbf{H:{\citet{Wang:2017}}}} & BiMPM & 0.802 & 0.875 \\ \hline

\multirow{3}{*}{\textbf{H:\citet{Tay:2018}}} & MCAN(SM) & 0.827 & 0.880 \\
& MCAN(NN) & 0.827 & 0.890\\
& MCAN(FM) & 0.838 & 0.904\\ \hline
 
 \multirow{4}{*}{\textbf{H: \citet{Yoon:2019}}} & Comp-Clip & 0.835 & 0.877  \\
 &  Comp-Clip + LM & 0.850 & 0.898  \\
 & Comp-Clip + LM + LC &  0.868 & 0.928  \\
 & Comp-Clip + LM + LC +TL & 0.875  & 0.940 \\ \hline
 
\hline

\end{longtable}
\end{small}
\end{center}


\vspace{-5mm}
\begin{table}[!h]
\caption{Results of Different Models on Yahoo! Dataset}
\label{tab:Results-Yahoo}
\begin{center}
\begin{small}
\begin{tabular}{| l | c | c | c |}
\hline\noalign{\smallskip}
Model & Setting & P@1 & MRR \\
\noalign{\smallskip}\hline\noalign{\smallskip}
\hline
 \hiderowcolors
Random Guess &  & 0.2 & 0.4570\\ \hline
Okapi BM-25 &  & 0.2250 & 0.4927 \\ \hline
CNN & & 0.4125 & 0.6323  \\ \hline
CNTN &  &0.4654 & 0.6687  \\ \hline
LSTM & & 0.4875 & 0.6829 \\  \hline
NTN-LSTM & & 0.5448 &  0.7309 \\ \hline
\citet{Robertson:1995} &BM25  & 0.579 & 0.726 \\  \hline
\citet{Socher:2011}& RAE & 0.398 & 0.652\\  \hline
\multirow{3}{*}{\citet{Hu:2014}} & ARC-1 & 0.581 & 0.756  \\  
&ARC-2 & 0.766 & 0.869 \\ 
& Deep Match & 0.452 & 0.679 \\  \hline
\citet{Qiu:2015} &CNTN   & 0.626 & 0.781 \\  \hline
\citet{yinAndSchutze:2015} &MultiGranCNN   & 0.725 & 0.840 \\  \hline
\citet{Palangi:2016} &LSTM-RNN  & 0.690 & 0.822\\  \hline
\citet{Pang:2016} &MatchPyramid-Tensor & 0.764 & 0.867\\  \hline
\hline

{\textbf{R:\citet{Tay:2017}}} & HD-LSTM & 0.5569 & 0.7347 \\  \hline
\hline

\multirow{2}{*}{\textbf{I:\citet{Wan:2016b}}} &Match-SRNN & 0.785 & \textbf{0.879} \\
 & Bi-Match-SRNN & \textbf{0.790} & 0.882\\  \hline
\hline

\multirow{3}{*}{\textbf{H:\citet{Wan:2016a}}} & MV-LSTM-Cosine & 0.739 & 0.852 \\
& MV-LSTM-Bilinear & 0.751  & 0.860 \\
&MV-LSTM-Tensor & 0.766 & 0.869 \\  \hline

\hline
\end{tabular}
\end{small}
\end{center}
\end{table}


\begin{table}[!h]
\caption{Results of Different Models on Insurance-QA Dataset}
\label{tab:Results-Insurance-QA}
\begin{center}
\begin{small}
\begin{tabular}{| l | c | c | c |}

\hline\noalign{\smallskip}
	 Model & Setting  & TEST1 & TEST2  \\
\noalign{\smallskip}\hline\noalign{\smallskip}

\hline
	\hiderowcolors
	 Bag-of-word & & 0.321 & 0.322 \\  \hline
	 Metzler-Bendersky IR model &  & 0.551 & 0.508 \\  \hline
	 \multirow{2}{*}{\cite{Feng:2015}} & CNN  & 0.628 & 0.592\\ 
	  & CNN with GESD  & 0.653 & 0.610 \\  \hline
	 \multirow{10}{*}{\textbf{R:{\cite{Tan:2016}}}}& QA-LSTM (head/tail)  & 0.536 & 0.510 \\  
	  & QA-LSTM (avg pooling,$k$=50)  & 0.557 & 0.524\\  
	  & QA-LSTM (max pooling,$k$=1)  & 0.631 & 0.580\\  
	  & QA-LSTM (max pooling,$k$=50)  & 0.666 & 0.637\\  
	 & Conv-pooling LSTM ($c$=4000,$k$=1) & 0.646 & 0.622 \\	  
	  & Conv-pooling LSTM ($c$=200,$k$=1) & 0.674 & 0.635 \\  
	& Conv-pooling LSTM ($c$=400,$k$=50) & 0.675 & 0.644\\  
	 & Conv-based LSTM ($|h|$=200,$k$=50) & 0.661 & 0.630 \\  
	 & Conv-based LSTM ($|h|$=400,$k$=50) & 0.676 & 0.644 \\  
	& QA-CNN (max-pooling, $k$ = 3) & 0.622 & 0.579 \\  \hline \hline
	
	 \multirow{4}{*}{\textbf{H:{ \cite{Tan:2016}}}}& Attentive CNN (max-pooling, $k$ = 3) & 0.633 & 0.602\\
	 & Attentive LSTM (avg-pooling $k$=1) & 0.681 & 0.622\\
	 & Attentive LSTM (avg-pooling $k$=50) & 0.678 & 0.632 \\
	 & Attentive LSTM (max-pooling $k$=50) & 0.690 & 0.648\\ \hline
	 
	 \multirow{7}{*}{\textbf{H:{\cite{Wang:2016b}}}} & IARNN-word  &  0.671 & 0.616  \\
	& IARNN-Occam (word)  & 0.696 & 0.637 \\
	 & IARNN-context  & 0.667 &  0.631\\
	 & IARNN-Occam (context)  & 0.689 & 0.651  \\
	 & IABRNN-GATE  & 0.701 & 0.628 \\
	 & GRU  & 0.532 & 0.581\\
	 & OARNN  & 0.661 & 0.602  \\ \hline

	 \multirow{6}{*}{\textbf{H:{ \cite{WangAndJiang:2017}}}}& NN  & 0.749 & 0.724 \\
	 & NTN  & 0.750 & 0.725\\
	 & EucCos  & 0.702 & 0.679 \\
	& Sub  & 0.713 & 0.682 \\
	 & Mult  & 0.752 &  \textbf{0.734} \\
	 & SUBMULT+NN &  \textbf{0.756} & 0.723 \\ \hline

	\hline
\end{tabular}
\end{small}
\end{center}
\end{table}


Performance of different models from Section 6 and their baselines on Yahoo! dataset is reported in Table \ref{tab:Results-Yahoo}. One representation-based model \citep{Tay:2017}, one interaction-based models \citep{Wan:2016b} , and one hybrid model \citep{Wan:2016a} are evaluated on the Yahoo! dataset.
As can be seen, Bi-Match-SRNN, proposed by \citet{Wan:2016b} outperforms other models in P@1 and MRR metrics. This is the only dataset in which an interaction-based model, namely Bi-Match-SRNN, performs better than the hybrid model. 


Table \ref{tab:Results-Insurance-QA} reports the results of different models on the Insurance-QA dataset. Among three categories of deep models (representation-based, interaction-based, and hybrid models) hybrid models achieves the best accuracy on the Insurance-QA dataset too. The proposed model by \cite{WangAndJiang:2017} with SUBMULT+NN comparison function on TEST1, and with Mult comparison function on TEST2 achieves the best accuracy.



\section{Discussion}

Using deep neural networks in QA eliminated the manual feature engineering. In representation-based models, contextual representation of each sentence is modeled separately and then they are compared. 

\citet{Yu:2014} and \citet{wang:2015} used distributional semantic model for generating a contextualized representation for text. \citet{Yu:2014} applied CNN for capturing local dependencies among phrases of a sentence. Average pooling over CNN helps to create a representation of meaning the whole sentence. CNNs are capable of capturing complex semantics of a sentence in the given window. \citet{wang:2015} utilized BiLSTM for creating a representation aware of the longer dependencies. 

Matching proper nouns and cardinal numbers in question and answer sentences is an important issue. That means if the proper nouns in two sentences do not match, it will be enough reason for rejecting the answer. But, using pre-trained word embeddings not only does not distinguish these names but also considers a close representation for them. For mitigating this problem, \citet{Yu:2014} used count of co-occurring words as a feature, and \citet{wang:2015} used GBDT for exact matching of question and answer sentences. 

\citet{Yin:2016} and \citet{Severyn:2015} used CNNs in their architecture. \citet{Severyn:2015} used CNN for embedding the input text and proposed an architecture capable of adding other additional features. \citet{Yin:2016} used wide convolution. \citet{Tay:2017} used holographic composition for better modeling of question and answer relation. They enriched the aggregation of two embeddings by using a circular correlation. \citet{Tan:2016} used different combinations of CNN and BiLSTM in different models. They used CNN on the top of BiLSTM for capturing richer information from the text and used BiLSTM on the top of CNN for capturing the long-range dependencies from local $n$-grams extracted by CNN. Using a combination of CNN and BiLSTM helps to mitigate the weakness of CNNs in capturing similarity between long dependencies in given texts. 

Representation-based models build an embedding for each sentence separately across a distinct component. Although they are simple and usually use shared parameters across two separate components, they are not able to capture the matching between each pair of tokens from question and answer sentences. 
Interaction-based models solve this problem by direct interaction between each term of given sentences. \citet{Yang:2016} generated a matching matrix by comparing each token of the question with each token of the answer and used attention for specifying the importance level of each question term. \citet{Wan:2016b} proposed a special recursive model for modeling the interaction between two sequences by using GRU. 

Using pre-trained language models like \citet {Peters:2017},  \citet{Radford:2018}, \citet{Devlin:2019} has been recently attracted the researchers and made significance advances in many downstream NLP tasks. Pre-trained language models have also been applied to the QA task including the current state-of-the-art models. TANDA is one of the text-based QA models which use Bert and Roberta for modeling the contextual relation between question and answer sentence. As transformers are used in these pre-trained models for modeling the contextual representation of each given token by using the multi-head attention mechanism, we classify these models as interaction-based models. 

\citet{Wan:2016a} and \citet{HeAndLin:2016} used BiLSTM for creating representation of each sentence and then built matching matrix on the top the new representation for each token. \citet{HeAndLin:2016} generated the interaction matrix by concatenating the different similarity measures. They utilized CNN for finding strong interactions among a pair of words. In their model, the interaction matrix is built by using three different similarity metrics for comparing each pair of tokens, and similarity focus are used for assigning weight to interactions based on their level of importance. \citet{Wan:2016a} captured rich contextualized local information of each token by using a BiLSTM and used three distinct similarity metrics for building an interaction tensor. 

Hybrid models combine both interaction and representation models. They consist of a representation component that combines a sequence of words into a fixed $d$-dimensional representation and an interaction component. The attention mechanism is used in most of the hybrid models for generating a richer representation for answer or question sentences by attending to the other sentence. 

We consider the attention mechanism as a type of interaction component because the attention weight for each token is calculated based on interacting with the two sentences. \citet{Tan:2016} used attention mechanism for generating a question aware representation for answer sentence in their proposed Attentive-LSTM model. They used output of LSTM for question sentence, for weighting each of the answer sentence tokens. The last hidden state of RNNs carries more information about the sentence and this biases the attention toward the later hidden states. 
\citet{Wang:2016b} utilized the attention mechanism in the word, the context, and the gate level of an RNN. \citet{Yin:2016} proposed the first model which incorporated the attention mechanism in CNN. 

Models proposed by \citet{Bian:2017}, \citet{Wang:2017}, \citet{WangAndJiang:2017}, \citet{Yoon:2019}, and \citet{Yang:2019} follow a compare-aggregate architecture. Usually, there is a fewer number of relevant words in question and answer sentences, and summing the small attention weight of these irrelevant tokens affects the impact of relevant tokens. For mitigating this problem, \citet{Bian:2017} introduced a new attention mechanism and adopted list-wise ranking instead of point-wise which better fits the nature of ranking. 

\citet{Wang:2017} matches the question and the answer sentences in two directions and from multiple perspectives. \citet{WangAndJiang:2017} used different comparison functions for matching the answer representation with an attentional representation of answer from question perspective. 

\citet{Yoon:2019} leveraged the performance of CompClip by using transfer learning. They also used the ELMo language model for capturing more meaningful contextual information from question and answer sentences.
\citet{Tay:2018} generates a new embedding for each word and used the concatenation of the new embedding and the main word embedding. The new word embedding is built by casting the co-attention multiple times. On the other words, it casts attention instead of using it as a pooling strategy. 
RE2 \citep{Yang:2019} extracts more informative features from two sentences by using a simple attention mechanism for aligning two sentences and using augmented residual connections. RE2 has a very simple and fast architecture by using the minimum number of parameters and does not use RNNs due to its slow speed. 

We discussed the main features of each model and found that currently, the state-of-the-art techniques include models that have used pre-trained language models like BERT. At the same time, representation models had the simplest architecture, and interaction models could extract richer information from the semantic relation of two sentences. Hybrid models, which are a combination of these two types of models, usually use the attention mechanism. The attention mechanism helps to create a better representation based on another sentence. Most of these models followed the compare-aggregate architecture. The attention mechanism has had a significant impact on QA. In the architecture of BERT also the multi-head attention is used. Using this mechanism and pre-training it with a large size dataset provides a powerful model for extracting rich semantic representation from the text. 


\section{Conclusion}
In this paper, we provided a comprehensive review of the state-of-the-art methods on text-based QA systems. We first introduced the general architecture of QA systems, and then proposed a categorization for existing publications in the field. In the first step, publications are divided into two classes: information retrieval-based techniques, and deep learning-based techniques. We reviewed the main methods from both categories and highlighted deep learning-based approaches in more detail by following the well-know categorization for neural text matching, namely representation-based, interaction-based, and hybrid models. The existing publications with the deep learning perspective are categorized in these classes.
We also reviewed available datasets that are widely used for training, validating, and testing text-based QA methods. The available results from different techniques on these datasets are presented in the paper to have a naive comparison of the techniques.



\bibliographystyle{spbasic}      
\bibliography{references}   


\begin{biography}[photo-momtazi]{S.~Momtazi}
is currently an assistant professor at the Amirkabir University of Technology, Iran. She completed her BSc and MSc education at the Sharif University of Technology, Iran. She received a Ph.D. degree in Artificial Intelligence from Saarland University, Germany. As part of her Ph.D., she was a visiting researcher at the Center of Language and Speech Processing at Johns Hopkins University, US. After finishing the Ph.D., she worked at the Hasso-Plattner Institute (HPI) at Potsdam University, Germany and the German Institute for International Educational Research (DIPF), Germany as a post-doctoral researcher. Natural language processing with a focus on question answering systems is her main research focus. She has worked in this area of research for more than 14 years.
\end{biography}

\begin{biography}[photo-abbasian]{Z.~Abbasiantaeb}
is a graduate student of Artificial Intelligence at the Amirkabir University of Technology, Iran. She received her B.Sc. degree in Software Engineering from the Amirkabir University of Technology, Iran. Natural language processing and information retrieval are her main research interests. Currently, she is working as a research assistant at NLP Lab under the supervision of Dr. Saeedeh Momtazi.  
\end{biography}

\graphicalabstract{./img/Taxonomy-text}{
Text-based Question Answering (QA) has been widely studied in Information Retrieval (IR) communities. By the advent of Deep Learning (DL) techniques, various DL-based methods have been used for this task which have not been studied and compared well. In this paper, we provide a comprehensive overview of different models proposed for QA, including both traditional IR perspective, and more recent DL perspective. 
}

\end{document}